\documentclass{aa}  

\usepackage{graphicx}
\usepackage{longtable}
\usepackage{txfonts}
\usepackage{amsmath,amssymb}
\begin{document}

   \title{Properties of quasi-periodic pulsations in solar flares from a single active region}


   \author{C. E. Pugh\inst{1}
          \and
          V. M. Nakariakov\inst{1,2}
          \and
          A.-M. Broomhall\inst{1,3}
          \and
          A. V. Bogomolov\inst{4}
          \and
          I. N. Myagkova\inst{4}
          }

   \institute{Department of Physics, University of Warwick,
              Coventry, CV4 7AL, UK\\
              \email{c.e.pugh@warwick.ac.uk}
         \and
             St. Petersburg Branch, Special Astrophysical Observatory, 
             Russian Academy of Sciences, 196140, St. Petersburg, Russia
         \and
             Institute of Advanced Study, University of Warwick, 
             Coventry, CV4 7HS, UK
         \and
             Skobeltsyn Institute of Nuclear Physics, Lomonosov Moscow State University, 119991, 
             Moscow, Russia
             }

   \date{Received September 15, 1996; accepted March 16, 1997}

 
  \abstract
   {Quasi-periodic pulsations (QPPs) are a common feature of solar and stellar flares, and so the nature of these pulsations should be understood in order to fully understand flares.}
   {We investigate the properties of a set of solar flares originating from a single active region that exhibit QPPs, and in particular look for any indication of the QPP periods relating to active region properties (namely photospheric area, bipole separation distance, and average magnetic field strength at the photosphere), as might be expected if the characteristic timescale of the pulsations corresponds to a characteristic length scale of the structure from which the pulsations originate. The active region studied, known as NOAA 12172/12192/12209, was unusually long-lived and persisted for over three Carrington rotations between September and November 2014. During this time a total of 181 flares were observed by GOES.}
   {Data from the GOES/XRS, SDO/EVE/ESP, \emph{Fermi}/GBM, \emph{Vernov}/DRGE and \emph{Nobeyama Radioheliograph} observatories were used to determine if QPPs were present in the flares. For the soft X-ray GOES/XRS and EVE/ESP data, the time derivative of the signal was used so that any variability in the impulsive phase of the flare was emphasised. Periodogram power spectra of the time series data (without any form of detrending) were inspected, and flares with a peak above the 95\% confidence level in the power spectrum were labelled as having candidate QPPs. The confidence levels were determined taking full account of data uncertainties and the possible presence of red noise. Active region properties were determined using SDO/HMI line of sight magnetogram data.}
   {A total of 37 flares (20\% of the sample) show good evidence of having stationary or weakly non-stationary QPPs, and some of the pulsations can be seen in data from multiple instruments and in different wavebands. Because the detection method used was rather conservative, this may be a lower bound for the true number of flares with QPPs. The QPP periods were found to show a weak correlation with the flare amplitude and duration, but this is likely due to an observational bias. A stronger correlation was found between the QPP period and duration of the QPP signal, which can be partially but not entirely explained by observational constraints. No correlations were found with the active region area, bipole separation distance, or average magnetic field strength.}
   {The fact that a substantial fraction of the flare sample showed evidence of QPPs using a strict detection method with minimal processing of the data demonstrates that these QPPs are a real phenomenon, which cannot be explained by the presence of red noise or the superposition of multiple unrelated flares. The lack of correlation between the QPP periods and active region properties implies that the small-scale structure of the active region is important, and/or that different QPP mechanisms act in different cases.}


   \maketitle
%

\section{Introduction}

Despite the large number of observations of quasi-periodic pulsations (QPPs) in solar and stellar flares \citep[e.g.][]{2010SoPh..267..329K, 2015SoPh..290.3625S, 2016MNRAS.459.3659P, 2016ApJ...827L..30H, 2016ApJ...833..284I}, their nature remains mysterious. While it is difficult to determine the exact cause of the QPPs for any given case, there are two groups of possible mechanisms that have been proposed: those based on magnetohydrodynamic (MHD) oscillations, and those based on regimes of repetitive magnetic reconnection that can be considered in terms of a ``magnetic dripping'' model \citep{2009SSRv..149..119N, 2010PPCF...52l4009N, 2016SoPh..291.3143V}.

When considering the MHD wave mechanisms, the oscillations could either originate from the flaring structure itself, or from a nearby structure. For the first case, MHD waves cause plasma parameters to vary periodically, which could either directly affect the emission, or could modulate the magnetic reconnection rate and acceleration of charged particles, and hence indirectly affect the emission. For example, for the case of the sausage mode of a coronal loop, the plasma density within the loop varies periodically, which in turn causes the magnetic field strength to vary. Hence the gyrosynchrotron emission in the microwave band would be modulated, and this variation could also modulate the acceleration of charged particles \citep[e.g.][]{2008PhyU...51.1123Z}, which would affect Bremsstrahlung emission from the loop foot-points. For the second case, where the MHD oscillations originate from an external source, the oscillations could leak into the intermediate plasma. Then as each wavefront reaches the flaring site, micro-instabilities could be formed which would result in anomalous resistivity and strong currents near the magnetic null point, hence triggering magnetic reconnection \citep{2004A&A...420.1129M}. \citet{2006A&A...452..343N} showed that even MHD waves that are low in amplitude when they approach the null point are able to cause strong spikes in the current, due to nonlinear effects. Another possibility specific to two ribbon flares is that slow magnetoacoustic waves propagate along the axis of the coronal arcade, periodically triggering reconnection in each of the arcade loops as they go \citep{2011ApJ...730L..27N}.

Numerous MHD simulations have shown regimes of repetitive reconnection which do not require a periodic driver. These regimes are often referred to as ``load/unload'' or ``magnetic dripping'' mechanisms when relating them to QPPs \citep{2010PPCF...52l4009N}. For example, \citet{2009A&A...494..329M} and \citet{2012ApJ...749...30M} performed 2.5D simulations of the emergence of a flux rope into a coronal hole (with a vertical magnetic field), and found that a current sheet formed. Reconnection occurred, with outflows from the ends of the current sheet resulting in a build up of gas pressure in a quasi-bound region. The increase in the pressure gradient caused the inflow field lines to move apart, stopping reconnection, and brought the outflow field lines together, causing reconnection to recommence in a different configuration. This process repeated in a periodic manner until eventually an equilibrium state was reached. Recently, \citet{2017ApJ...844....2T} extended this work by demonstrating that oscillatory reconnection can also occur at a 3D magnetic null point. \citet{2000A&A...360..715K} instead focussed on a long current sheet above a soft X-ray coronal loop, and found that instabilities caused anomalous resistivity and the formation of magnetic islands as a result of reconnection, which could then coalesce, forming one or more plasmoids which would then be ejected. The ejection of multiple plasmoids (and hence resulting flare emission) could either be sporadic or quasi-periodic. A more recent work by \citet{2016ApJ...820...60G} investigates charged particle acceleration in flares, based on the mechanism for electron acceleration via interactions with magnetic islands proposed by \citet{2006Natur.443..553D}. They find that this mechanism is capable of explaining the observed electron energies in flares, and also find that it results in sporadic flare emission because of the intermittent magnetic island formation, which could relate to observed flare pulsations. 

Some of the proposed QPP mechanisms relate the characteristic timescale of the QPPs to a spatial scale: for example, the period of an MHD oscillation of a coronal loop relates to the length of the loop. Hence this motivates looking for correlations between the QPP periods and spatial scales of the region from which the flare originates. To do this we chose a set of flares from a single active region (AR), so that any evolution of the QPP properties corresponding to the evolution of the AR properties can be checked for. In addition, focussing on just one AR means that we can utilise the automatic boundary detection and tracking algorithm from \citet{2011AdSpR..47.2105H}, which means that calculating AR properties around the time of a particular flare can largely be automated. The AR studied in this work, known as NOAA 12172/12192/12209, was chosen because it produced a large number of flares (a total of 181 GOES class flares), it existed at a time when many high-quality solar observation instruments were operating, and also because it was very long lived, persisting for around three solar rotations. This AR has been the subject of several other studies due to its highly active nature, but also because it is unusual in that none of the X-class flares were accompanied by coronal mass ejections (CMEs), and the few CMEs that did emerge from the AR were relatively small considering the amount of flare activity \citep{2015ApJ...801L..23T, 2016ApJ...822L..23P, 2016ApJ...826..119L, 2016ApJ...828...62J, 2016IAUS..320..196D}.

In terms of the detection of the QPPs, in the past many different approaches have been taken. Some examples of these include manual identification \citep[e.g.][]{1983ApJ...271..376K}, searching for a peak in the periodogram or wavelet power spectrum (usually after doing some form of detrending of the flare time series data, e.g. \citealt{2011A&A...525A.112R, 2013SoPh..284..559K, 2017ApJ...836...84D}), and empirical mode decomposition \citep{2015A&A...574A..53K}. More recently questions have been raised regarding the potential for false detections with some of the detrending methods. This is especially an issue if the data contains red noise, where the data is correlated in time and therefore the spectral power is related to the frequency. For example, \citet{2011A&A...533A..61G} and \citet{2015ApJ...798..108I} showed that if red noise is present in the flare time series data, then detrending the data can lead to the overestimation of the significance of a signal. In addition, \citet{2016ApJ...825..110A} showed that if a signal containing red noise is detrended by subtracting a boxcar-smoothed version of the signal from the original signal before calculating the power spectrum, then the power spectrum will contain what looks like a broad peak, but this apparent spectral feature is completely artificial. While the trends in flare time series data cannot be considered to be entirely ``random walk'' red noise, since there seems to be a general characteristic shape that flare light curves follow (a rapid rise followed by a more gradual decay), finding the true trend of the flare is a huge challenge in itself when many flares show deviations from the characteristic shape. For this reason \citet{2017A&A...602A..47P} demonstrated two related methods, based on the method of \citet{2005A&A...431..391V}, to assess the significance of periodic signals in flares, accounting for the presence of red noise and data uncertainties and without any form of detrending. In this work we apply the methods of \citet{2017A&A...602A..47P} to the set of flares from the AR NOAA 12172/12192/12209. The methods could complement that used by \citep{2016ApJ...833..284I}, which also avoids detrending and accounts for the presence of red noise, but instead involves a power spectrum model comparison.

This paper is structured as follows. In Sect. \ref{sec:obs} we describe the solar flare and AR magnetogram data used. Sect. \ref{sec:dat} summarises the QPP detection method, including details of the use of time derivative data and how it impacts on the power spectrum, and also describes how the AR properties were obtained. The results and discussion of the search for the QPPs themselves along with any correlations with flare or AR properties are given in Sect. \ref{sec:res}, and finally conclusions are given in Sect. \ref{sec:con}.


\section{Observations}
\label{sec:obs}

Data were used from the X-ray sensor (XRS) aboard the \emph{Geostationary Operational Environmental Satellite} (GOES), which makes near continuous observations of the Sun in two soft X-ray (SXR) wavebands, 1--8\,\AA\ (1.5--12.4\,keV) and 0.5--4\,\AA\ (3.1--24.8\,keV), with a cadence of 2.047\,s. \citet{2015SoPh..290.3625S} showed that the irradiance steps due to the digitisation of the data are greater than the Poisson noise from counting statistics. Therefore, we used half of the irradiance step size as a function of the irradiance as an estimate of the uncertainty for each measurement.

We also made use of the \emph{Extreme ultraviolet SpectroPhotometer} (ESP) channel of the \emph{Extreme ultraviolet Variability Experiment} (EVE) aboard NASA's \emph{Solar Dynamics Observatory} (SDO), which observes the 1--70\,\AA\ (0.18--12.4\,keV) extreme ultraviolet (EUV) and SXR waveband \citep{2012SoPh..275..179D}. This waveband overlaps with the GOES wavebands, and since all of the flares in this study were observed by GOES, the EVE/ESP data can be used to rule out the possibility that a periodic signal in the GOES data is due to an artefact and is unrelated to the flare \citep{2016ApJ...827L..30H, 2017ApJ...836...84D}. The time cadence of ESP is 0.25\,s, but in order to estimate the uncertainties these measurements were binned down to a 1\,s cadence, and the standard deviation of the measurements within each 1\,s time bin was used as the uncertainty. The disadvantage of this instrument when searching for QPPs is that the waveband is very broad, so much of the fine structure of the flare is smeared out due to the Neupert effect.

The \emph{Gamma-ray Burst Monitor} (GBM) aboard NASA's \emph{Fermi} satellite \citep{2009ApJ...702..791M} measures X- and gamma-ray photons with energies between 4\,keV and around 40\,MeV. We made use of the CSPEC data, which has 128 energy channels that we combined into three energy ranges: 6--25\,keV, 25--50\,keV, and 50--100\,keV. These energy ranges were chosen so that comparisons could be made with data from NASA's \emph{Reuven Ramaty High Energy Solar Spectroscopic Imager} (RHESSI), however the RHESSI data for the sample of flares examined in this study did not show any significant QPP signals using the method described in Sect. \ref{sec:dat}. The time cadence of the CSPEC observations is 4.096\,s, or 1.024\,s when the count rate exceeds a certain threshold and GBM goes into ``trigger'' mode for a set amount of time.  A better time resolution can be obtained from the CTIME data, with a cadence of 0.256/0.064\,s, however this data is noisier and hence we did not find that it offered much of a benefit over the CSPEC data for the study of QPPs. GBM consists of 2 BGO and 12 NaI detectors which all point in different directions, therefore the angle between the detectors and the Sun must be checked, which we did using the IDL Solar Software routine \emph{gbm\_get\_det\_cos}. Data from the most sunward NaI detector at the time of the flare were used with the exception of flares greater than M5 class, where data from the most sunward detector may be subject to discontinuities due to gain changes. Therefore the second most sunward detector was chosen for these flares. Because of the orbit of the \emph{Fermi} satellite, it cannot observe solar flares while in the Earth's shadow.

Hard X-ray (HXR) data were also obtained with the \emph{Detector of the Roentgen and Gamma-ray Emissions} (DRGE) instrument aboard the Russian satellite \emph{Vernov} \citep{2016SoPh..291.3439M}. The spacecraft had a solar-synchronous orbit with the following parameters: an apogee of 830\,km, perigee of 640\,km, inclination of 98.4$^{\circ}$, and an orbital period of 100\,min. It was launched on 2014 July 8 and operated until 2014 December 10. The DRGE instrument included four identical detector blocks (DRGE11, DRGE12, DRGE21 and DRGE22), based on a NaI(Tl)/CsI(Tl) phoswich. The diameter of both scintillators was 13\,cm, while the NaI(Tl) thickness was 0.3\,cm, and the CsI(Tl) thickness 1.7\,cm. These detector blocks were designed for measuring terrestrial gamma flashes and other atmospheric phenomena, so they were directed towards the Earth. The Sun was to the side of the detectors ($\sim$90$^{\circ}$ from the zenith angle) during the whole period of flare observations, so the effective area of the detectors was only a few cm$^2$. A more detailed description of the experiment along with a catalogue of HXR solar flares from the active regions NOAA 12172 and 12192 observed by \emph{Vernov} is given in \citet{2016SoPh..291.3439M}. In the present work we processed the data of six flares from this catalogue: those where the possibility of detecting QPPs appeared most evident. The monitoring parameter ``count rates of all events in NaI'' was used, and for solar flare emission this refers to the integral channel of photons with energy >30\,keV. The time resolution of the measurements was 1\,s. \emph{Vernov} was a polar low-altitude satellite, thus the background conditions for solar flares were far from optimal, hence two methods of background rejection were used. In the equatorial regions and in the polar caps the background was estimated from the count rates shortly before and after a flare, and in the regions close to the Earth's radiation belts we also took into account count rates from the previous orbits. Poisson counting statistics was assumed, so the uncertainties for each count rate measurement were taken to be equal to the square root of the count rate.

Correlation data with a microwave frequency of 17\,GHz and a cadence of 1\,s from the \emph{Nobeyama Radioheliograph} (NoRH) \citep{1994IEEEP..82..705N}, which is more sensitive to emission from small spatial-scale structures on the Sun rather than the global emission, were also used. Because NoRH is ground-based, solar observations are only made between 22:45 and 06:30 UT each day. The uncertainty of the data was estimated to be $1.1911749 \times 10^{-5}$, which is the standard deviation of a flat section of data \citep{2017A&A...602A..47P}.

\begin{figure}
	\centering
	\includegraphics[width=\linewidth]{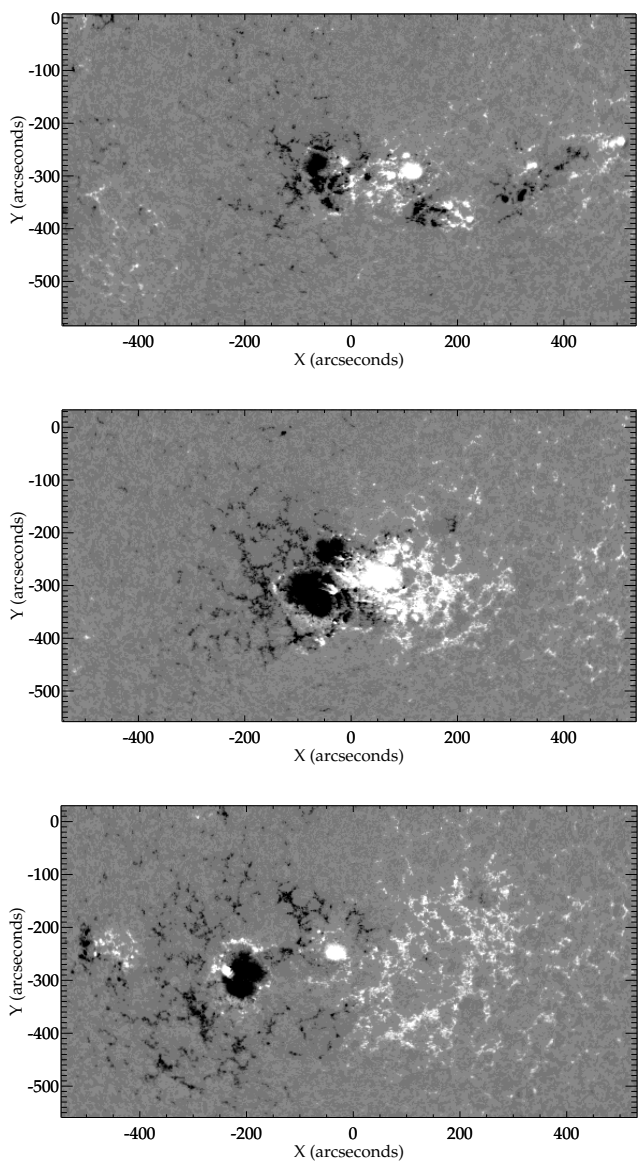}
	\caption{HMI magnetogram images of the active region during its three crossings of the solar disk, at 2014 September 26 22:01:30 (top), 2014 October 23 15:01:30 (middle), and 2014 November 19 07:01:30 (bottom).}
	\label{fig:ar}
\end{figure}

Finally, properties of the active region (AR) were determined using data from the \emph{Helioseismic and Magnetic Imager} (HMI) aboard SDO \citep{2012SoPh..275..207S}. Line of sight magnetogram images were used with a reduced cadence of one hour, and resolution of 1024 $\times$ 1024 pixels. The timescale of the evolution of an AR is typically much greater than an hour, and the reduced resolution does not significantly affect the properties calculated, while it does greatly speed up the calculations. The three time intervals corresponding to the AR's three crossings of the solar disk were chosen to be 2014 September 22 15:00:34 until September 30 08:00:33, 2014 October 19 15:00:30 until October 27 02:00:30, and 2014 November 16 15:00:27 until November 23 08:00:26, where the AR labels during its three crossings are NOAA 12172, NOAA 12192, and NOAA 12209 respectively. Fig. \ref{fig:ar} shows magnetogram images of the AR during these three time intervals. Note that times when the AR was close to the solar limb were omitted, as line of sight effects mean the AR magnetogram images near the limb are highly distorted, and so AR properties cannot be obtained reliably. Unfortunately many of the flares from the AR occurred outside of these time ranges, so these flares had to be excluded when looking for relationships between the QPP periods and AR properties (see Table \ref{tab:qppflares}).

\section{Data analysis}
\label{sec:dat}

The basic outline of our approach to identifying candidate QPPs in the sample of solar flares is as follows. All available light curves for the flares were manually shortened to focus on sections showing the most variability. Shortening the light curves when searching for QPP signals is often helpful due to the transient nature of QPPs, and also because of the presence of background trends which influence the shape of the power spectrum. Some more complex flares showed variability in more than one section, so these different sections of light curve were analysed separately. The longest allowed light curve durations were the same as the flare durations, and the shortest were 16 times the data time cadence. Because the methods for calculating the confidence levels (see Sect. \ref{sec:cl}) require even time sampling, any gaps in the data were avoided when manually choosing the time intervals. Additionally, times when there was a switch between trigger mode and non-trigger mode in the \emph{Fermi} data were also avoided. For the light curves from SXR observations, the time derivative was calculated and used for further analysis (see Sect. \ref{sec:deriv}). 

Lomb-Scargle periodograms were calculated for all of these shortened light curves. No window function was applied to the shortened light curve prior to calculating the periodogram, as doing so would alter the distribution of the noise and additionally may not be beneficial for the detection of low amplitude transient signals. A broken power law model was fitted to the periodograms (see \citealt{2017A&A...602A..47P}), and model uncertainties at each frequency index were estimated by performing Monte Carlo simulations. The 95\% and 99\% confidence levels were then calculated (see Sect. \ref{sec:cl}) taking account of any power law dependence of the spectrum and uncertainties associated with the model fits. Additionally, rebinned power spectra were calculated in order to better assess the significance of any broader peaks that spanned more than one frequency bin in the regular power spectra \citep{2004A&A...428.1039A}.

We did not consider peaks with a period less than four times the cadence or greater than a quarter of the duration of the time series data as candidate QPP signals, as we do not believe periods in these ranges can be detected reliably without supporting data from another instrument with a higher time resolution.

The start and end times of the light curves were refined manually in order to maximise the confidence level of any periodic component of the signal. This was done simply by decreasing, then increasing the start time by one data point at a time, to search for a start time which maximised the significance of the peak in the power spectrum. This iterative process was repeated with the end time, then again with the start time, in order to find the maximum confidence level of the peak. If multiple significant spectral peaks were found then these would be listed separately in the results, although there were no cases where we found multiple periods in the same section of light curve and same waveband. Flares with a peak in the power spectrum above the 95\% confidence level were included in the sample of flares with strong candidate QPPs, and used to study the QPP properties.

\subsection{Time derivative data}
\label{sec:deriv}

Making use of the time derivative of high-precision SXR observations has been shown to be useful for the study of QPPs \citep{2015SoPh..290.3625S, 2016ApJ...827L..30H, 2017ApJ...836...84D}. Taking the derivative will have a significant impact on the power spectrum of time series data, however, and so it is extremely important to understand this impact when assessing the significance of peaks in the power spectrum. One of the most basic numerical approximations to the derivative of time series data is a three-point finite difference, defined as:
\begin{equation}
	\dot{x}_n = \frac{x_{n+1}-x_{n-1}}{2h},
\end{equation}
where $x$ represents intensity, $n$ is the time index, and $h$ is the time cadence. Then the discrete Fourier transform of $\dot{x}$ as a function of time can be written:
\begin{equation}
	\mathcal{F}_k(\dot{x}) = \frac{1}{2h}\displaystyle\sum_{n=0}^{N-1} (x_{n+1}-x_{n-1}) e^{-2\pi ikn/N},
\end{equation}
where $k$ is the frequency index, ranging from 0 to $N-1$, and $N$ is the number of data points. This expression can be rearranged to:
\begin{align}
	\mathcal{F}_k(\dot{x}) &= \frac{1}{2h}\left(\displaystyle\sum_{n=0}^{N-1} x_{n+1} e^{-2\pi ikn/N} - \displaystyle\sum_{n=0}^{N-1} x_{n-1} e^{-2\pi ikn/N}\right)\\
	&= \frac{1}{2h}\left(\displaystyle\sum_{n=1}^{N} x_{n} e^{-2\pi ik(n-1)/N} - \displaystyle\sum_{n=-1}^{N-2} x_{n} e^{-2\pi ik(n+1)/N}\right).
\end{align}
In the above expression the sums go outside of the data range (the derivative cannot be calculated at the first and last points in the time series), so instead the following expression must be considered:
\begin{equation}
	\mathcal{F}_k(\dot{x}) = \frac{1}{2h}\left(\displaystyle\sum_{n=1}^{N-2} x_{n} e^{-2\pi ik(n-1)/N} - \displaystyle\sum_{n=1}^{N-2} x_{n} e^{-2\pi ik(n+1)/N}\right), 
\end{equation}
which can then be rearranged to:
\begin{align}
	\mathcal{F}_k(\dot{x}) &= \frac{1}{2h}\left(\displaystyle\sum_{n=1}^{N-2} x_{n} e^{-2\pi ikn/N}e^{2\pi ik/N} - \displaystyle\sum_{n=1}^{N-2} x_{n} e^{-2\pi ikn/N}e^{-2\pi ik/N}\right)\\
	&= \frac{1}{2h}\left(e^{2\pi ik/N} - e^{-2\pi ik/N}\right)\displaystyle\sum_{n=1}^{N-2} x_{n}e^{-2\pi ikn/N}\\
	&= \frac{i}{h}\sin\left(\frac{2\pi k}{N}\right)\mathcal{F}_k(x) = \frac{i}{h}\sin(\omega)\mathcal{F}_k(x),
\end{align}
where $\omega$ is an angular frequency which ranges from 0 to $2\pi$. The Fourier power spectrum is the square of the absolute value of the Fourier transform, so for the power spectrum we have:
\begin{equation}
	\left|\mathcal{F}_k(\dot{x})\right| ^2 = \frac{1}{h^2}\sin^2(\omega)\left|\mathcal{F}_k(x)\right| ^2. 
\end{equation}
The periodogram of evenly spaced data with no oversampling is equivalent to the discrete Fourier power spectrum with additional normalisation, and so this $\sin^2(\omega)$ multiplying term will appear when the periodogram of time derivative data is calculated. For a perfectly periodic signal, the periodogram of the time derivative of the signal will be equal to $\frac{1}{h^2}\sin^2(\omega)$ multiplied by the periodogram of the original signal. Flare time series data is not completely periodic, however, and the presence of background trends will have a substantial impact on the power spectra. Taking the time derivative will suppress slowly-varying background trends, and hence if a periodic component of the signal is present it will be more visible in the time derivative power spectrum. Taking the time derivative is most beneficial for SXR flare observations, since the impulsive phase of a flare is best seen in the HXR and microwave/radio wavebands, and the Neupert effect means this phase corresponds to a rise in the SXR emission. Hence QPPs which are most often seen in the impulsive phase of a flare will appear during the rise of the SXR emission, and this rising trend will make QPPs less visible in the power spectrum. For all SXR observations used in this study, from GOES/XRS and EVE/ESP, the time derivatives of the signals have been used, and the power spectra have been divided by $\sin^2(\omega)$ before proceeding to calculate the confidence levels. Note that because $\omega$ varies between 0 (at the lowest frequency sampled) and $\pi$ (at the highest frequency sampled) for the positive frequencies, this means that $\sin^2(\omega)$ is equal to zero at the edges of the power spectrum, hence the powers at the lowest and highest frequencies of the power spectrum cannot be calculated. In addition, where $\omega$ is close to 0 and $\pi$, $\sin^2(\omega)$ is very small and therefore numerical uncertainties will have a bigger impact. To avoid this, all derivative power spectra have had the first and final 2\% of frequencies removed, with the exception of power spectra with less than 50 data points, which have the first and last points removed.

\subsection{Confidence levels}
\label{sec:cl}

The methods used to calculate the 95\% and 99\% confidence levels on the power spectra are described in detail in \citet{2017A&A...602A..47P}, and are based on the test described by \citet{2005A&A...431..391V}. The first method addresses regular power spectra, and the second rebinned power spectra. In this section we briefly summarise the methods, but direct readers to \citet{2017A&A...602A..47P} for a more thorough description.

The confidence level corresponding to a particular false alarm probability, $\gamma _{\epsilon _j}$, is the level where the probability of having one or more points, $\hat{\gamma _j}$, in the power spectrum above that level is approximately equal to the false alarm probability divided by the number of data points in the power spectrum, $N'$, if the original time series signal is due to random noise:
	\begin{equation}
		\text{Pr}\left\{\hat{\gamma _j} > \gamma _{\epsilon _j}\right\} \approx \frac{\epsilon _{N'}}{N'}\,.
	\label{eq:prob}
	\end{equation}
This probability is equal to the probability density function integrated between $\gamma _{\epsilon _j}$ and infinity, and the overall probability density function can be found by combining those of the chi-squared distribution of the noise and the log-normal distribution of the uncertainty of a fit to the power spectrum at a particular frequency index. Because the combined probability density function is an integral, the probability must be found by solving a double integral. For regular power spectra (where the noise follows a chi-squared 2 degrees of freedom distribution) one of the integrals can be solved analytically, simplifying the equation for the probability to:
	\begin{equation}
	 \text{Pr}\left\{\hat{\gamma _j} > \gamma _{\epsilon _j}\right\} = \int_{0}^{\infty}\frac{1}{\sqrt{2\pi\;}S_jw} \exp \left\{-\frac{\left(\ln\,w\right)^2}{2S^2_j} - \frac{\gamma _{\epsilon _j}w}{2}\right\} \mathrm{d}w\,,
	\label{eq:probreg}
	\end{equation}
where $w$ is a dummy variable representing power and $S_j = \text{err}\left\{\log\left[\hat{\mathcal{P}}(f_j)\right]\right\} \times \ln[10]$, where $\text{err}\left\{\log\left[\hat{\mathcal{P}}(f_j)\right]\right\}$ is the uncertainty corresponding to the logarithm of the model power spectrum $\hat{\mathcal{P}}$ at a particular frequency index $f_j$.

Rebinning the power spectrum by summing together the powers in every $n$ frequency bins, in order to better assess the power contained in a peak that spans more than one frequency bin, will alter the distribution of the noise. Fortunately this is straightforward to account for. The noise in the rebinned power spectrum will follow a chi-squared $2n$ degrees of freedom distribution rather than chi-squared 2 degrees of freedom, which is the case for the regular power spectrum \citep{2004A&A...428.1039A}. The value of $n$ can be chosen depending on how broad the spectral peak of interest is; for the flares considered in this study either $n=2$ or $n=3$ gave the best results. The probability density function corresponding to this different distribution of the noise can be combined with that of the log-normal distribution of the uncertainty of the power law fit to the power spectrum to give the overall probability density function, which can then be integrated between $\gamma _{\epsilon _j}$ and infinity, as before, to give the probability of having a value in the rebinned power spectrum above a certain threshold. This probability equation for the rebinned power spectra can be simplified to:
	\begin{equation}
	 	\text{Pr}\left\{\hat{\gamma _j} > \gamma _{\epsilon _j}\right\} = \int_{0}^{\infty}\frac{1}{\sqrt{2\pi\;}S_jw} \exp \left\{-\frac{\left(\ln\,w\right)^2}{2S^2_j}\right\} \frac{\Gamma (n, w\gamma _{\epsilon _j}/2)}{\Gamma (n)}\mathrm{d}w\,,
	\label{eq:probrebin}
	\end{equation}
where $n$ is the number of frequency bins summed over, and $\Gamma (n, w\gamma _{\epsilon _j}/2)$ is the upper incomplete gamma function. 

Equations \ref{eq:probreg} and \ref{eq:probrebin} should then be equated to Eq. \ref{eq:prob} and solved numerically in order to determine $\gamma _{\epsilon _j}$. The confidence level would be equal to $\gamma _{\epsilon _j}$ if the spectral powers were independent of the frequency and normalised so that the mean power were equal to the number of degrees of freedom of the chi-squared distribution of the noise (i.e. 2 for the regular power spectra, or $2n$ for the rebinned spectra). Instead we are dealing with power spectra that are not normalised and have a power law dependence. The regular power spectra could be normalised by multiplying by $2/\langle I_j / \hat{\mathcal{P}}_j\rangle$, or alternatively $\gamma _{\epsilon _j}$ can be multiplied by $\langle I_j / \hat{\mathcal{P}}_j\rangle /2$ to account for the lack of normalisation, where $I_j$ is the original power spectrum, $\hat{\mathcal{P}}_j$ is a broken power law fit to the power spectrum, and $\langle I_j / \hat{\mathcal{P}}_j\rangle$ is the mean of the ``flattened'' power spectrum (with the power law dependence removed). Finally the power law dependence needs to be accounted for by multiplying $\gamma _{\epsilon _j}\langle I_j / \hat{\mathcal{P}}_j\rangle /2$ by the power law fit, or adding if working in log space. Hence the confidence level for the regular power spectrum is equal to $\log[\hat{\mathcal{P}}_j] + \log[\gamma _{\epsilon _j}\langle I_j / \hat{\mathcal{P}}_j\rangle /2]$. Similarly for the rebinned power spectrum, the confidence level can be found by calculating $\log[\hat{\mathcal{P}}_j] + \log[\gamma _{\epsilon _j}\langle I_j / \hat{\mathcal{P}}_j\rangle /2n]$, where here $I_j$ is the rebinned power spectrum and $\hat{\mathcal{P}}_j$ is the corresponding fitted model.

\subsection{Active region properties}

The processing of the HMI line of sight magnetograms and calculation of some AR properties was done using the \emph{SolarMonitor Active Region Tracking} (SMART) routines provided by \citet{2011AdSpR..47.2105H}. For each AR crossing of the solar disk (while within around $\pm 60^{\circ}$ of the central meridian line, since the AR images are highly distorted when close to the limb), the processing technique is as follows. First the magnetogram frame where the AR is approximately half way across the solar disk was used to determine a bounding box around the AR. ARs visible on the disk at that time were detected automatically according to \citet{2011AdSpR..47.2105H}. After selecting the AR of interest, the X and Y coordinate ranges of the SMART detection outline were used to define a box around the AR. Next, the line of sight projection effect of features closer to the limb appearing smaller compared to when closer to the centre of the solar disk is accounted for by differentially rotating the other magnetogram frames to the time where the AR is approximately at the central meridian, using the IDL Solar Software routine \emph{drot\_map}. The previously defined box was then used to crop all frames to include only the AR of interest.

In order to estimate the AR photospheric area as a function of time, pixels within the bounding box with an absolute magnetic field strength greater than a threshold value of 70\,G were selected. This threshold was chosen as quiet Sun regions tend to have magnetic field values less than this \citep{2011AdSpR..47.2105H}. For each selected pixel, the area of the solar surface that the pixel would correspond to if it were located at the disk centre was multiplied by a cosine correction factor, to account for the spherical nature of the Sun meaning that different pixels correspond to different surface areas \citep{2005SoPh..228...55M}, then the resulting values were summed together to obtain an AR area for a particular magnetogram frame.

The bipole separation is defined as \citep{2011ApJ...729...97M}:
\begin{equation}
	S = |\mathbf{S}(t)| = \left|\frac{\sum_{B_z>+70\,\text{G}}B_z(i,j)\mathbf{R}_{i,j}}{\sum_{B_z>+70\,\text{G}}B_z(i,j)} - \frac{\sum_{B_z<-70\,\text{G}}B_z(i,j)\mathbf{R}_{i,j}}{\sum_{B_z<-70\,\text{G}}B_z(i,j)}\right|,
\end{equation}
where $\mathbf{S}(t)$ is the vector pointing from the centre of one pole to the other, $B_z(i,j)$ is the line of sight magnetic field at pixel position $(i, j)$, and $\mathbf{R}_{i,j}$ is the position vector pointing from the origin to the pixel at $(i, j)$. Once the bipole separation, $S$, had been calculated it was then converted to a great circle distance in Mm.

Finally, the average magnetic field strength of the active region at the photosphere as a function of time was calculated by summing together the absolute magnetic field strength values of all pixels in a particular magnetogram frame with a magnitude greater than the threshold value of 70\,G, then dividing by the number of pixels with absolute values greater than the threshold.


\section{Results and discussion}
\label{sec:res}

\subsection{The set of flares with QPPs}

\begin{figure*}
	\centering
	\includegraphics[width=0.9\linewidth]{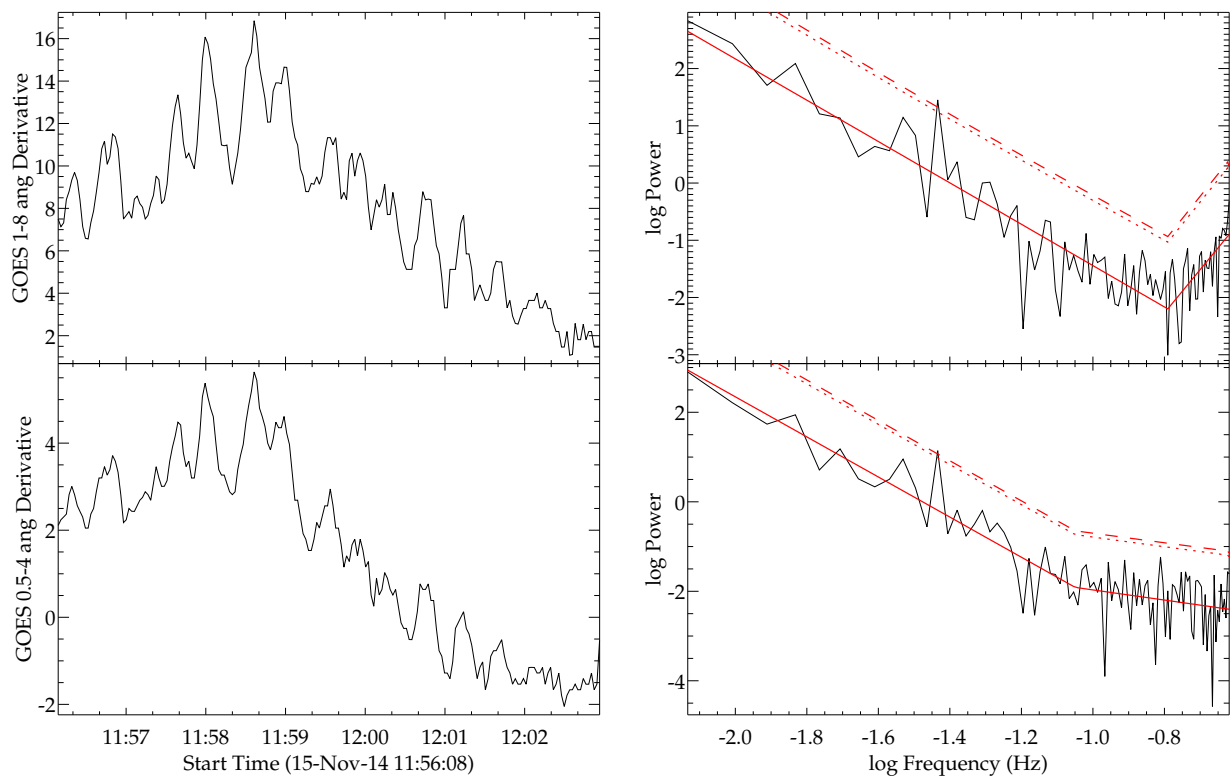}
	\caption{Left: the time derivatives of a section of flare 152 observed by GOES/XRS, where the top panel shows the 1--8$\AA$ emission and the bottom panel the 0.5--4\,\AA\ emission. Right: the corresponding power spectra, where the red solid lines are broken power law fits to the spectra, the red dotted lines represent the 95\% confidence levels, and the red dashed lines the 99\% levels. One peak in each is above the 99\% level, at a period of $27.2 \pm 0.9$\,s.}
	\label{152goes}
\end{figure*}

\begin{figure*}
	\centering
	\includegraphics[width=0.9\linewidth]{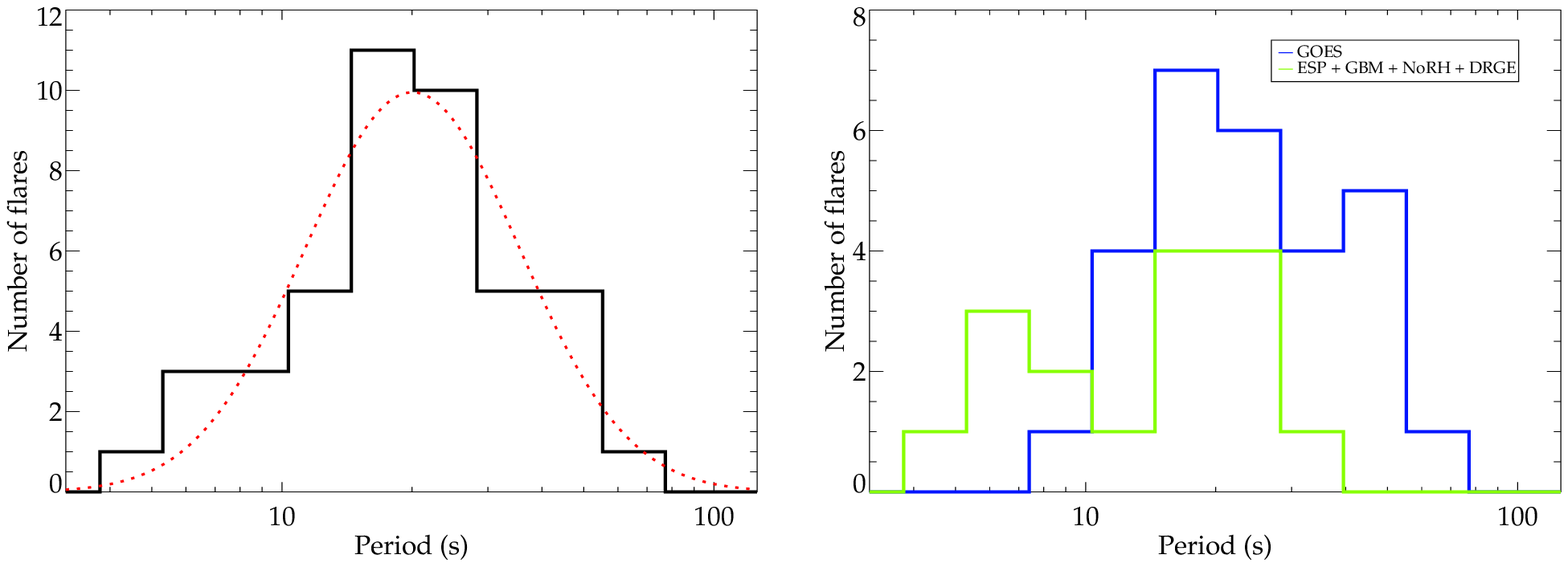}
	\caption{Histograms of the QPP periods. Left: the black solid line shows all QPP periods combined, and the dotted red line shows a Gaussian fit to the overall distribution which corresponds to an average QPP period of $20^{+16}_{-9}$\,s for the set of flares examined. Right: the same histogram but with the QPP periods separated based on which instrument was used. The blue line shows the QPP periods detected in the GOES/XRS wavebands with a 2\,s cadence, and the green line shows those detected by EVE/ESP, \emph{Fermi}/GBM, NoRH, and \emph{Vernov}/DRGE with a 1\,s cadence. The distribution for GOES/XRS appears to be shifted slightly towards longer periods than the other instruments.}
	\label{fig:hist}
\end{figure*}

Details of all 181 flares used in the analysis are given in Table \ref{tab:allflares}. These flares were selected from the list of automatically detected flares provided by the NOAA Space Weather Prediction Centre (SWPC)\footnote{\url{http://www.swpc.noaa.gov}}, and the spatial location of the flares was checked using SDO Atmospheric Imaging Assembly (AIA) 94\,\AA\ difference images provided by SolarMonitor\footnote{\url{https://www.solarmonitor.org}} \citep{2002SoPh..209..171G}, to ensure that the flares originated from the active region of interest. After searching each of the flares for evidence of QPPs using the methods described in Sect. \ref{sec:dat}, a total of 37 flares with convincing candidate QPPs were identified, corresponding to 20\% of flares in the sample. These flares are summarised in Table \ref{tab:qppflares}, where the upper and lower uncertainties for each period are taken to be plus or minus half of the corresponding frequency bin width in the power spectrum. Note that for some of these flares, QPPs were found in more than one section of the flare light curve, while there were no cases where multiple significant periods were found in the same section of light curve observed in a particular waveband. The vast majority of the QPPs occurred during the impulsive phase of the flare, with the only exception being the flare labelled ``022'' in Tables \ref{tab:allflares} and \ref{tab:qppflares}, where the QPPs were predominantly in the decay phase. Plots showing the time series data and power spectra of one of these flares as an example are given in Figs. \ref{152goes} and \ref{152eve}, while similar plots for the other 36 flares are shown by Figs. \ref{008goes}--\ref{177goes} in the Appendix. A histogram of the QPP periods is given in the left-hand panel of Fig. \ref{fig:hist}, and if a log-normal distribution is assumed (more data is needed to confirm if this model is a good approximation, but similar histograms shown by \citet{2016ApJ...833..284I} also appear to have a log-normal distribution), then the average QPP period for this set of flares is $20^{+16}_{-9}$\,s. This seems to be consistent with the results of \citet{2016ApJ...833..284I}. The right-hand panel of Fig. \ref{fig:hist} shows separate histograms for the QPP periods detected by GOES/XRS, and those detected by EVE/ESP, \emph{Fermi}/GBM, NoRH, and \emph{Vernov}/DRGE. The distribution for GOES/XRS appears to be shifted slightly towards longer periods than the other instruments, which could be explained by the other instruments having a higher time resolution and also only capturing the impulsive phase of the flare. GOES/XRS has a lower time resolution and observes both the impulsive and gradual phases of the flare, meaning that the detection of shorter periods is limited by the time resolution, whereas longer periods can be seen more easily.

Seven flares (those labelled 056, 072, 104, 106, 135, 142, and 152 in Tables \ref{tab:allflares} and \ref{tab:qppflares}) have a QPP signal from two different instruments above the 95\% confidence level in their power spectrum, which rules out the possibility that these signals are due to some instrumental artefact. A further two flares (010 and 024) have peaks just below the 95\% level in the EVE/ESP power spectra at the same period as those seen above the 95\% level in the GOES/XRS data. On the other hand, three of the flares (010, 037, and 038) have QPPs observed in two different wavebands over the same time range, but with different periods. According to the standard flare model, different wavebands of the emission originate from different positions within the flaring region, so these could be unrelated periodic signals originating from different places, or alternatively they could result from the same process, but be shifted from one another due to changes in the local physical parameters.

It is slightly surprising that the majority of significant QPP detections made by GOES/XRS are not supported by EVE/ESP, considering both instruments observe the Sun near continuously and have overlapping observational wavebands. A possible reason for this is that the EVE/ESP waveband is so wide. QPPs are often more visible in a particular waveband than others (which could relate to the mechanism or spatial origin of the signal), or they could be phase shifted across different wavelengths. These would result in the signal being hidden or blurred in wide waveband observations. The visibility of QPP signals in different wavebands would also explain the other cases where there is a detection in one instrument but not another over the same time range. Alternatively the cause may simply be that the flare signal-to-noise ratio is lower for EVE/ESP than GOES/XRS, thus making any QPP signals in the EVE/ESP data more difficult to detect above the noise level. Also of note are the differences between the two overlapping GOES/XRS wavebands. In most cases a signal can be seen in both wavebands even if it is not above the 95\% confidence level threshold, while there are a few cases where the signal can only be seen in the power spectrum of one of the wavebands. An explanation of this could be a combination of the Neupert effect resulting in a steeper trend in one of the waveband light curves compared to the other, and that the optimal choice of time interval for one waveband might not be the same as for the other waveband. \citet{2017A&A...602A..47P} demonstrated that steeper trends in light curves result in QPP signals having a lower significance in the power spectra. Alternatively there could be a physical reason for the QPP signal appearing stronger in one of the wavebands over the other, based on the QPP mechanism.

\citet{2016ApJ...833..284I} looked for QPPs in all M- and X-class flares in the GOES/XRS and \emph{Fermi}/GBM data between 2011 February 01 and 2015 December 31, meaning that 44 of those flares are included in this study (flares included in the \citet{2016ApJ...833..284I} sample are indicated in Table \ref{tab:allflares}). Rather than shortening the flare time series to focus on a section showing a potential QPP signal like in this work, \citet{2016ApJ...833..284I} use the flare start and end times from the GOES catalogue in order to automate their method. Their method involves a model comparison, where three different models are fitted to the flare power spectra, compared by calculating the Bayesian Information Criterion (BIC) for each, and checked for a reasonable goodness of fit. The three models are a single power law plus constant (model $S_0$), a power law plus Gaussian bump and constant (model $S_1$), and a broken power law plus constant (model $S_2$). A lower BIC value means a more favoured model, and \citet{2016ApJ...833..284I} imposes a selection criterion that model $S_1$ should have a BIC value that is at least 10 less than those for models $S_0$ and $S_2$, so only cases where model $S_1$ is strongly favoured over models $S_0$ and $S_2$ are considered. They also require the models to be fit to the power spectra sufficiently well, based on a goodness-of-fit statistic. We find similar periods (within the $1\sigma$ uncertainties) to \citet{2016ApJ...833..284I} for six flares: 029, 056, 135, 140, 152, 153, and a further seven flares when the selection criteria of \citet{2016ApJ...833..284I} are relaxed: 049, 054, 085, 104, 105, 117, 161. We consider periods identified with relaxed selection criteria here (all cases where model $S_1$ is preferred over model $S_0$) because these are cases where a period identified by the automated method of \citet{2016ApJ...833..284I} does not quite match their selection criteria, while this study finds the same period to be significant. Therefore we regard these cases as promising and worthy of mention. In addition we find the same periodic signals identified by \citet{2016SoPh..291.3439M} in flares 010, 056, and 135. We find different significant periods than \citet{2016ApJ...833..284I} from the same instrument for one flare: 092, and different significant periods from different instruments for one flare: 098. The three flares where there is a significant period identified by \citet{2016ApJ...833..284I} but not the present study are 075, 115, and 139, and the two flares where the present study finds a significant period whereas \citet{2016ApJ...833..284I} does not are 008 and 072. For the remaining 24 flares both this work and \citet{2016ApJ...833..284I} find no convincing evidence of QPPs. We believe that the majority of cases where the results of this work differ from \citet{2016ApJ...833..284I} can be explained by the different time intervals used or different detection criteria. For example, we neglect any periods that are less than four times the cadence or greater than a quarter of the length of the time series data as these are difficult to detect reliably, whereas \citet{2016ApJ...833..284I} requires a model containing a QPP signal to be sufficiently favoured over two alternative models.

\longtab{
\begin{longtable}{c c c c c c c}
	\caption{Summary table of the 181 flares from the chosen active region. The first column contains a numerical label for each flare, the second and third columns are the flare start and end times (UT), the fourth column is the flare GOES class, the fifth the approximate position on the solar disk. The sixth column shows which instruments, other than GOES/XRS and EVE/ESP, observed the flare, where \emph{R}, \emph{F}, \emph{N}, \emph{V} correspond to RHESSI, \emph{Fermi}/GBM, NoRH, and \emph{Vernov}/DRGE respectively, and \emph{(part)} indicates that only part of the flare was observed by that instrument. Finally, the seventh column contains references to other studies that include the flare.}\\
	\hline\hline
	Flare no. & Start time & End time & GOES class & Coordinates & Other instruments & References\\
	\hline\\
	\endfirsthead
	\multicolumn{7}{c}
	{\tablename\ \thetable\ -- \textit{Continued from previous page}}\\
	\hline\hline
	Flare no. & Start time & End time & GOES class & Coordinates & Other instruments & References\\
	\hline\\
	\endhead
	\hline \multicolumn{7}{r}{\textit{Continued on next page}}\\
	\endfoot
	\endlastfoot
	001 & 2014-09-19 18:03 & 2014-09-19 19:52 & C3.3 & S10E84 & & \\ 
	002 & 2014-09-20 12:44 & 2014-09-20 13:50 & C1.1 & S13E88 & R(part), F(part) & \\
	003 & 2014-09-20 17:10 & 2014-09-20 17:38 & C1.1 & S12E88 & R(part), F(part) & \\
	004 & 2014-09-21 01:46 & 2014-09-21 02:14 & C1.2 & S13E85 & R, F, N & \\
	005 & 2014-09-21 07:10 & 2014-09-21 07:43 & C2.0 & S12E78 & & \\
	006 & 2014-09-22 19:28 & 2014-09-22 20:30 & C1.6 & S16E61 & & \\
	007 & 2014-09-23 15:29 & 2014-09-23 15:44 & C1.1 & S14E47 & F(part) & \\
	008 & 2014-09-23 23:05 & 2014-09-24 00:20 & M2.3 & S14E32 & R, F(part), N & 1 \\
	009 & 2014-09-24 16:04 & 2014-09-24 16:18 & C1.8 & S13E23 & & \\
	010 & 2014-09-24 17:48 & 2014-09-24 18:01 & C7.0 & S13E23 & R, V & 2 \\
	011 & 2014-09-25 19:20 & 2014-09-25 21:51 & C3.2 & S16E17 & R(part), F(part) & \\
	012 & 2014-09-26 03:54 & 2014-09-26 10:00 & C8.7 & S08E08 & N(part) & \\
	013 & 2014-09-26 13:53 & 2014-09-26 14:01 & C4.2 & S11E06 & F & \\
	014 & 2014-10-14 18:12 & 2014-10-14 19:01 & M1.1 & S12E88 & F(part) & 1 \\
	015 & 2014-10-14 19:07 & 2014-10-15 08:30 & M2.2 & S11E88 & & 1 \\
	016 & 2014-10-16 07:34 & 2014-10-16 08:02 & C7.7 & S14E88 & R(part) & \\
	017 & 2014-10-16 08:38 & 2014-10-16 10:02 & C6.4 & S11E88 & R(part), F(part) & \\
	018 & 2014-10-16 13:01 & 2014-10-16 13:09 & M4.3 & S13E88 & R(part), V & 1,2 \\
	019 & 2014-10-16 18:04 & 2014-10-16 18:23 & C2.0 & S15E88 & F & \\
	020 & 2014-10-16 20:12 & 2014-10-16 21:14 & C2.9 & S13E88 & R(part), F(part) & \\
	021 & 2014-10-17 03:12 & 2014-10-17 03:45 & C3.9 & S17E87 & R(part), N & 2 \\
	022 & 2014-10-17 04:58 & 2014-10-17 05:48 & C6.6 & S15E88 & F, N & 2 \\
	023 & 2014-10-17 12:45 & 2014-10-17 13:11 & C3.6 & S13E88 & R(part) & \\
	024 & 2014-10-17 15:35 & 2014-10-17 15:46 & C6.7 & S13E75 & R, F(part) & \\
	025 & 2014-10-17 19:30 & 2014-10-17 19:42 & C6.3 & S13E72 & F & \\
	026 & 2014-10-17 20:45 & 2014-10-17 21:29 & C2.3 & S13E83 & F(part) & \\
	027 & 2014-10-18 00:55 & 2014-10-18 01:38 & C5.0 & S12E80 & R, F(part), N & \\
	028 & 2014-10-18 06:45 & 2014-10-18 07:00 & C3.7 & S13E68 & F & \\
	029 & 2014-10-18 07:02 & 2014-10-18 10:15 & M1.6 & S14E82 & R(part), F(part) & 1 \\
	030 & 2014-10-18 13:03 & 2014-10-18 13:36 & C2.6 & S14E66 & R(part), F & \\
	031 & 2014-10-18 14:23 & 2014-10-18 14:34 & C1.7 & S08E61 & & \\
	032 & 2014-10-18 16:01 & 2014-10-18 16:13 & C2.6 & S12E69 & F(part) & \\
	033 & 2014-10-18 17:03 & 2014-10-18 17:10 & C2.1 & S12E69 & R, F(part) & \\
	034 & 2014-10-18 17:11 & 2014-10-18 17:47 & C2.7 & S11E68 & R(part) & \\
	035 & 2014-10-18 19:01 & 2014-10-18 19:53 & C3.8 & S16E66 & F(part), V & 2 \\
	036 & 2014-10-18 19:54 & 2014-10-18 20:35 & C6.7 & S07E58 & R, F(part) & \\
	037 & 2014-10-19 01:17 & 2014-10-19 03:42 & C5.7 & S12E65 & R(part), F(part), N	& \\
	038 & 2014-10-19 04:00 & 2014-10-19 09:11 & X1.1 & S14E64 & F(part), N(part) & 1,2 \\
	039 & 2014-10-19 11:07 & 2014-10-19 11:49 & C4.2 & S12E57 & R(part), F(part) & \\
	040 & 2014-10-19 12:11 & 2014-10-19 12:42 & C5.8 & S12E57 & R(part) & \\
	041 & 2014-10-19 15:52 & 2014-10-19 16:18 & C3.9 & S16E52 & R(part), F(part) & \\
	042 & 2014-10-19 17:31 & 2014-10-19 17:49 & C4.7 & S12E52 & R(part), F(part) & \\
	043 & 2014-10-19 20:22 & 2014-10-19 20:45 & C2.1 & S13E50 & & \\
	044 & 2014-10-20 00:51 & 2014-10-20 01:07 & C2.5 & S18E37 & R(part) & \\
	045 & 2014-10-20 02:00 & 2014-10-20 02:25 & C2.8 & S18E36 & R, F, N & \\
	046 & 2014-10-20 02:26 & 2014-10-20 02:49 & C3.2 & S15E43 & R(part), F(part), N & \\
	047 & 2014-10-20 03:31 & 2014-10-20 04:27 & C5.4 & S15E43 & R, F(part), N & \\
	048 & 2014-10-20 05:37 & 2014-10-20 06:17 & C9.0 & S12E46 & R(part), N & \\
	049 & 2014-10-20 09:01 & 2014-10-20 09:51 & M3.9 & S16E42 & F(part) & 1 \\
	050 & 2014-10-20 11:20 & 2014-10-20 11:33 & C2.8 & S11E44 & R(part), F & \\
	051 & 2014-10-20 13:58 & 2014-10-20 14:14 & C2.7 & S11E43 & F(part) & \\
	052 & 2014-10-20 14:41 & 2014-10-20 14:48 & C3.1 & S15E36 & R, F & \\
	053 & 2014-10-20 14:57 & 2014-10-20 15:59 & C8.6 & S14E40 & R(part), F(part) & \\
	054 & 2014-10-20 16:01 & 2014-10-20 17:38 & M4.5 & S14E39 & R(part), F(part) & 1 \\
	055 & 2014-10-20 18:44 & 2014-10-20 18:55 & C6.2 & S19E45 & & \\
	056 & 2014-10-20 18:56 & 2014-10-20 19:09 & M1.4 & S15E46 & R(part), F(part), V	& 1,2 \\
	057 & 2014-10-20 19:55 & 2014-10-20 21:05 & M1.7 & S14E36 & R(part), F(part) & 1,3 \\
	058 & 2014-10-20 22:45 & 2014-10-21 00:28 & M1.2 & S14E36 & R(part), F(part), N(part) & 1,3 \\
	059 & 2014-10-21 02:13 & 2014-10-21 02:35 & C4.2 & S10E36 & R(part), F(part), N	& \\
	060 & 2014-10-21 06:01 & 2014-10-21 06:33 & C5.7 & S10E34 & R(part), F(part), N	& \\
	061 & 2014-10-21 06:55 & 2014-10-21 07:13 & C2.9 & S14E27 & R(part), F & \\
	062 & 2014-10-21 08:09 & 2014-10-21 08:15 & C3.1 & S10E29 & R, F & \\
	063 & 2014-10-21 12:26 & 2014-10-21 12:34 & C4.4 & S18E36 & R(part), F(part) & \\
	064 & 2014-10-21 13:37 & 2014-10-21 13:42 & M1.2 & S14E35 & & 1 \\
	065 & 2014-10-21 18:54 & 2014-10-21 19:56 & C4.0 & S13E23 & R(part), F(part) & \\
	066 & 2014-10-21 20:12 & 2014-10-21 21:00 & C6.5 & S20E30 & R(part), F(part) & \\
	067 & 2014-10-21 21:56 & 2014-10-21 22:44 & C3.4 & S16E21 & R, F(part) & \\
	068 & 2014-10-22 01:15 & 2014-10-22 04:44 & M8.7 & S13E21 & R(part), F(part), N	& 1,4 \\
	069 & 2014-10-22 05:13 & 2014-10-22 05:42 & M2.7 & S15E14 & F(part), N & 1 \\
	070 & 2014-10-22 09:07 & 2014-10-22 09:49 & C4.6 & S16E12 & R(part), F(part) & \\
	071 & 2014-10-22 12:01 & 2014-10-22 12:58 & C3.2 & S18E13 & R(part) & \\
	072 & 2014-10-22 14:02 & 2014-10-22 15:50 & X1.6 & S14E13 & R(part), F(part) & 1,4,5,6,7 \\
	073 & 2014-10-22 16:54 & 2014-10-22 17:07 & C5.7 & S19E17 & R(part) & \\
	074 & 2014-10-23 04:16 & 2014-10-23 05:07 & C3.7 & S13E06 & R(part), F, N & \\
	075 & 2014-10-23 09:46 & 2014-10-23 10:14 & M1.1 & S16E03 & F(part) & 1 \\
	076 & 2014-10-23 15:02 & 2014-10-23 16:45 & C4.6 & S08W01 & R(part), F(part) & \\
	077 & 2014-10-23 17:42 & 2014-10-23 18:43 & C5.9 & S13W01 & R(part), F(part) & \\
	078 & 2014-10-23 19:13 & 2014-10-23 19:23 & C3.3 & S20E04 & F(part) & \\
	079 & 2014-10-24 02:35 & 2014-10-24 02:52 & C4.2 & S19W01 & R, F, N & \\
	080 & 2014-10-24 02:54 & 2014-10-24 03:24 & C3.4 & S20W02 & R(part), F, N & \\
	081 & 2014-10-24 03:56 & 2014-10-24 04:30 & C3.6 & S21W00 & R, F(part), N & \\
	082 & 2014-10-24 07:39 & 2014-10-24 08:31 & M4.0 & S19W05 & R(part), F(part) & 1,4 \\
	083 & 2014-10-24 09:53 & 2014-10-24 12:32 & C3.6 & S14W07 & F(part) & \\
	084 & 2014-10-24 14:47 & 2014-10-24 17:05 & C5.1 & S16W15 & R(part), F(part) & \\
	085 & 2014-10-24 20:46 & 2014-10-25 01:25 & X3.1 & S22W21 & R(part), F(part) & 1,5 \\
	086 & 2014-10-25 04:07 & 2014-10-25 04:25 & C4.4 & S09W07 & R(part), F, N & \\
	087 & 2014-10-25 07:22 & 2014-10-25 09:10 & C9.2 & S16W25 & R(part), F(part) & \\
	088 & 2014-10-25 09:44 & 2014-10-25 10:11 & C4.6 & S11W22 & R(part) & \\
	089 & 2014-10-25 12:14 & 2014-10-25 13:04 & C3.2 & S13W22 & R(part), F & 2 \\
	090 & 2014-10-25 14:57 & 2014-10-25 15:18 & C5.1 & S13W23 & R, F(part) & 8 \\
	091 & 2014-10-25 15:47 & 2014-10-25 16:24 & C9.7 & S13W23 & R(part), F(part) & 8 \\
	092 & 2014-10-25 16:35 & 2014-10-25 20:00 & X1.0 & S10W22 & R(part), F(part) & 1,8 \\
	093 & 2014-10-25 23:21 & 2014-10-26 00:06 & C8.4 & S13W28 & R(part), F, N & \\
	094 & 2014-10-26 01:10 & 2014-10-26 01:26 & C3.1 & S14W29 & F(part), N & \\
	095 & 2014-10-26 05:10 & 2014-10-26 05:37 & C2.8 & S13W30 & R, F, N & \\
	096 & 2014-10-26 05:42 & 2014-10-26 06:08 & C4.0 & S14W31 & F, N & 2 \\
	097 & 2014-10-26 06:09 & 2014-10-26 06:55 & C9.5 & S14W31 & R(part), F(part), N(part) & \\
	098 & 2014-10-26 10:35 & 2014-10-26 12:10 & X2.0 & S14W37 & R(part), F(part) & 1 \\
	099 & 2014-10-26 13:05 & 2014-10-26 13:54 & C9.2 & S13W35 & R, F(part) & \\
	100 & 2014-10-26 15:06 & 2014-10-26 15:52 & C5.2 & S16W42 & R(part), F & \\
	101 & 2014-10-26 15:52 & 2014-10-26 16:09 & C3.5 & S13W39 & R & \\
	102 & 2014-10-26 17:07 & 2014-10-26 17:46 & M1.0 & S13W38 & R(part), F(part) & 1 \\
	103 & 2014-10-26 17:56 & 2014-10-26 18:07 & C7.8 & S14W38 & R, F(part) & \\
	104 & 2014-10-26 18:08 & 2014-10-26 18:40 & M4.2 & S14W37 & R(part), F & 1 \\
	105 & 2014-10-26 18:41 & 2014-10-26 19:28 & M1.9 & S13W38 & R(part) & 1 \\
	106 & 2014-10-26 19:54 & 2014-10-26 21:35 & M2.4 & S15W45 & R(part), F(part) & 1 \\
	107 & 2014-10-26 21:45 & 2014-10-26 23:00 & C8.3 & S14W43 & R(part), F(part) & \\
	108 & 2014-10-27 00:03 & 2014-10-27 01:03 & M7.1 & S12W42 & R(part), F(part), N	& 1,2,9 \\
	109 & 2014-10-27 01:45 & 2014-10-27 02:16 & M1.0 & S13W45 & R, F(part), N & 1,3 \\
	110 & 2014-10-27 03:34 & 2014-10-27 04:13 & M1.3 & S13W45 & R(part), F(part), N	& 1,3 \\
	111 & 2014-10-27 05:02 & 2014-10-27 05:16 & C3.4 & S13W44 & R, N & 2 \\
	112 & 2014-10-27 05:21 & 2014-10-27 06:11 & C4.9 & S16W50 & R(part), F, N & \\
	113 & 2014-10-27 06:57 & 2014-10-27 07:12 & C9.6 & S14W46 & R(part), F & \\
	114 & 2014-10-27 07:12 & 2014-10-27 07:30 & C9.6 & S13W46 & F & \\
	115 & 2014-10-27 09:55 & 2014-10-27 11:05 & M6.7 & S15W51 & R(part), F(part) & 1 \\
	116 & 2014-10-27 14:05 & 2014-10-27 16:56 & X2.0 & S18W57 & R(part) & 1,2 \\
	117 & 2014-10-27 17:30 & 2014-10-27 18:20 & M1.4 & S21W51 & R, F(part) & 1 \\
	118 & 2014-10-27 21:18 & 2014-10-27 21:45 & C5.4 & S13W53 & F & \\
	119 & 2014-10-27 22:52 & 2014-10-27 23:30 & C4.6 & S13W54 & F, N & \\
	120 & 2014-10-28 02:02 & 2014-10-28 03:14 & M3.4 & S17W65 & R(part), F(part) & 1 \\
	121 & 2014-10-28 06:07 & 2014-10-28 06:37 & C4.2 & S11W56 & R, N & \\
	122 & 2014-10-28 08:21 & 2014-10-28 08:34 & C6.5 & S19W62 & R(part), F & \\
	123 & 2014-10-28 11:03 & 2014-10-28 11:21 & C5.3 & S15W61 & R & \\
	124 & 2014-10-28 13:55 & 2014-10-28 15:15 & M1.6 & S16W74 & R & 1 \\
	125 & 2014-10-29 03:11 & 2014-10-29 03:21 & C3.3 & S13W69 & R, N & 2 \\
	126 & 2014-10-29 03:31 & 2014-10-29 05:12 & C8.4 & S16W79 & R(part), F(part), N	& 2 \\
	127 & 2014-10-29 06:04 & 2014-10-29 08:02 & M1.0 & S14W77 & R(part), N(part) & 1 \\
	128 & 2014-10-29 08:12 & 2014-10-29 09:22 & C6.5 & S19W62 & & \\
	129 & 2014-10-29 09:54 & 2014-10-29 11:25 & M1.2 & S18W77 & R(part), F(part) & 1 \\
	130 & 2014-10-29 13:57 & 2014-10-29 14:07 & C5.5 & S14W75 & R & \\
	131 & 2014-10-29 14:25 & 2014-10-29 15:21 & M1.4 & S13W77 & R(part), F(part) & 1 \\
	132 & 2014-10-29 16:07 & 2014-10-29 17:03 & M1.0 & S14W82 & F(part) & 1 \\
	133 & 2014-10-29 18:49 & 2014-10-29 18:57 & M1.3 & S13W77 & R & 1,10 \\
	134 & 2014-10-29 19:32 & 2014-10-29 20:19 & C6.8 & S08W89 & R(part), F(part) & 10 \\
	135 & 2014-10-29 21:21 & 2014-10-29 21:44 & M2.3 & S09W88 & R(part), F(part), V	& 1,2,10 \\
	136 & 2014-10-29 23:01 & 2014-10-29 23:16 & C2.7 & S14W86 & R, F(part), N & 2 \\
	137 & 2014-10-29 23:20 & 2014-10-29 23:35 & C3.6 & S14W88 & R, N & \\
	138 & 2014-10-29 23:40 & 2014-10-30 00:34 & C7.1 & S13W88 & F(part), N & \\
	139 & 2014-10-30 00:35 & 2014-10-30 01:00 & M1.3 & S14W81 & R, F(part), N, V & 1,2 \\
	140 & 2014-10-30 01:20 & 2014-10-30 03:40 & M3.5 & S14W88 & R(part), F(part), N	& 1,3 \\
	141 & 2014-10-30 04:18 & 2014-10-30 05:15 & M1.2 & S08W89 & R(part), F(part), N	& 1 \\
	142 & 2014-10-30 05:42 & 2014-10-30 05:53 & C3.5 & S14W88 & R, N & \\
	143 & 2014-10-30 12:38 & 2014-10-30 13:05 & C2.9 & S13W88 & R(part), F(part) & \\
	144 & 2014-10-30 15:19 & 2014-10-30 16:37 & C9.7 & S13W88 & R(part), F(part) & \\
	145 & 2014-10-31 00:32 & 2014-10-31 02:30 & C8.2 & S15W88 & R(part), F(part), N	& \\
	146 & 2014-11-12 09:21 & 2014-11-12 09:35 & C1.4 & S13E88 & F(part) & \\
	147 & 2014-11-13 05:39 & 2014-11-13 06:30 & C8.4 & S10E80 & F, N & \\
	148 & 2014-11-13 19:20 & 2014-11-13 19:27 & C1.4 & S14E78 & R & \\
	149 & 2014-11-14 07:45 & 2014-11-14 08:35 & C5.4 & S12E77 & R(part), F(part) & \\
	150 & 2014-11-14 23:27 & 2014-11-15 00:20 & C2.0 & S11E67 & R(part), F(part), N	& \\
	151 & 2014-11-15 06:11 & 2014-11-15 06:24 & C1.1 & S12E65 & R, F(part), N & \\
	152 & 2014-11-15 11:47 & 2014-11-15 13:20 & M3.2 & S11E62 & R(part), F(part) & 1 \\
	153 & 2014-11-15 20:41 & 2014-11-15 21:55 & M3.7 & S14E63 & & 1 \\
	154 & 2014-11-15 23:31 & 2014-11-16 00:49 & C2.7 & S12E45 & R(part), F(part), N	& \\
	155 & 2014-11-16 01:12 & 2014-11-16 01:28 & C1.4 & S10E45 & R, F(part), N & \\
	156 & 2014-11-16 07:29 & 2014-11-16 09:08 & C3.9 & S11E51 & R(part), F(part) & \\
	157 & 2014-11-16 09:09 & 2014-11-16 09:31 & C2.0 & S12E50 & R, F(part) & \\
	158 & 2014-11-16 10:00 & 2014-11-16 10:21 & C2.4 & S12E50 & R(part) & \\
	159 & 2014-11-16 13:40 & 2014-11-16 13:56 & C1.5 & S15E38 & F(part) & \\
	160 & 2014-11-16 16:41 & 2014-11-16 17:19 & C3.9 & S12E47 & R(part) & \\
	161 & 2014-11-16 17:35 & 2014-11-16 19:17 & M5.7 & S12E46 & R(part), F(part) & 1 \\
	162 & 2014-11-16 22:02 & 2014-11-16 22:16 & C2.4 & S12E43 & F(part) & \\
	163 & 2014-11-17 06:03 & 2014-11-17 06:11 & C1.7 & S16E43 & R(part), N & \\
	164 & 2014-11-17 18:20 & 2014-11-17 18:56 & C1.6 & S11E33 & R(part), F(part) & \\
	165 & 2014-11-18 08:02 & 2014-11-18 08:45 & C1.8 & S10E26 & R(part) & \\
	166 & 2014-11-18 14:15 & 2014-11-18 18:05 & C1.9 & S12E21 & & \\
	167 & 2014-11-18 19:23 & 2014-11-18 19:45 & C1.3 & S12E05 & R, F & \\
	168 & 2014-11-19 18:57 & 2014-11-19 19:20 & C2.4 & S12E05 & R, F & \\
	169 & 2014-11-20 19:37 & 2014-11-20 20:50 & C2.5 & S12W08 & R(part), F(part) & \\
	170 & 2014-11-20 21:01 & 2014-11-20 22:00 & C1.3 & S12W27 & F(part) & \\
	171 & 2014-11-21 01:54 & 2014-11-21 02:39 & C1.4 & S11W13 & F(part), N & \\
	172 & 2014-11-21 04:44 & 2014-11-21 04:54 & C1.9 & S12W13 & R, F, N & \\
	173 & 2014-11-21 23:29 & 2014-11-21 23:45 & C1.1 & S13W20 & R, F(part), N & \\
	174 & 2014-11-22 00:08 & 2014-11-22 00:51 & C2.3 & S14W26 & R(part), F(part), N	& \\
	175 & 2014-11-22 00:56 & 2014-11-22 02:35 & C8.1 & S12W25 & R(part), F(part), N	& \\
	176 & 2014-11-22 03:36 & 2014-11-22 04:02 & C2.4 & S12W26 & R(part), F(part), N	& \\
	177 & 2014-11-22 05:58 & 2014-11-22 06:34 & C6.5 & S13W23 & R(part), F(part), N	& \\
	178 & 2014-11-22 09:51 & 2014-11-22 11:05 & C1.5 & S11W31 & R(part), F(part) & \\
	179 & 2014-11-22 13:22 & 2014-11-22 14:03 & C2.1 & S23W43 & R(part) & \\
	180 & 2014-11-22 17:09 & 2014-11-22 17:36 & C3.1 & S12W33 & R(part), F(part) & \\
	181 & 2014-11-23 05:30 & 2014-11-23 06:08 & C2.4 & S16W43 & R, N & \\
	\hline
	\multicolumn{7}{l}{\textbf{References.} (1) \citet{2016ApJ...833..284I}; (2) \citet{2016SoPh..291.3439M}; (3) \citet{2016ApJ...830..110C}; (4) \citet{2016ApJ...816....6K};} \\
	\multicolumn{7}{l}{(5) \citet{2016SoPh..291.3385K}; (6) \citet{2017ApJ...840..116B}; (7) \citet{2017ApJ...836..150L}; (8) \citet{2017ApJ...838..134B}; (9) \citet{2017A&A...597L...4L};} \\
	\multicolumn{7}{l}{(10) \citet{2017ApJ...838...32Y}.}
	\label{tab:allflares}
\end{longtable}
}

\longtab{
\begin{longtable}{c c c c c c c}	
	\caption{List of flares with a peak above the 95\% confidence level in the power spectrum. The first column contains a numerical label for the flares (see also Table \ref{tab:allflares}), the second and third columns give the start and end times of the section of the flare where the QPP signal is most visible in the power spectrum, the fourth column is the instrument used, the fifth column is the type of power spectrum in which the signal was detected (the numbers for the rebinned power spectra refer to the number of frequency bins that were summed over, see Sect. \ref{sec:dat}), the sixth column is the QPP period, and finally the seventh column gives references to other studies that find pulsations in the flare. Flare numbers marked with an asterisk indicate those which occurred while the active region was far enough away from the solar limb so that active region properties could be determined.}\\
	\hline\hline
	Flare no. & Start time & End time & Instrument & Method & Period (s) & References \\[2.3pt]
	\hline\\
	\endfirsthead
	\multicolumn{7}{c}
	{\tablename\ \thetable\ -- \textit{Continued from previous page}}\\
	\hline\hline
	Flare no. & Start time & End time & Instrument & Method & Period (s) & References \\[2.3pt]
	\hline\\
	\endhead
	\hline \multicolumn{7}{r}{\textit{Continued on next page}}\\
	\endfoot
	\endlastfoot
	008* & 2014-09-23 23:08:20 & 2014-09-23 23:13:52 & GOES 0.5--4\,\AA & Regular & $41.2^{+2.7}_{-2.4}$ & \\[2.3pt] 
	010* & 2014-09-24 17:49:04 & 2014-09-24 17:50:18 & GOES 1--8\,\AA & Rebinned (2) & $9.6^{+1.4}_{-1.1}$ & 2 \\[2.3pt]
	010* & 2014-09-24 17:49:01 & 2014-09-24 17:49:50 & Vernov & Rebinned (2) & $5.8^{+0.8}_{-0.6}$ & 2 \\[2.3pt]
	022 & 2014-10-17 05:23:18 & 2014-10-17 05:26:35 & NoRH & Rebinned (3) & $39.2^{+16.8}_{-9.0}$ & \\[2.3pt]
	024 & 2014-10-17 15:35:30 & 2014-10-17 15:37:54 & GOES 1--8\,\AA & Rebinned (2) & $22.1^{+4.0}_{-3.0}$ & \\[2.3pt]
	027 & 2014-10-18 01:04:10 & 2014-10-18 01:08:30 & NoRH & Rebinned (3) & $8.0^{+1.4}_{-1.0}$ & \\[2.3pt]
	029 & 2014-10-18 07:36:14 & 2014-10-18 07:48:48 & GOES 1--8\,\AA & Regular & $50.1^{+1.7}_{-1.6}$ & 1 \\[2.3pt]
	030 & 2014-10-18 13:14:38 & 2014-10-18 13:15:52 & GOES 0.5--4\,\AA & Regular & $14.3^{+1.6}_{-1.3}$ & \\[2.3pt]
	035 & 2014-10-18 19:02:30 & 2014-10-18 19:06:58 & GOES 1--8\,\AA & Regular & $13.3^{+0.4}_{-0.3}$ & 2 \\[2.3pt]
	037 & 2014-10-19 01:35:00 & 2014-10-19 01:43:00 & GOES 0.5--4\,\AA & Regular & $79.5^{+7.2}_{-6.1}$ & \\[2.3pt]
	037 & 2014-10-19 01:36:28 & 2014-10-19 01:42:42 & NoRH & Regular & $24.9^{+0.9}_{-0.8}$ & \\[2.3pt]
	038 & 2014-10-19 04:20:24 & 2014-10-19 04:24:56 & GOES 1--8\,\AA & Regular & $53.7^{+5.9}_{-4.9}$ & \\[2.3pt]
	038 & 2014-10-19 04:20:24 & 2014-10-19 04:24:56 & GOES 0.5--4\,\AA & Regular & $24.4^{+1.2}_{-1.1}$ & \\[2.3pt]
	038 & 2014-10-19 04:41:39 & 2014-10-19 04:50:00 & GOES 0.5--4\,\AA & Regular & $45.4^{+2.2}_{-2.0}$ & \\[2.3pt]
	049* & 2014-10-20 09:05:45 & 2014-10-20 09:08:14 & GOES 0.5--4\,\AA & Rebinned (3) & $14.7^{+2.6}_{-1.9}$ & 1 \\[2.3pt]
	052* & 2014-10-20 14:41:47 & 2014-10-20 14:43:33 & GOES 1--8\,\AA & Regular & $26.1^{+3.7}_{-2.9}$ & \\[2.3pt]
	054* & 2014-10-20 16:23:02 & 2014-10-20 16:31:20 & GOES 1--8\,\AA & Regular & $35.4^{+1.3}_{-1.2}$ & \\[2.3pt]
	056* & 2014-10-20 18:57:51 & 2014-10-20 18:59:01 & Fermi 25--50\,keV & Rebinned (3) & $13.9^{+6.0}_{-3.2}$ & 1,2 \\[2.3pt]
	056* & 2014-10-20 18:57:53 & 2014-10-20 18:59:02 & Vernov & Regular & $17.2^{+2.5}_{-1.9}$ & 1,2 \\[2.3pt]
	058* & 2014-10-20 22:45:18 & 2014-10-20 22:49:46 & GOES 1--8\,\AA & Rebinned (2) & $48.4^{+10.8}_{-7.5}$ & 1,3 \\[2.3pt]
	058* & 2014-10-20 22:45:18 & 2014-10-20 22:49:46 & GOES 0.5--4\,\AA & Rebinned (2) & $48.4^{+10.8}_{-7.5}$ & 1,3 \\[2.3pt]
	068* & 2014-10-22 01:43:04 & 2014-10-22 01:46:36 & GOES 1--8\,\AA & Regular & $21.1^{+1.1}_{-1.0}$ & \\[2.3pt]
	068* & 2014-10-22 01:43:04 & 2014-10-22 01:46:36 & GOES 0.5--4\,\AA & Regular & $21.1^{+1.1}_{-1.0}$ & \\[2.3pt]
	072* & 2014-10-22 14:06:55 & 2014-10-22 14:12:02 & Fermi 50--100\,keV & Regular & $30.6^{+1.6}_{-1.4}$ & 5 \\[2.3pt]
	072* & 2014-10-22 14:06:56 & 2014-10-22 14:09:30 & GOES 1--8\,\AA & Regular & $30.3^{+3.4}_{-2.8}$ & 5 \\[2.3pt]
	072* & 2014-10-22 14:15:24 & 2014-10-22 14:23:40 & GOES 0.5--4\,\AA & Regular & $49.4^{+2.6}_{-2.4}$ & 5 \\[2.3pt]
	079* & 2014-10-24 02:38:30 & 2014-10-24 02:41:20 & NoRH & Rebinned (2) & $7.9^{+0.4}_{-0.3}$ & \\[2.3pt]
	081* & 2014-10-24 03:59:30 & 2014-10-24 04:01:00 & NoRH & Rebinned (3) & $14.8^{+5.0}_{-3.0}$ & \\[2.3pt]
	085* & 2014-10-24 21:19:38 & 2014-10-24 21:23:47 & GOES 0.5--4\,\AA & Regular & $49.6^{+5.5}_{-4.5}$ & \\[2.3pt]
	092* & 2014-10-25 17:02:11 & 2014-10-25 17:10:10 & GOES 1--8\,\AA & Regular & $25.1^{+0.7}_{-0.6}$ & 1 \\[2.3pt]
	092* & 2014-10-25 17:02:11 & 2014-10-25 17:10:10 & GOES 0.5--4\,\AA & Regular & $25.1^{+0.7}_{-0.6}$ & 1 \\[2.3pt]
	098* & 2014-10-26 10:48:52 & 2014-10-26 10:50:34 & Fermi 25--50\,keV & Regular & $20.3^{+2.2}_{-1.9}$ & \\[2.3pt]
	104* & 2014-10-26 18:11:18 & 2014-10-26 18:15:24 & GOES 0.5--4\,\AA & Regular & $20.3^{+0.9}_{-0.8}$ & 1 \\[2.3pt]
	104* & 2014-10-26 18:12:40 & 2014-10-26 18:15:28 & Fermi 25--50\,keV & Regular & $20.9^{+1.4}_{-1.2}$ & 1 \\[2.3pt]
	105* & 2014-10-26 18:45:04 & 2014-10-26 18:48:02 & GOES 1--8\,\AA & Regular & $25.2^{+1.9}_{-1.7}$ & 1 \\[2.3pt]
	106* & 2014-10-26 20:03:42 & 2014-10-26 20:11:18 & GOES 1--8\,\AA & Rebinned (2) & $36.4^{+3.2}_{-2.7}$ & 1 \\[2.3pt]
	106* & 2014-10-26 20:03:42 & 2014-10-26 20:11:18 & GOES 0.5--4\,\AA & Rebinned (2) & $36.4^{+3.2}_{-2.7}$ & 1 \\[2.3pt]
	106* & 2014-10-26 20:03:38 & 2014-10-26 20:09:02 & Fermi 25--50\,keV & Regular & $31.9^{+1.7}_{-1.5}$ & 1 \\[2.3pt]
	117* & 2014-10-27 17:36:36 & 2014-10-27 17:37:26 & GOES 1--8\,\AA & Regular & $12.3^{+1.8}_{-1.4}$ & 1 \\[2.3pt]
	129 & 2014-10-29 09:58:00 & 2014-10-29 10:01:45 & EVE & Rebinned (2) & $26.4^{+3.5}_{-2.8}$ & 1 \\[2.3pt]
	135 & 2014-10-29 21:21:33 & 2014-10-29 21:22:19 & Fermi 25--50\,keV & Rebinned (3) & $7.5^{+2.5}_{-1.5}$ & 1,2 \\[2.3pt]
	135 & 2014-10-29 21:21:28 & 2014-10-29 21:22:52 & Vernov & Rebinned (3) & $7.6^{+1.2}_{-0.9}$ & 1,2 \\[2.3pt]
	138 & 2014-10-29 23:44:16 & 2014-10-29 23:45:48 & NoRH & Regular & $10.1^{+0.6}_{-0.5}$ & \\[2.3pt]
	140 & 2014-10-30 01:28:38 & 2014-10-30 01:30:30 & GOES 1--8\,\AA & Rebinned (2) & $24.6^{+7.0}_{-4.5}$ & 1,3 \\[2.3pt]
	141 & 2014-10-30 04:22:08 & 2014-10-30 04:28:00 & GOES 1--8\,\AA & Regular & $21.9^{+0.7}_{-0.7}$ & 1 \\[2.3pt]
	142 & 2014-10-30 05:44:27 & 2014-10-30 05:45:45 & GOES 1--8\,\AA & Regular & $18.9^{+2.7}_{-2.1}$ & \\[2.3pt]
	142 & 2014-10-30 05:44:27 & 2014-10-30 05:45:43 & NoRH & Regular & $18.7^{+2.7}_{-2.1}$ & \\[2.3pt]
	147 & 2014-11-13 06:04:59 & 2014-11-13 06:07:15 & NoRH & Regular & $9.0^{+0.3}_{-0.3}$ & \\[2.3pt]
	152 & 2014-11-15 11:56:08 & 2014-11-15 12:02:58 & GOES 1--8\,\AA & Regular & $27.2^{+0.9}_{-0.9}$ & 1 \\[2.3pt]
	152 & 2014-11-15 11:56:08 & 2014-11-15 12:02:58 & GOES 0.5--4\,\AA & Regular & $27.2^{+0.9}_{-0.9}$ & 1 \\[2.3pt]
	152 & 2014-11-15 11:56:08 & 2014-11-15 12:02:58 & EVE & Regular & $27.3^{+0.9}_{-0.9}$ & 1 \\[2.3pt]
	153 & 2014-11-15 20:42:57 & 2014-11-15 20:44:28 & GOES 0.5--4\,\AA & Rebinned (2) & $20.0^{+5.7}_{-3.6}$ & 1 \\[2.3pt]
	161* & 2014-11-16 17:42:46 & 2014-11-16 17:45:24 & GOES 0.5--4\,\AA & Regular & $19.5^{+1.3}_{-1.2}$ & \\[2.3pt]
	177* & 2014-11-22 06:02:16 & 2014-11-22 06:04:48 & GOES 1--8\,\AA & Regular & $18.7^{+1.2}_{-1.1}$ & \\[2.3pt]
	177* & 2014-11-22 06:02:16 & 2014-11-22 06:04:48 & GOES 0.5--4\,\AA & Regular & $18.7^{+1.2}_{-1.1}$ & \\[2.3pt]
	\hline
	\multicolumn{7}{l}{\textbf{References.} (1) \citet{2016ApJ...833..284I}; (2) \citet{2016SoPh..291.3439M}; (3) \citet{2016ApJ...830..110C}; (5) \citet{2016SoPh..291.3385K}} \\
	\label{tab:qppflares}
\end{longtable}
}

\begin{figure*}
	\centering
	\includegraphics[width=0.9\linewidth]{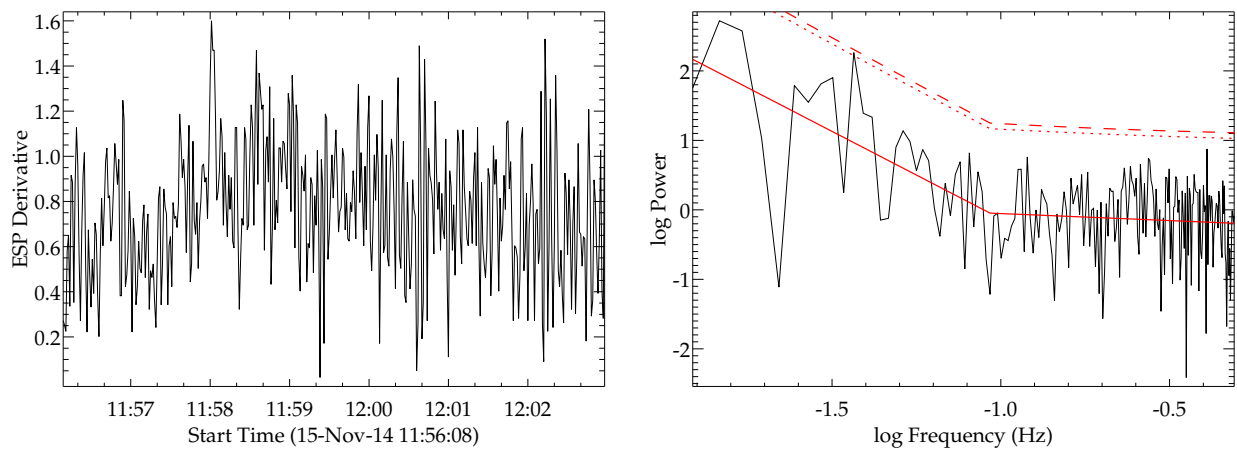}
	\caption{Left: the time derivative of a section of flare 152 observed by EVE/ESP. Right: the corresponding power spectrum, where the red solid line is a broken power law fit to the spectrum, the red dotted line represents the 95\% confidence level, and the red dashed line the 99\% level. One peak is above the 95\% level, at a period of $27.3 \pm 0.9$\,s.}
	\label{152eve}
\end{figure*}

\subsection{Comparing QPP and flare properties}

\begin{figure}
	\centering
	\includegraphics[width=0.9\linewidth]{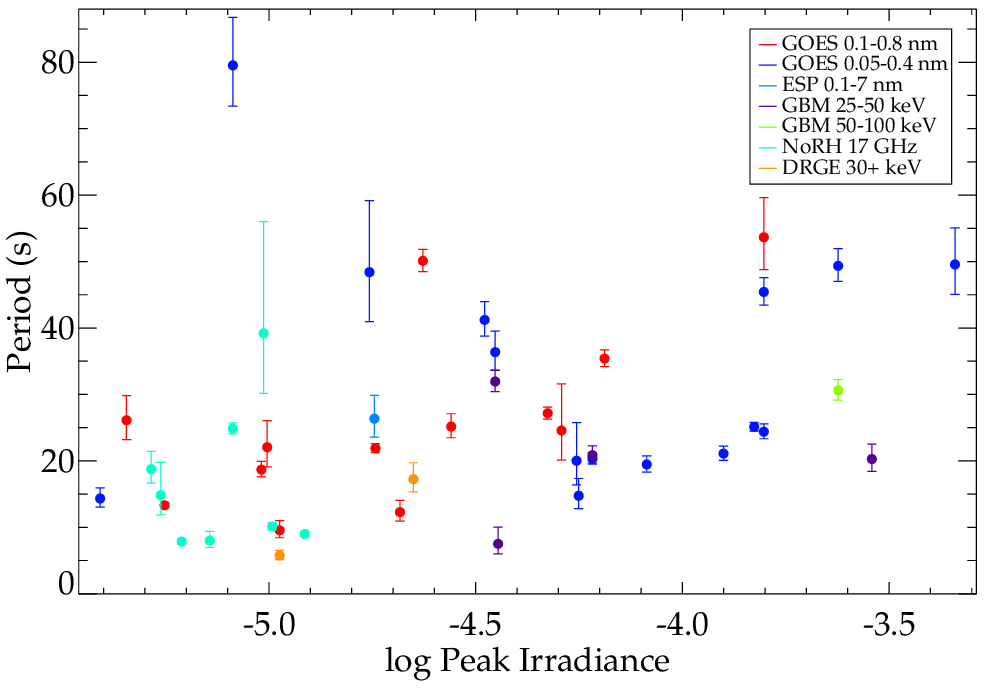}
	\caption{QPP periods plotted against the peak GOES/XRS 1-8\,$\AA$ irradiance, where the different colours correspond to the different instruments and wavebands used to observe the flares. The Pearson correlation coefficient is 0.33, suggesting a very slight positive correlation.}
	\label{fig:ampl}
\end{figure}

\begin{figure}
	\centering
	\includegraphics[width=0.9\linewidth]{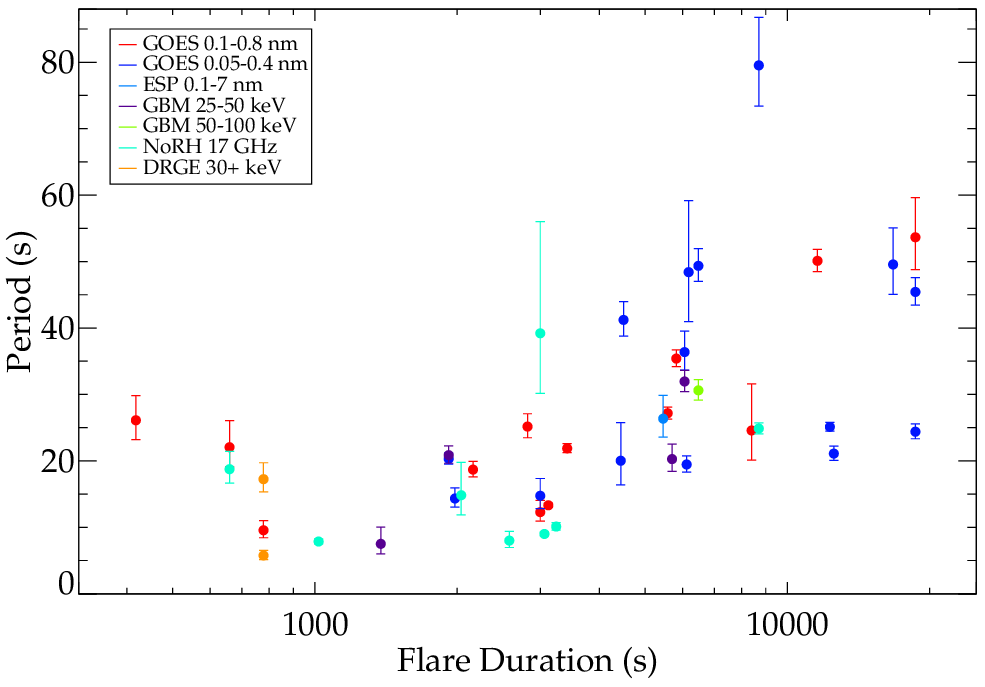}
	\caption{QPP periods plotted against the duration of the flares. The Pearson correlation coefficient is 0.59, suggesting a positive correlation.}
	\label{fig:dur}
\end{figure}

Figures \ref{fig:ampl} and \ref{fig:dur} show a weak correlation of the QPP period the flare amplitude in the GOES 1-8\,\AA\ waveband and a moderate correlation with the flare duration, which is consistent with the flare amplitude being correlated with the duration \citep[e.g.][]{1993ApJ...412..841L, 2002A&A...382.1070V}. The flare durations were estimated from the GOES 1-8\,\AA\ data, and were taken as the time between when the intensity begins increasing above the base level and when the intensity returns to the base level. Duplicate points, where the same QPP signal can be seen in multiple instruments or wavelength ranges, are omitted. This apparent correlation may be due to observational constraints, however, since the short periods tend to be more easily visible in the shorter duration flares, while the detection of long periods is only possible in the longer duration flares. In addition, \citet{2016ApJ...833..284I} found no correlation between the period and GOES class for a larger sample of flares.

\begin{figure}
	\centering
	\includegraphics[width=0.9\linewidth]{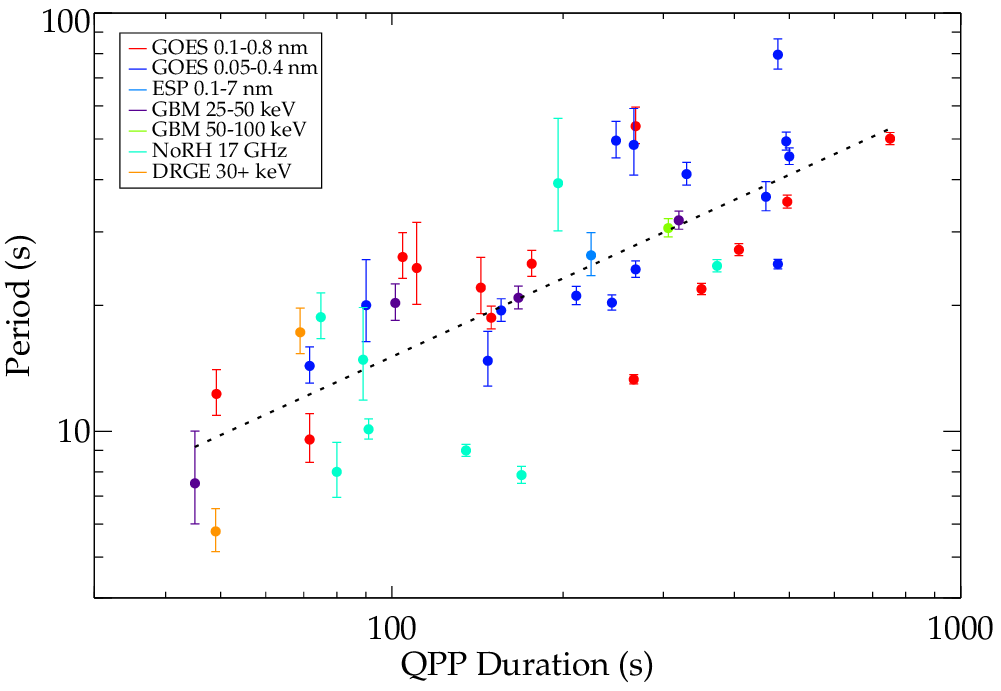}
	\caption{QPP periods plotted against the duration of the QPP signal. The Pearson correlation coefficient of 0.76 shows a positive correlation. The black dotted line shows a linear fit.}
	\label{fig:qppdur}
\end{figure}

Figure \ref{fig:qppdur} shows a positive correlation between the QPP period and the duration of the QPP signal, for which we use the time interval that gave the most significant peak in the power spectrum (see Sect.~\ref{sec:dat}). Fitting a linear model gives a relationship:
\begin{equation}
	\log P = (0.62 \pm 0.03)\log \tau - (0.07 \pm 0.07),
\end{equation}
where $P$ is the period and $\tau$ is the QPP signal duration time. Observational constraints mean that the maximum detectable period will depend on the time interval being used, so longer periods will require longer durations, however this does not explain why short periods with longer durations are not seen. We also note that the relationship is different from the period versus decay time relationships found by \citet{2016ApJ...830..110C} and \citet{2016MNRAS.459.3659P}, although these studies focussed on decaying QPPs whereas many of the QPP signals included in the present study do not show a clear decay, and appear more like a set of periodic pulses rather than a harmonic signal.

\begin{figure}
	\centering
	\includegraphics[width=0.9\linewidth]{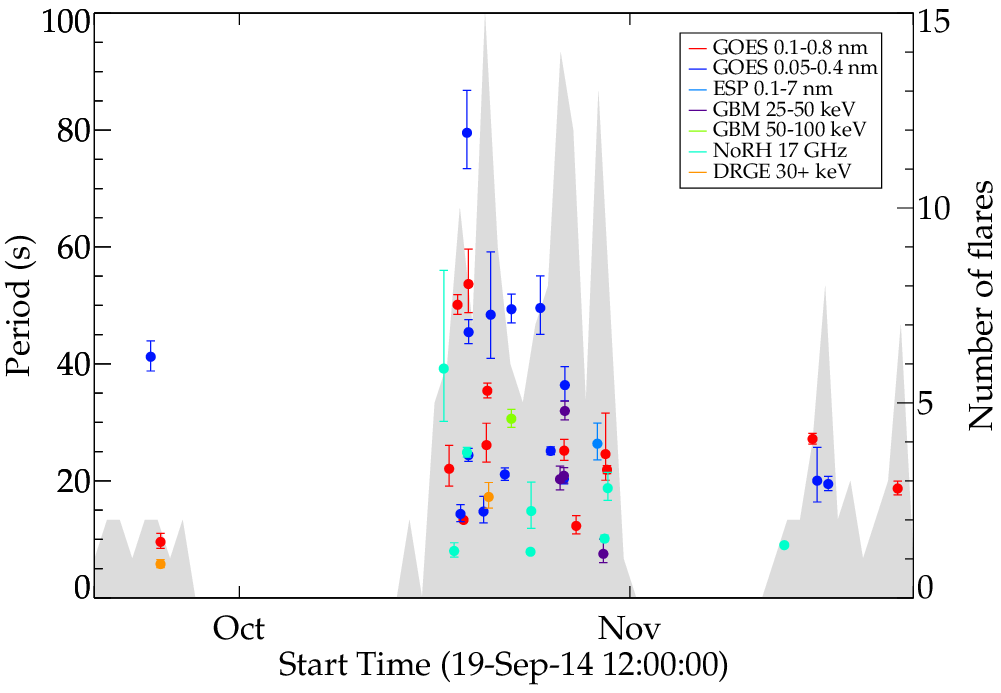}
	\caption{QPP periods plotted against the approximate time at which the QPP signal begins. There is no obvious trend suggesting that there is no characteristic timescale which is evolving with time. The grey shaded region indicates the number of flares that occurred on a particular day.}
	\label{fig:time}
\end{figure}

A plot of QPP period against the time at which the QPP signal occurs is shown in Fig. \ref{fig:time}, where there is no clear trend in how the periods change with time. It can be seen that the majority of the QPPs were found during the AR's second crossing of the solar disk (as NOAA 12192), but this is simply because most of the flares occurred during this time as shown by the grey shaded region.

\subsection{Comparing QPP and active region properties}

\begin{figure}
	\centering
	\includegraphics[width=0.9\linewidth]{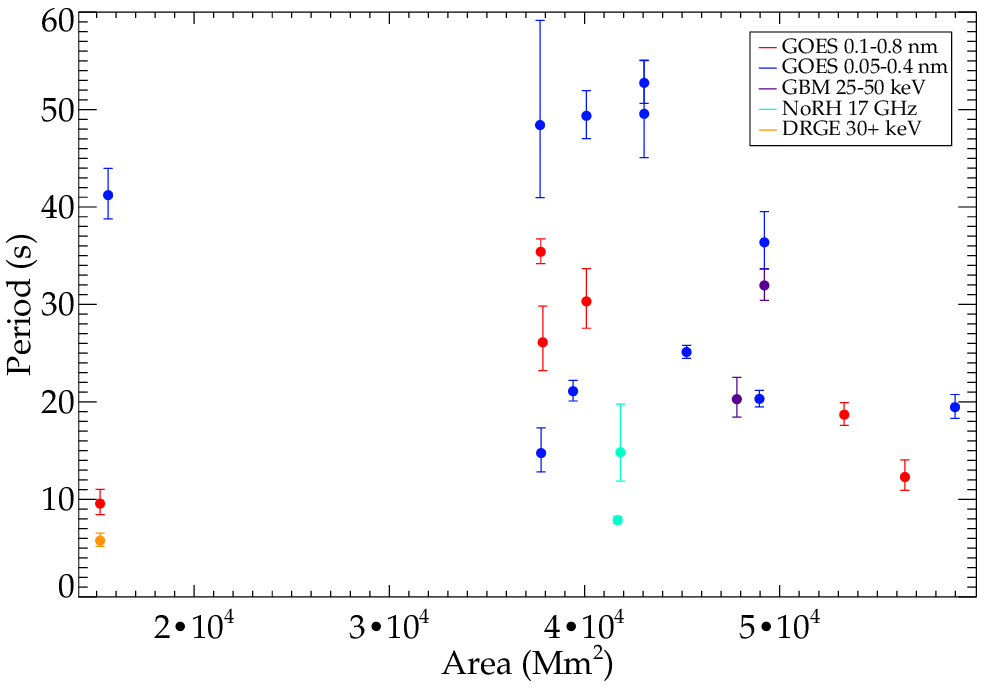}
	\caption{QPP periods plotted against the area of the AR at the time of the flare. The Pearson correlation coefficient is 0.05, suggesting no correlation.}
	\label{fig:area}
\end{figure}

\begin{figure}
	\centering
	\includegraphics[width=0.9\linewidth]{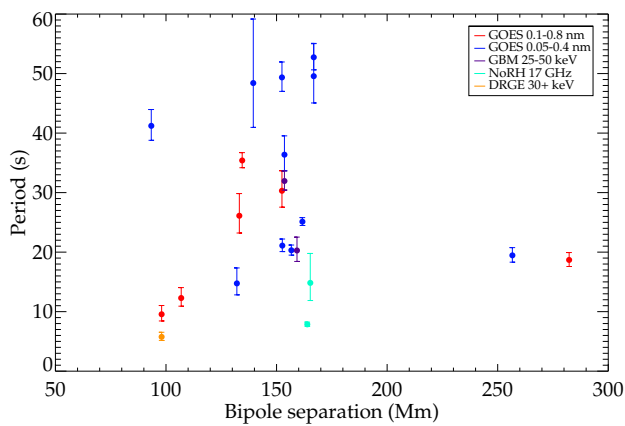}
	\caption{QPP periods plotted against the separation of the centres of positive and negative magnetic flux in the AR at the time of the flare. The Pearson correlation coefficient is 0.01, suggesting no correlation.}
	\label{fig:sep}
\end{figure}

\begin{figure}
	\centering
	\includegraphics[width=0.9\linewidth]{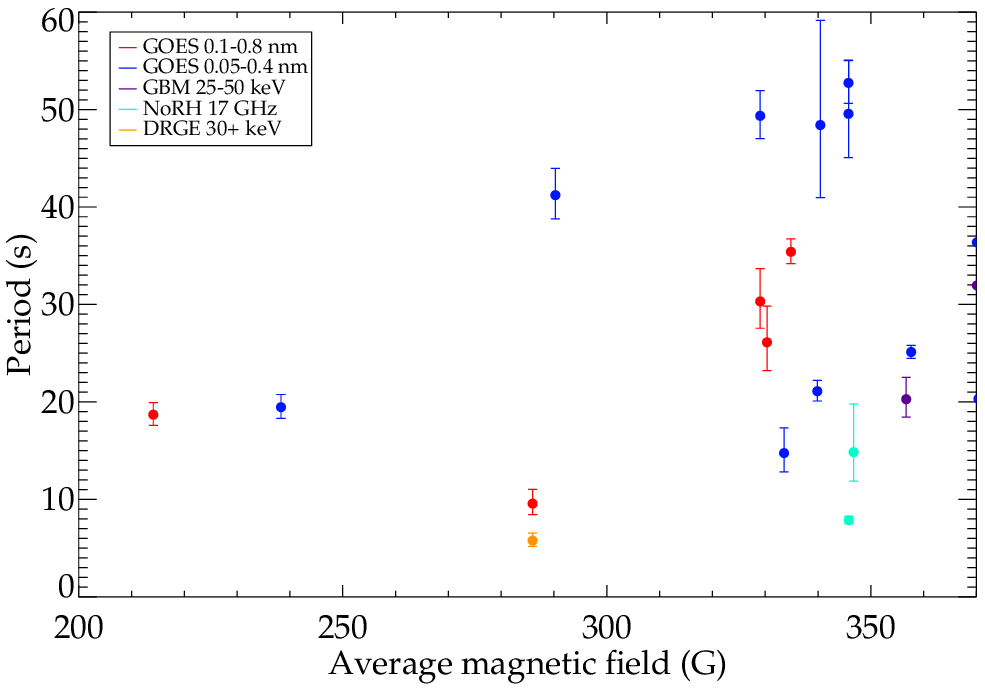}
	\caption{QPP periods plotted against the average magnetic field strength of the AR at the time of the flare. The Pearson correlation coefficient is 0.19, suggesting no correlation.}
	\label{fig:flux}
\end{figure}

Figures \ref{fig:area}--\ref{fig:flux} show scatter plots of the QPP period with total area, bipole separation distance, and average photospheric magnetic field strength of the AR around the time of the QPP signal onset, respectively. These plots show no correlation between the QPP periods and AR properties, which, if the characteristic timescale of the QPPs is assumed to be related to a characteristic length scale, suggests that the fine structure of the AR may be important since different structures within the AR will have different length scales. Another possibility is that for different QPP mechanisms, the characteristic length scale has a different relationship with the characteristic timescale, and that different mechanisms are responsible for different QPP signals.

\section{Conclusions}
\label{sec:con}

In this paper we have shown that 37 out of 181 flares from a single active region show good evidence of having stationary or weakly non-stationary QPPs using methods that limit the potential for false detections, and using data from several different instruments observing different wavelength ranges. On the other hand, this may be a lower limit for the true number of flares in the sample with QPPs. For example, the presence of background trends due to the flare itself can mask QPP signals in the power spectrum \citep{2017A&A...602A..47P}, the QPPs could be non-stationary (where the period drifts with time, e.g. \citealt{2010SoPh..267..329K}) and hence would not have a well defined peak in the power spectrum, the QPPs could be too low amplitude or have the wrong period to be detectable with the instruments operating at the time, or lower quality QPP signals could have been missed during the manual search stage. Additionally we show how taking the time derivative of light curve data, which has previously been shown to be useful when searching for QPPs \citep{2015SoPh..290.3625S, 2016ApJ...827L..30H, 2017ApJ...836...84D}, impacts the power spectrum, and suggest how this can be accounted for when searching for periodic signals. Out of the 44 flares in this sample that overlap with those included by \citet{2016ApJ...833..284I}, we find the same periods in 6 (or 13 if the selection criteria used by \citet{2016ApJ...833..284I} are relaxed) and agree with the lack of evidence of a QPP signal in a further 24. For the other flares either only one method identifies a periodic signal, or the periods identified by the different methods are different. The mean period for the QPPs in our sample is $20^{+16}_{-9}$\,s. We also find a significant correlation between the period and QPP duration, and while we acknowledge that observational constraints may be part of the cause we believe this cannot fully explain the strong correlation.

Three properties of the AR from which the flares originate (namely area, bipole separation and average photospheric magnetic field strength) have been tracked over time, to test for any correlation with the QPP periods. No correlations were found, which could either suggest that the small-scale structure within the AR is more important, that different mechanisms act in different cases, or that the sausage mode is responsible for the QPPs, since the oscillation period may be only weakly dependent on the length of the hosting coronal loop for the sausage mode \citep{2012ApJ...761..134N}. For this reason we aim to make use of spatially resolved observations of the flares themselves in future work.


\begin{acknowledgements}
We thank the anonymous referee for the detailed review that helped improve this paper. C.E.P. \& V.M.N.: This work was supported by the European Research Council under the \textit{SeismoSun} Research Project No. 321141. A.-M.B. thanks the Institute of Advanced Study, University of Warwick for their support. C.E.P. would like to thank Paulo Sim\~{o}es and Hugh Hudson for advice on handling the \emph{Fermi}/GBM and GOES/XRS data. The authors would like to acknowledge the ISSI International Team led by A.-M.B. for many productive discussions, and are grateful to the GOES/XRS, SDO/EVE and HMI, \emph{Fermi}/GBM, RHESSI, and Nobeyama teams for providing the data used, as well as Paul A. Higgins for making the SMART routines available online at https://github.com/pohuigin/smart\_library. Flare information was provided courtesy of SolarMonitor.org.
\end{acknowledgements}


\bibliographystyle{aa}
\bibliography{ms}

\begin{appendix}

\section{Additional figures}
\label{sec:figs}

\begin{figure}
	\centering
	\includegraphics[width=\linewidth]{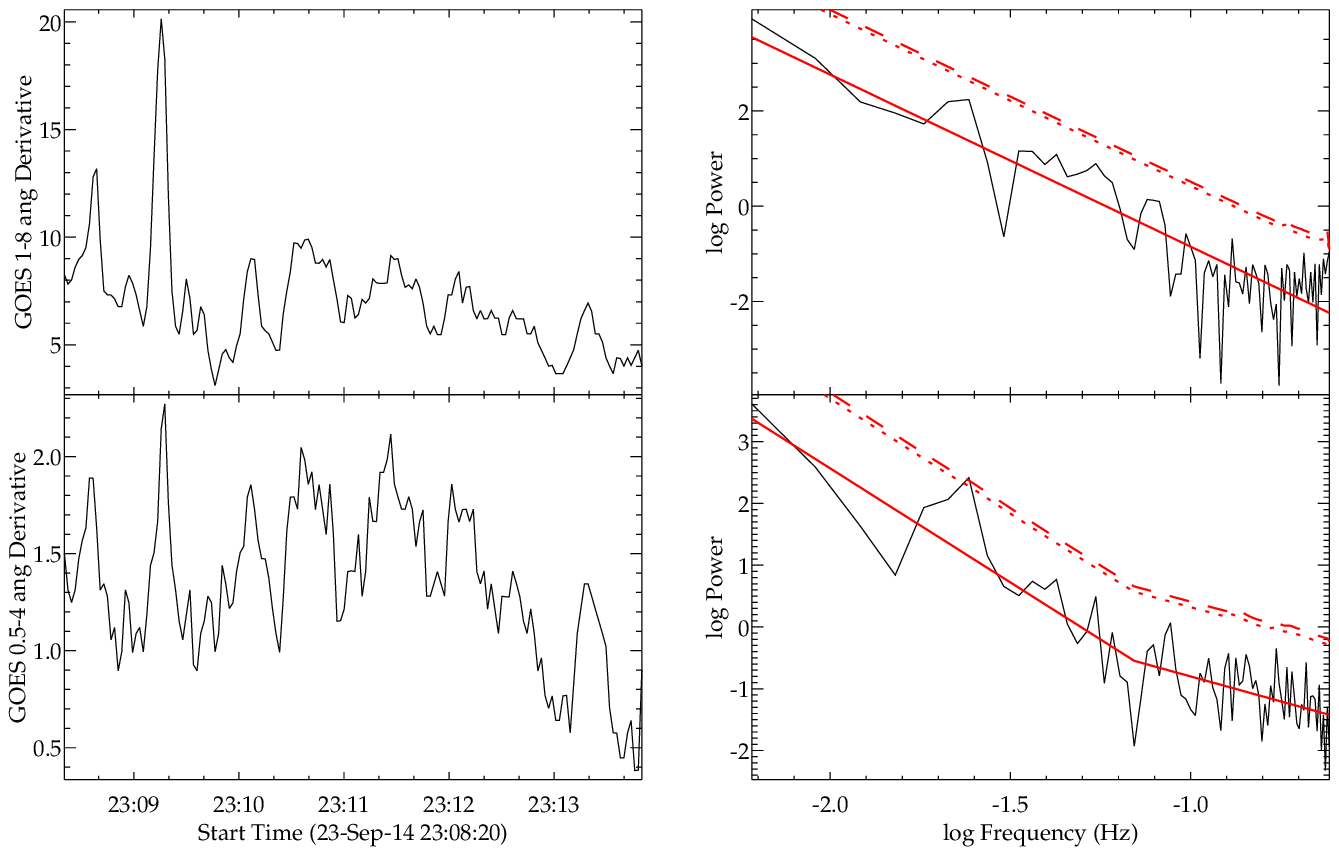}
	\caption{Similar to Fig. \ref{152goes}, with GOES/XRS data for flare 008.}
	\label{008goes}
\end{figure}

\begin{figure}
	\centering
	\includegraphics[width=\linewidth]{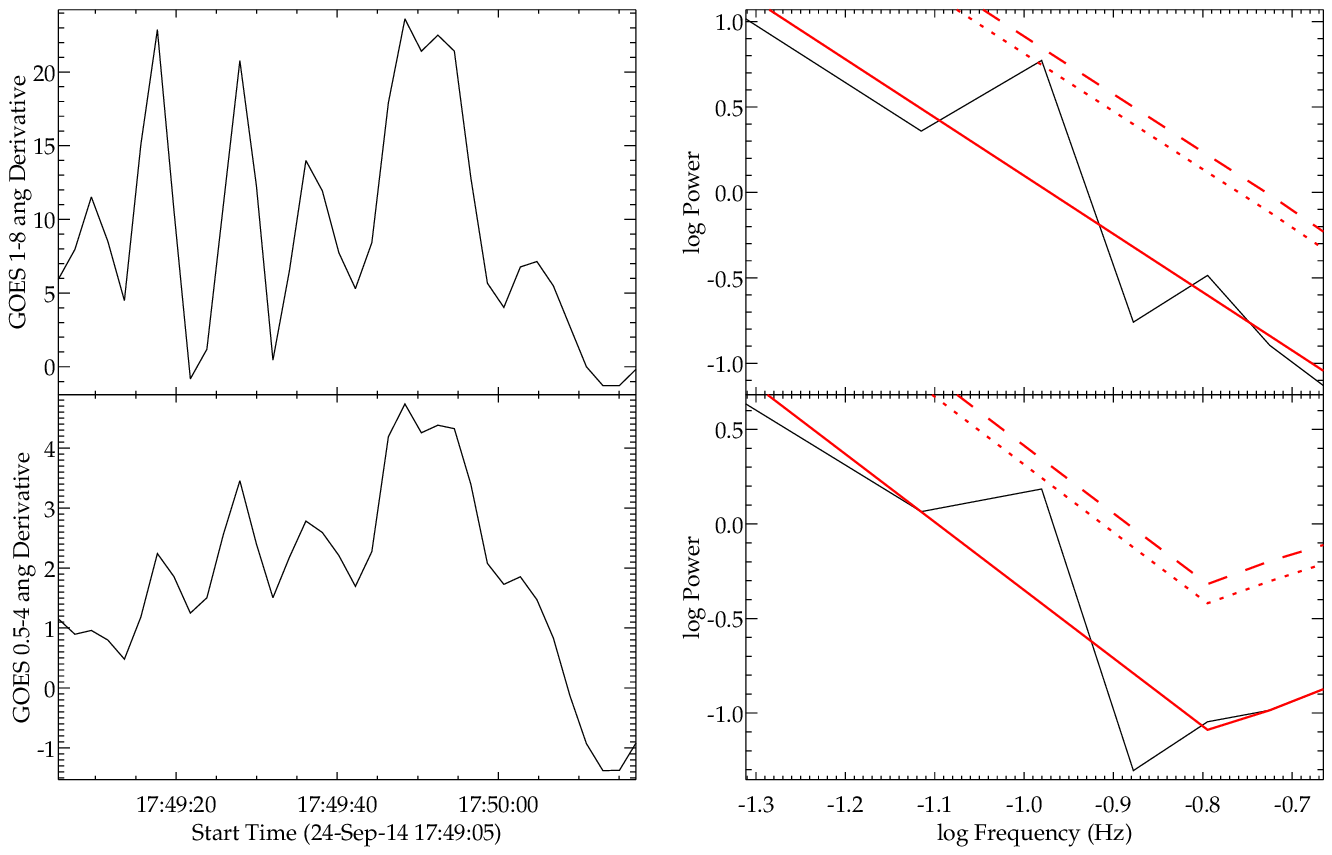}
	\caption{Similar to Fig. \ref{152goes}, with GOES/XRS data for flare 010.}
	\label{010goes}
\end{figure}

\begin{figure}
	\centering
	\includegraphics[width=\linewidth]{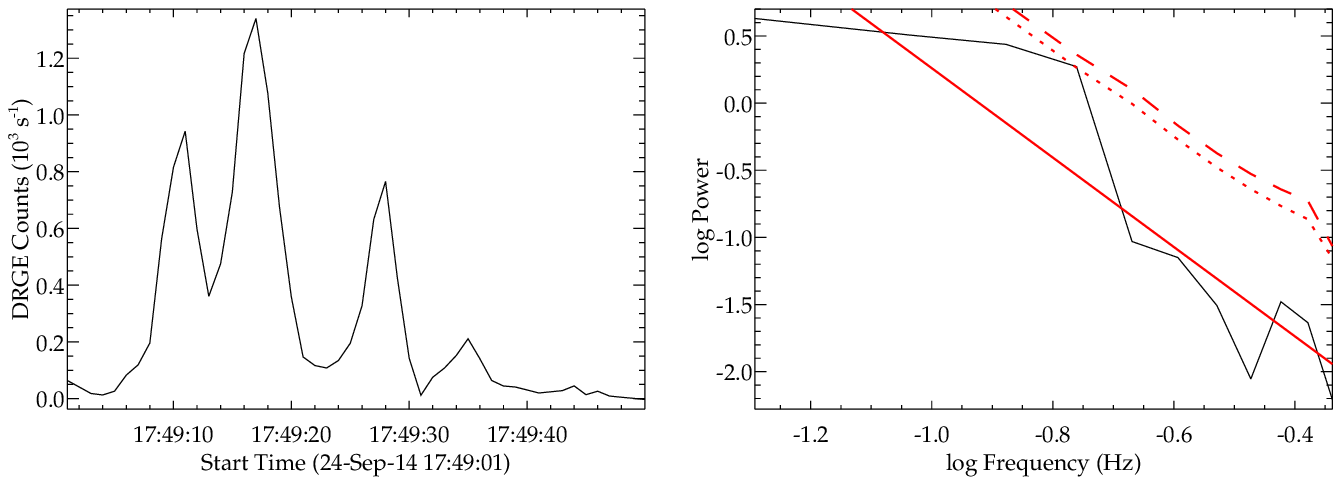}
	\caption{Similar to Fig. \ref{152goes}, with \emph{Vernov}/DRGE data for flare 010.}
	\label{010vern}
\end{figure}

\begin{figure}
	\centering
	\includegraphics[width=\linewidth]{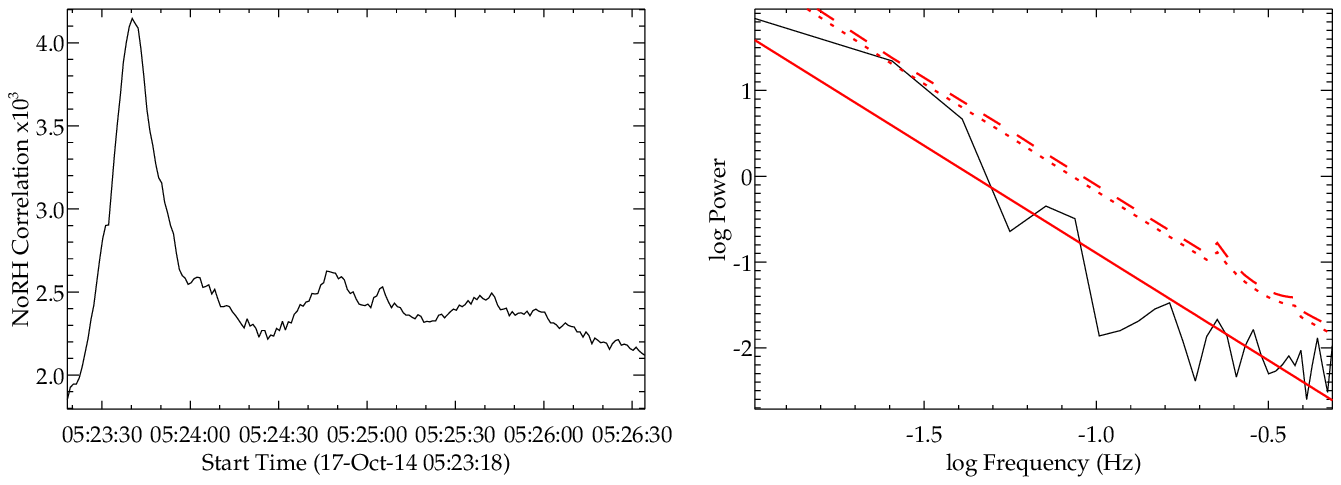}
	\caption{Similar to Fig. \ref{152goes}, with NoRH data for flare 022.}
	\label{022norh}
\end{figure}

\begin{figure}
	\centering
	\includegraphics[width=\linewidth]{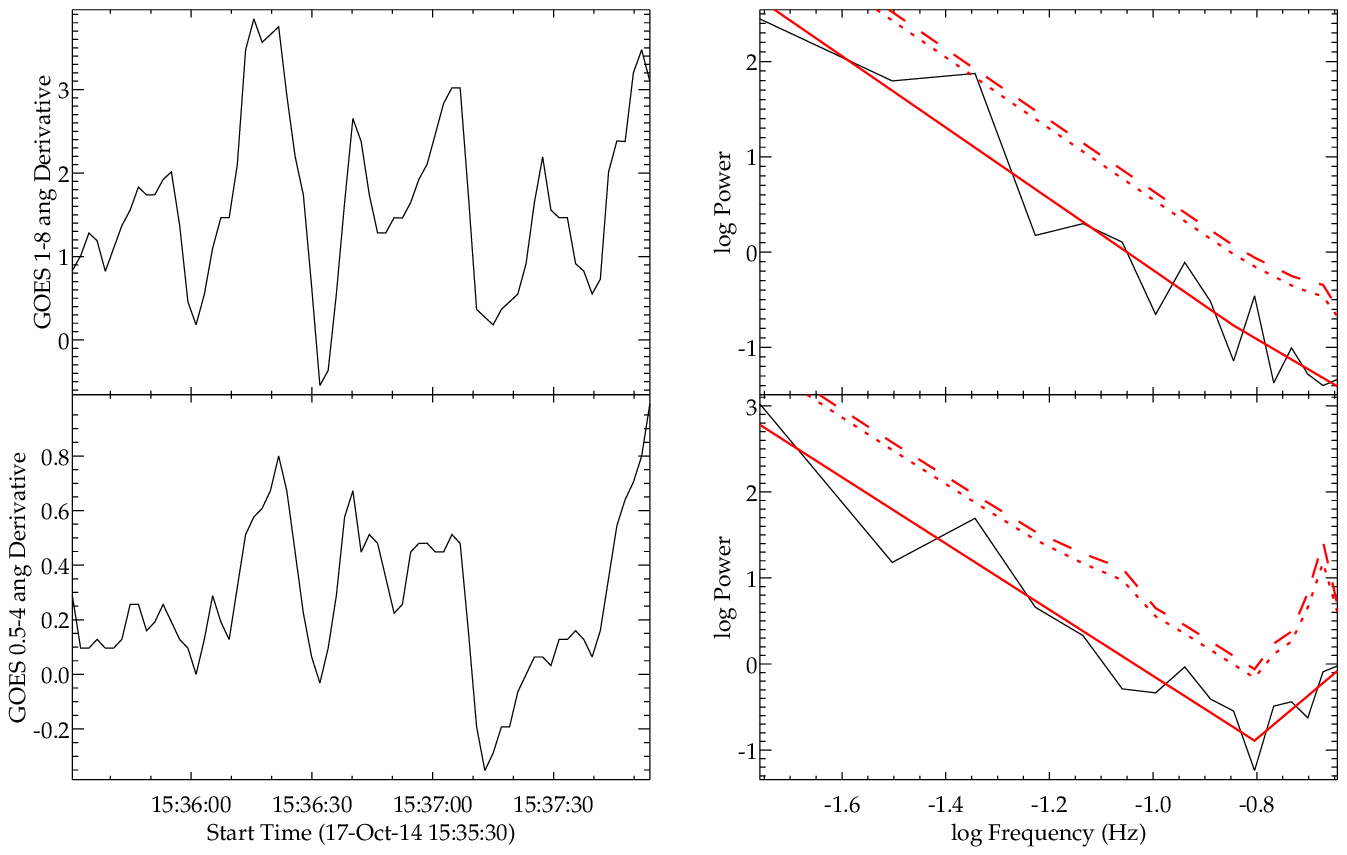}
	\caption{Similar to Fig. \ref{152goes}, with GOES/XRS data for flare 024.}
	\label{024goes}
\end{figure}

\begin{figure}
	\centering
	\includegraphics[width=\linewidth]{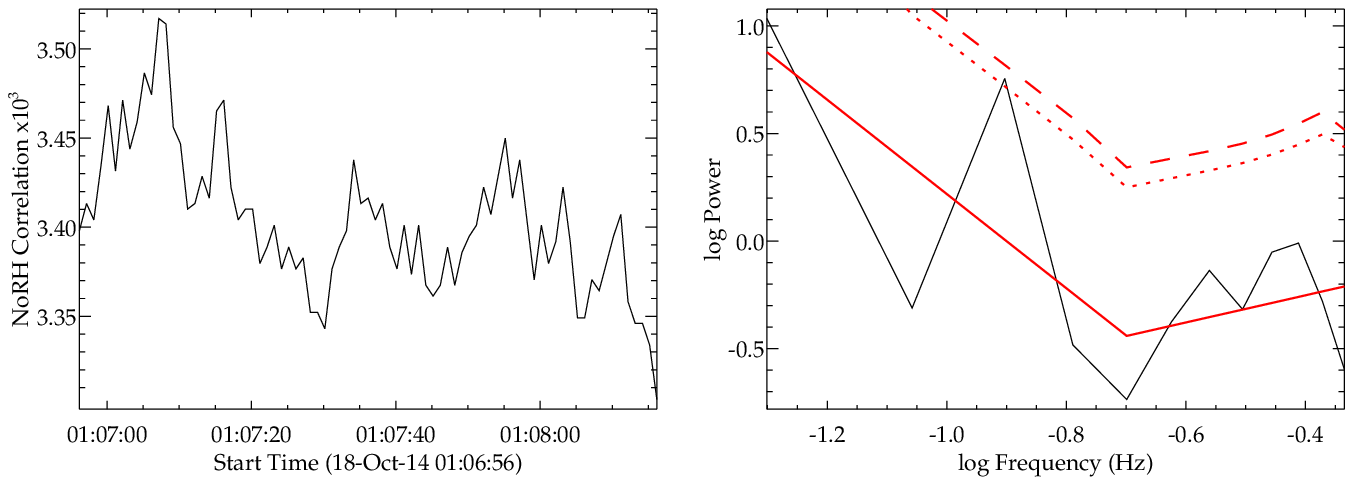}
	\caption{Similar to Fig. \ref{152goes}, with NoRH data for flare 027.}
	\label{027norh}
\end{figure}

\begin{figure}
	\centering
	\includegraphics[width=\linewidth]{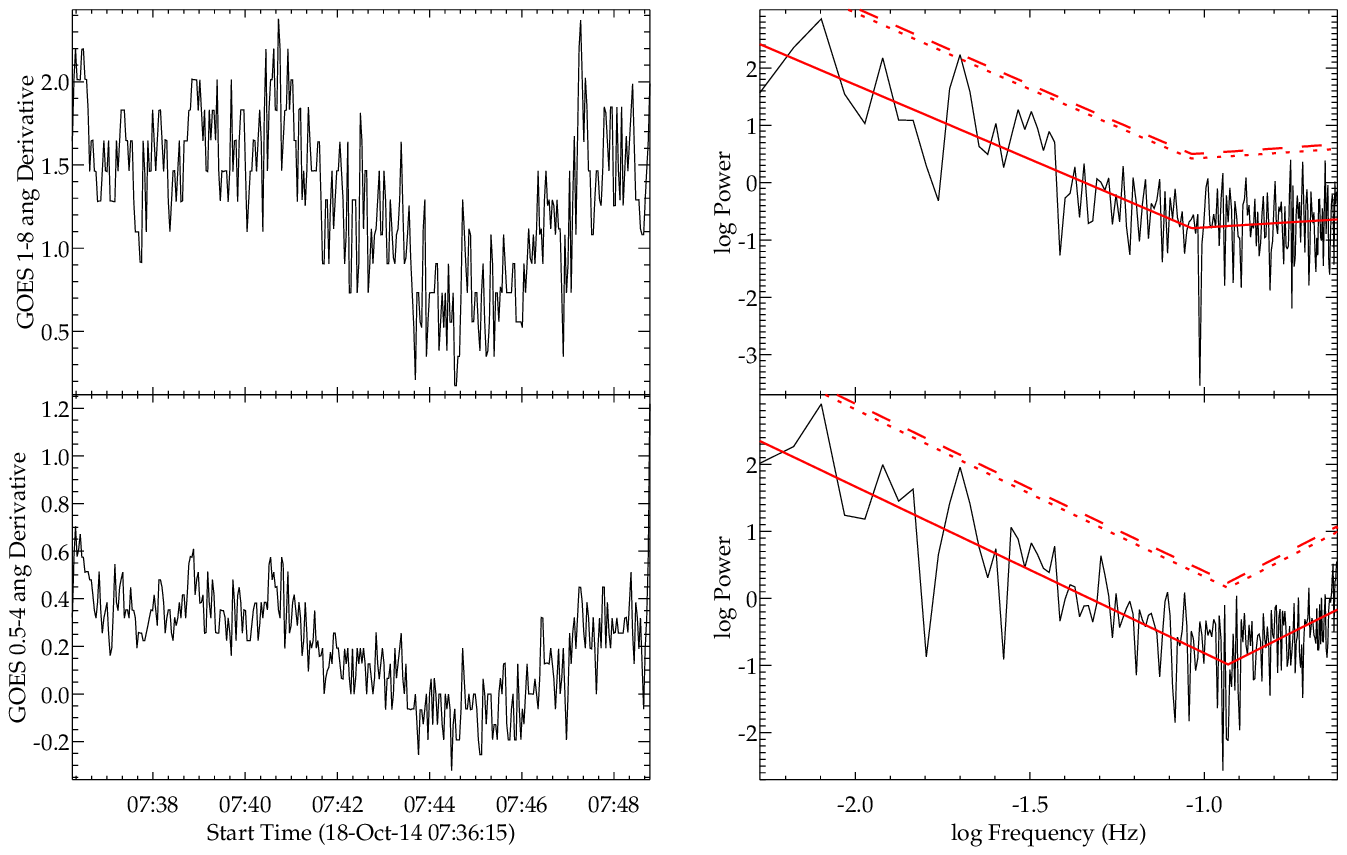}
	\caption{Similar to Fig. \ref{152goes}, with GOES/XRS data for flare 029.}
	\label{029goes}
\end{figure}

\begin{figure}
	\centering
	\includegraphics[width=\linewidth]{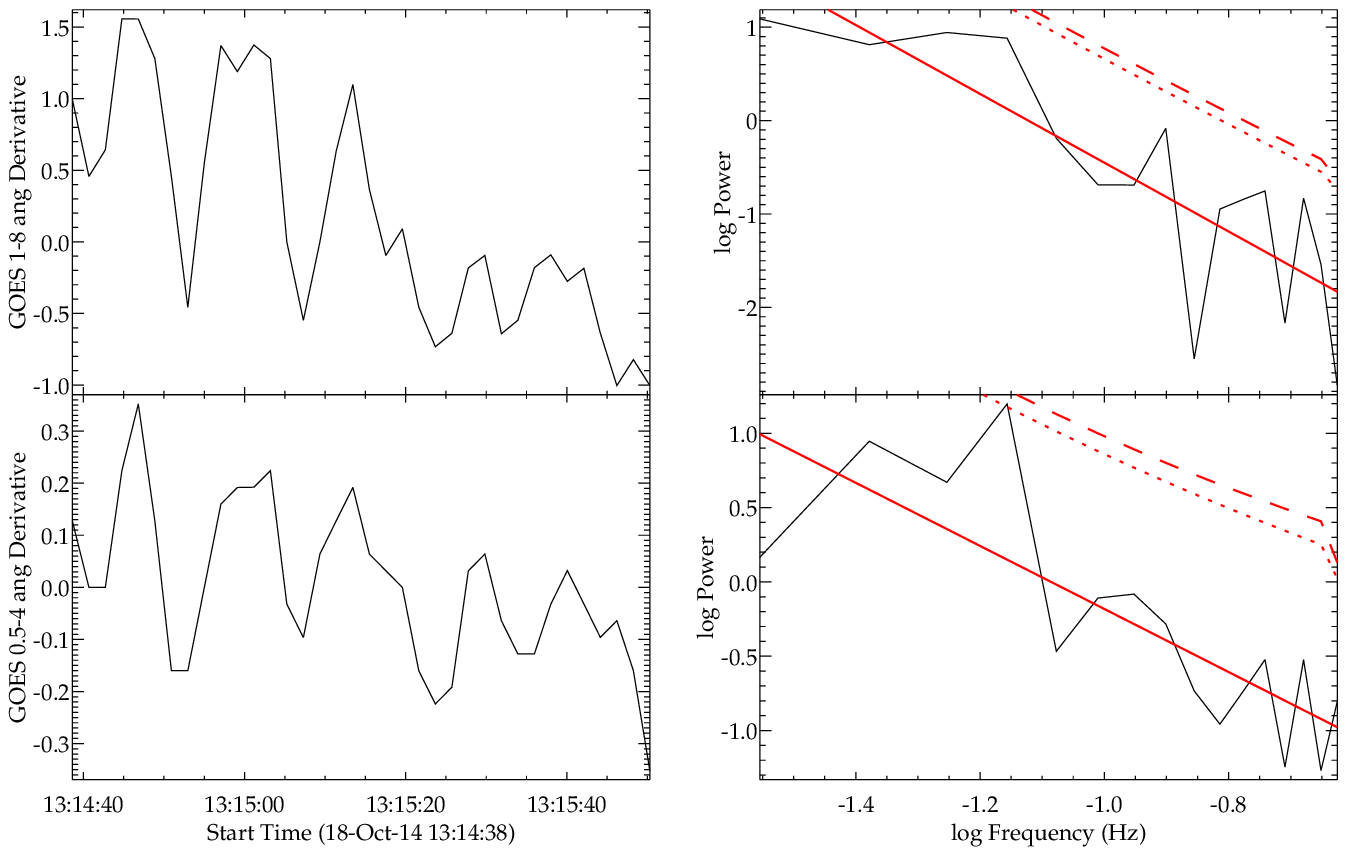}
	\caption{Similar to Fig. \ref{152goes}, with GOES/XRS data for flare 030.}
	\label{030goes}
\end{figure}

\begin{figure}
	\centering
	\includegraphics[width=\linewidth]{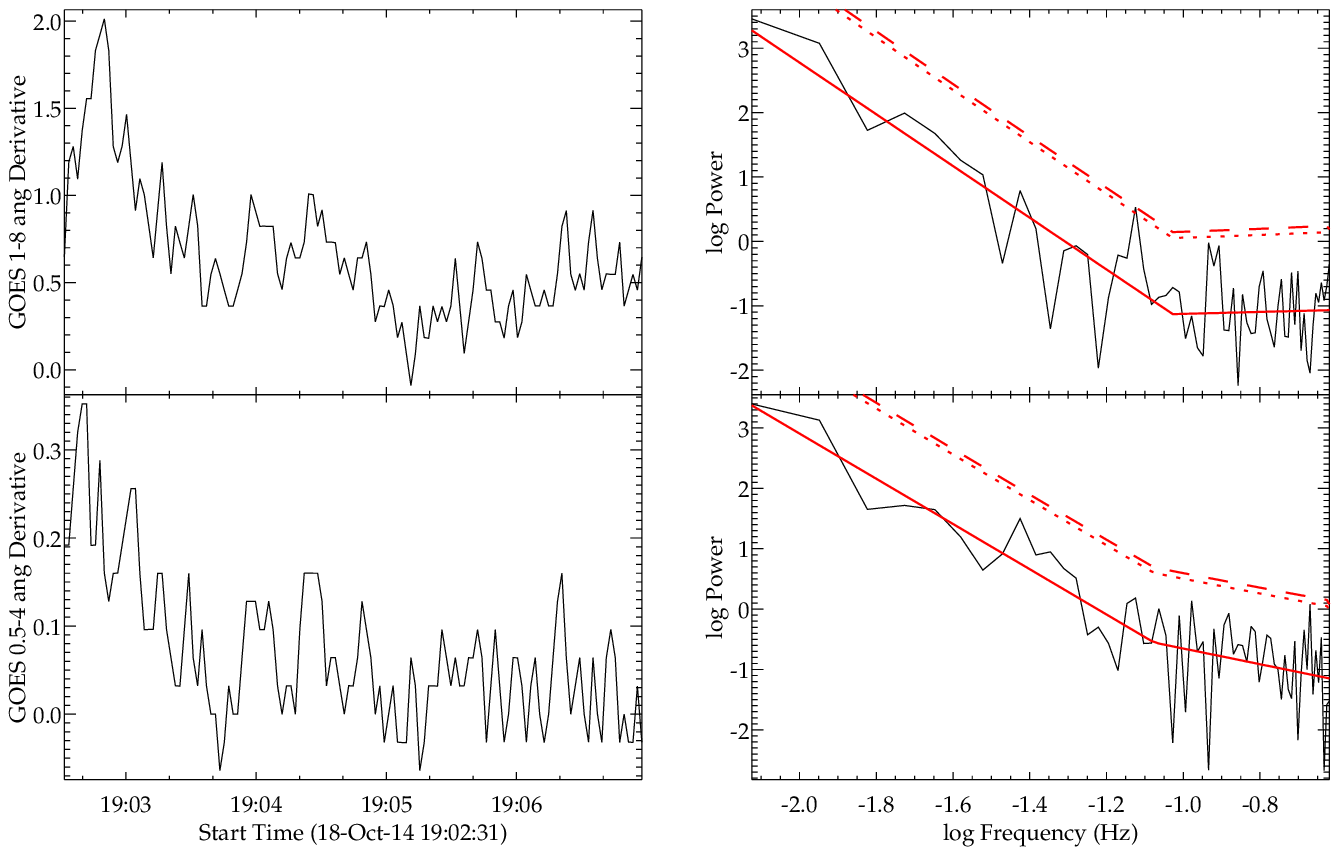}
	\caption{Similar to Fig. \ref{152goes}, with GOES/XRS data for flare 035.}
	\label{035goes}
\end{figure}

\begin{figure}
	\centering
	\includegraphics[width=\linewidth]{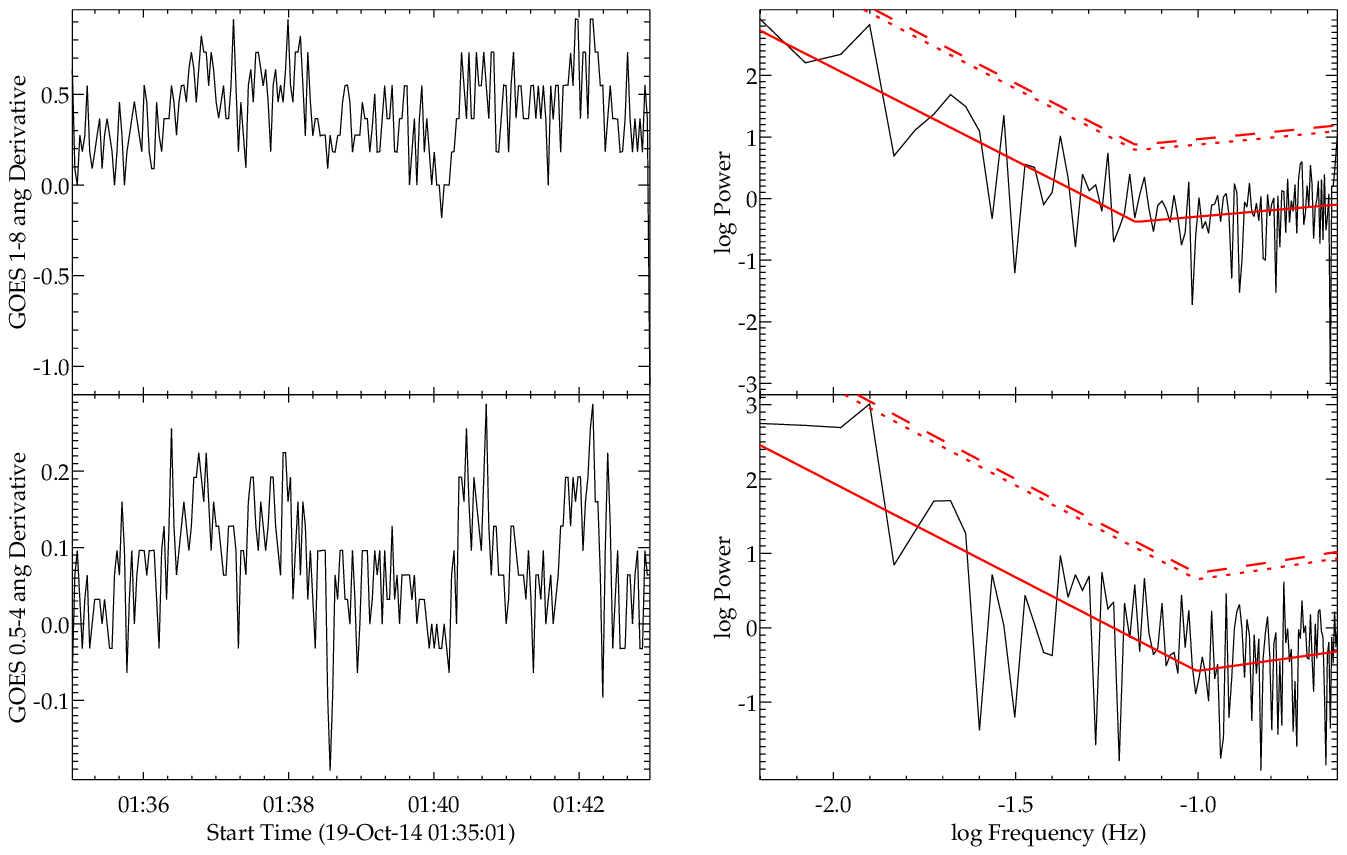}
	\caption{Similar to Fig. \ref{152goes}, with GOES/XRS data for flare 037.}
	\label{037goes}
\end{figure}

\begin{figure}
	\centering
	\includegraphics[width=\linewidth]{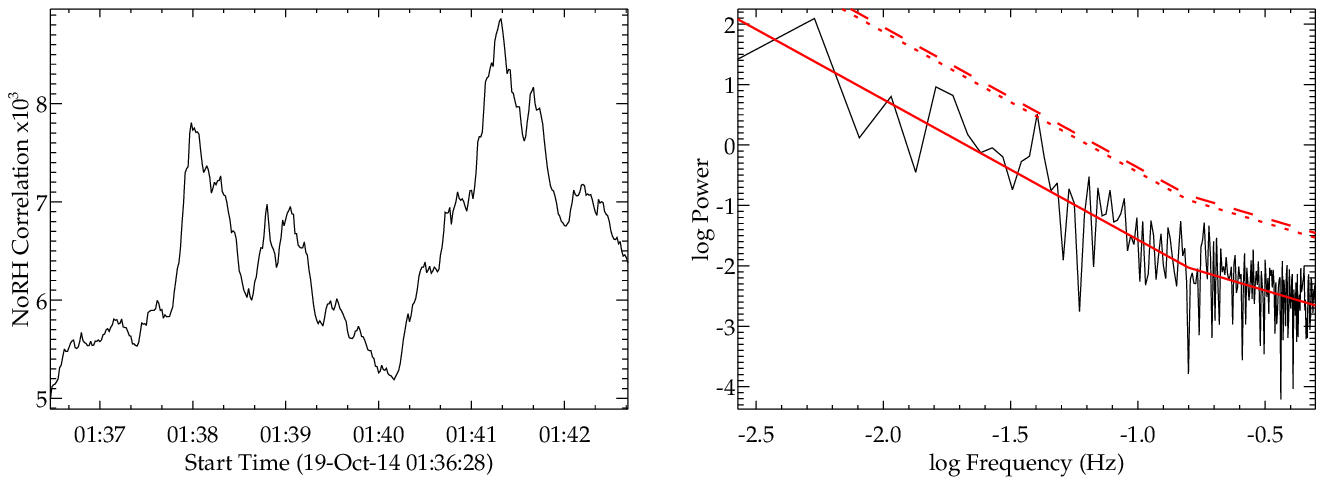}
	\caption{Similar to Fig. \ref{152goes}, with NoRH data for flare 037.}
	\label{037norh}
\end{figure}

\begin{figure}
	\centering
	\includegraphics[width=\linewidth]{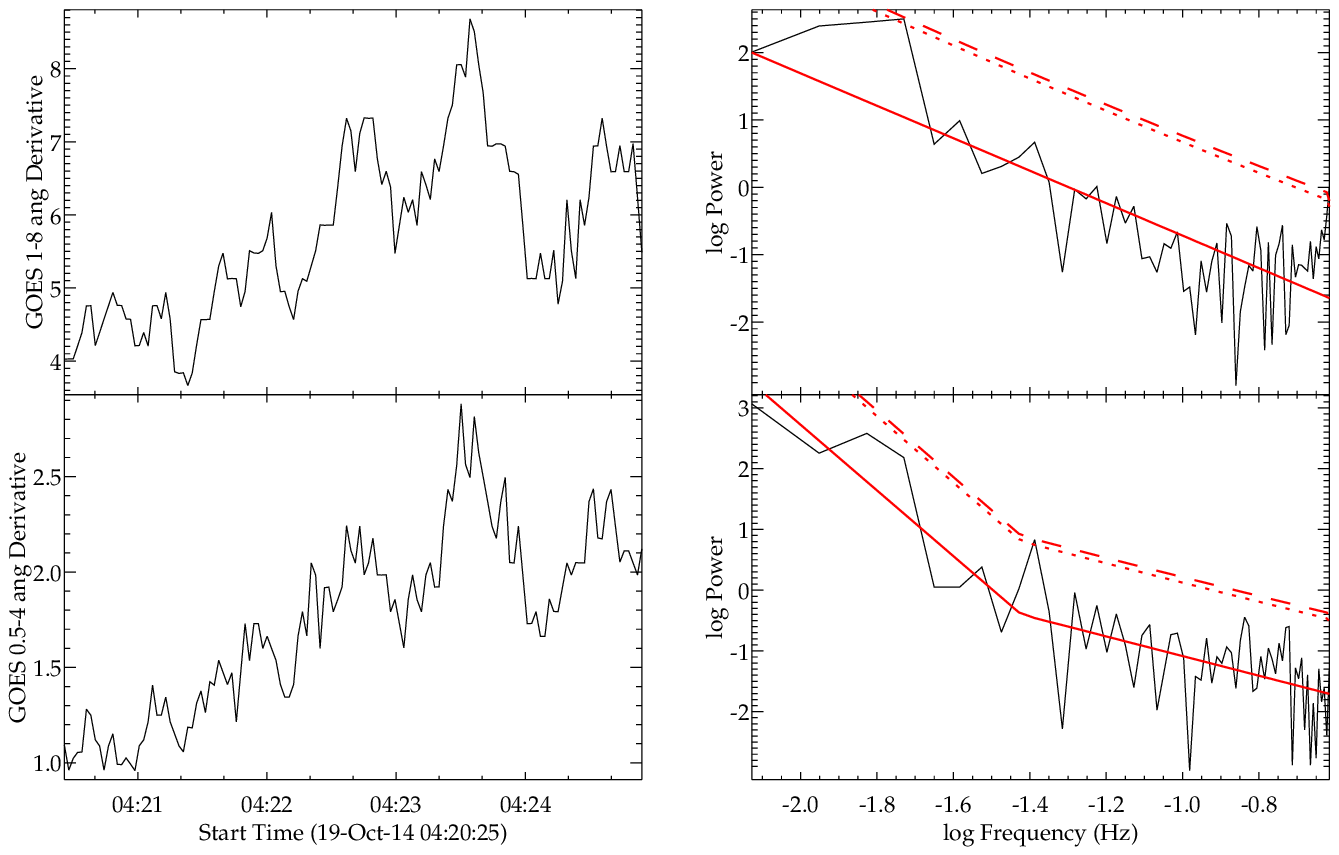}
	\caption{Similar to Fig. \ref{152goes}, with GOES/XRS data for flare 038.}
	\label{038goes0}
\end{figure}

\begin{figure}
	\centering
	\includegraphics[width=\linewidth]{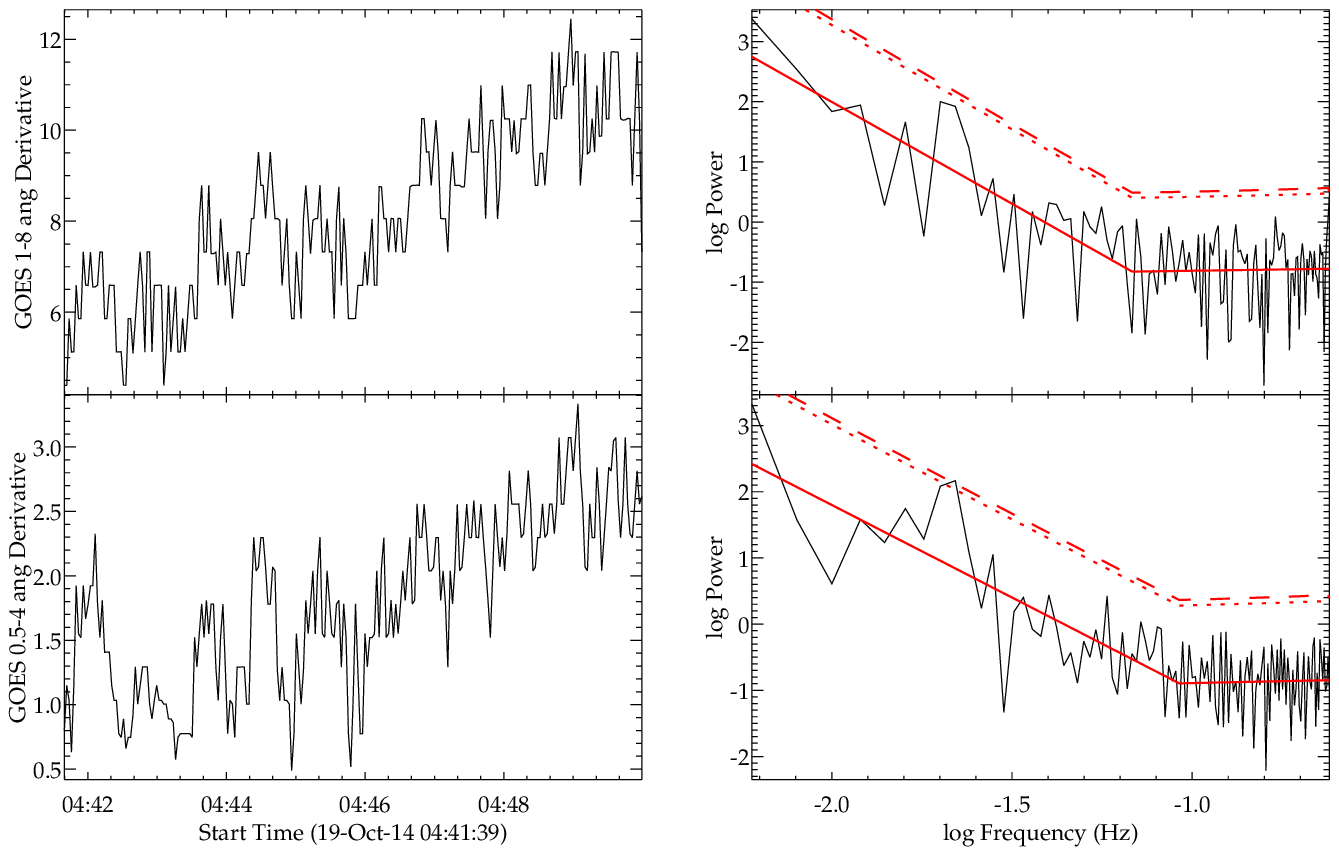}
	\caption{Similar to Fig. \ref{152goes}, with GOES/XRS data for flare 038.}
	\label{038goes1}
\end{figure}

\begin{figure}
	\centering
	\includegraphics[width=\linewidth]{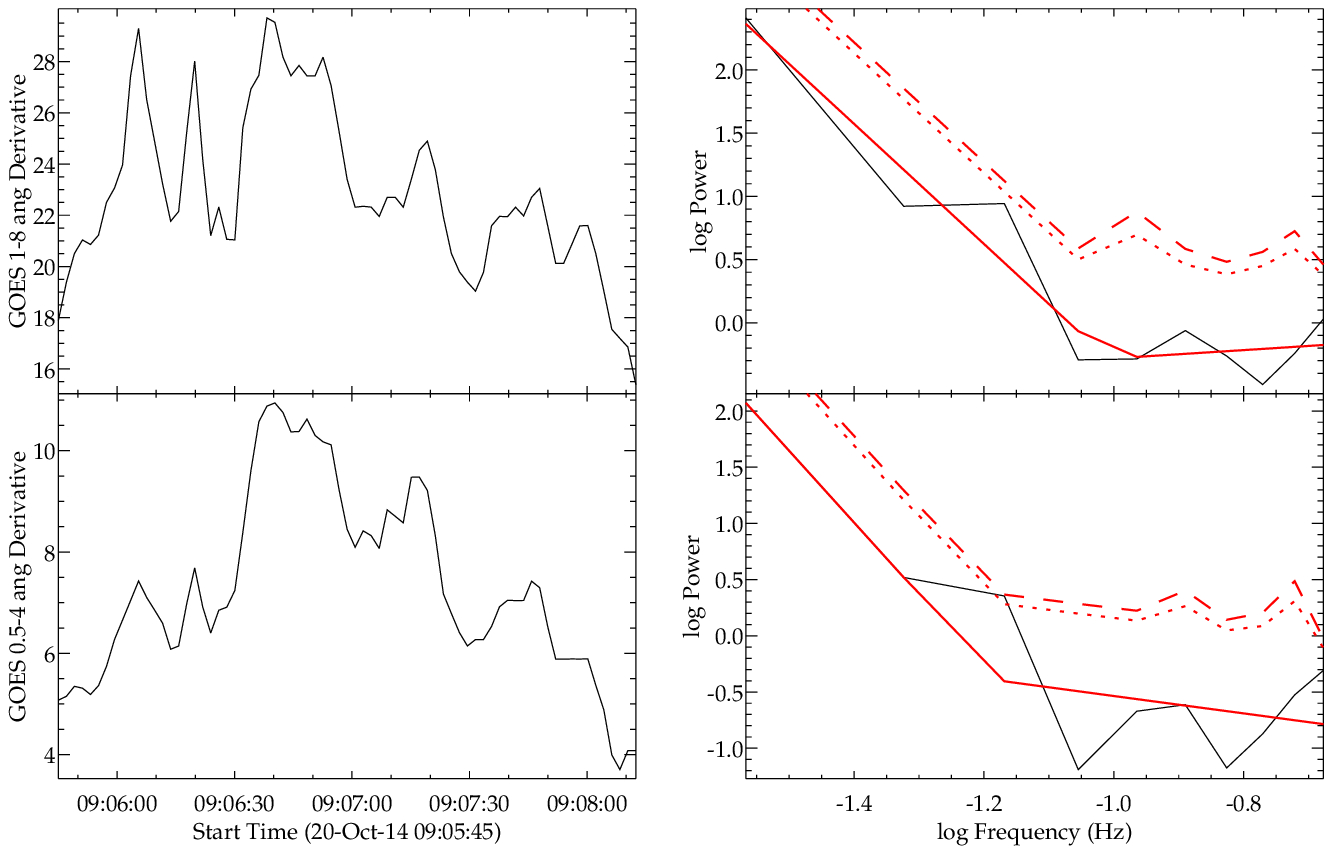}
	\caption{Similar to Fig. \ref{152goes}, with GOES/XRS data for flare 049.}
	\label{049goes}
\end{figure}

\begin{figure}
	\centering
	\includegraphics[width=\linewidth]{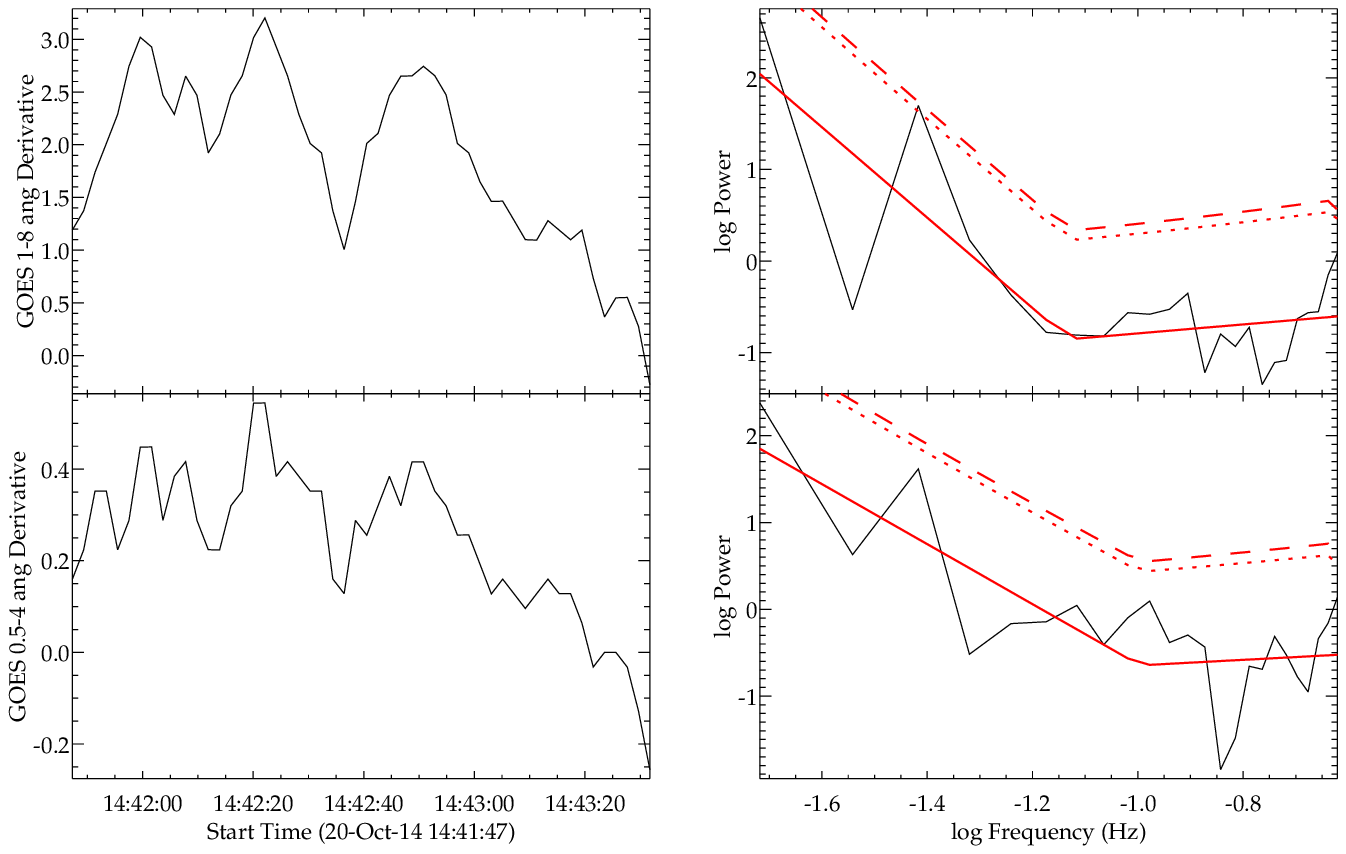}
	\caption{Similar to Fig. \ref{152goes}, with GOES/XRS data for flare 052.}
	\label{052goes}
\end{figure}

\clearpage

\begin{figure}
	\centering
	\includegraphics[width=\linewidth]{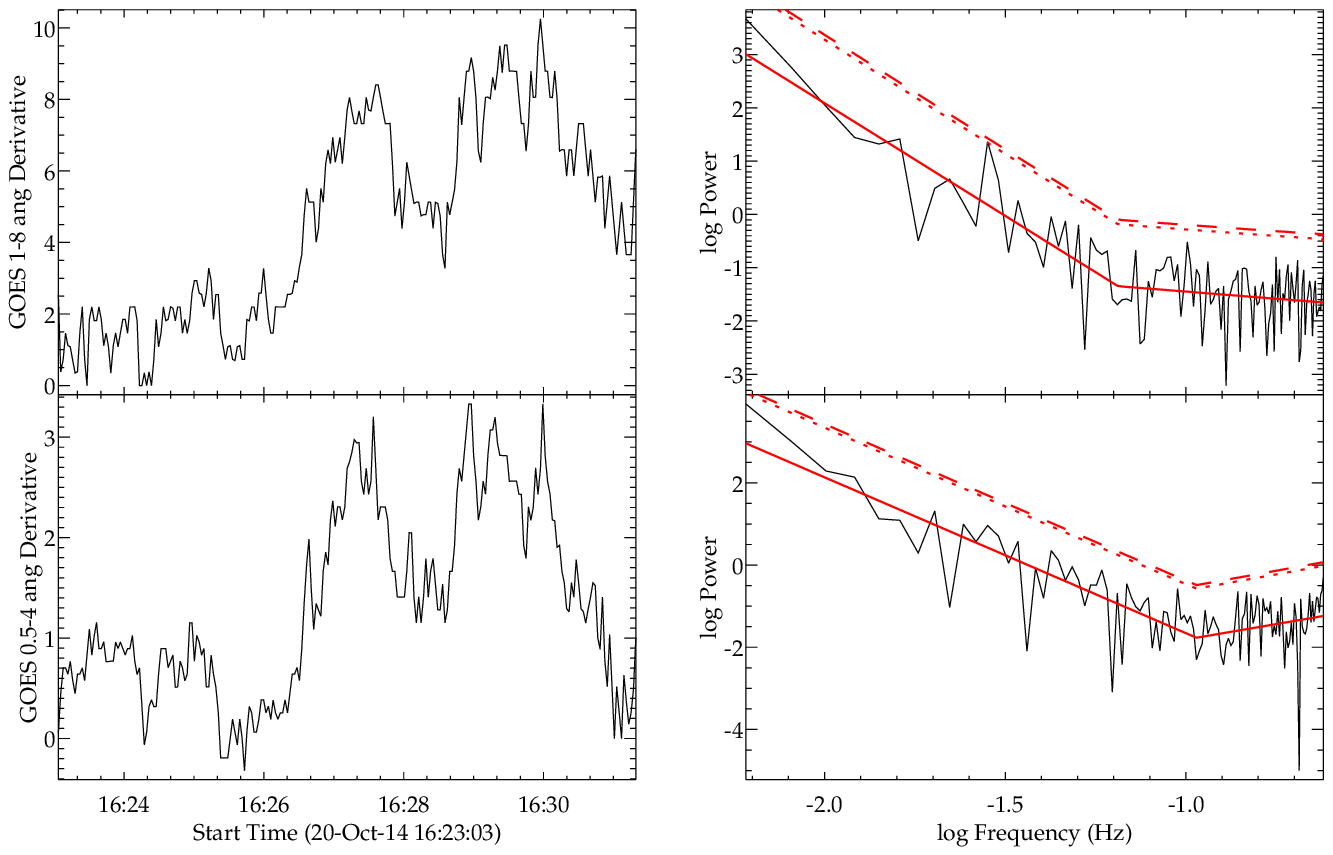}
	\caption{Similar to Fig. \ref{152goes}, with GOES/XRS data for flare 054.}
	\label{054goes}
\end{figure}

\begin{figure}
	\centering
	\includegraphics[width=\linewidth]{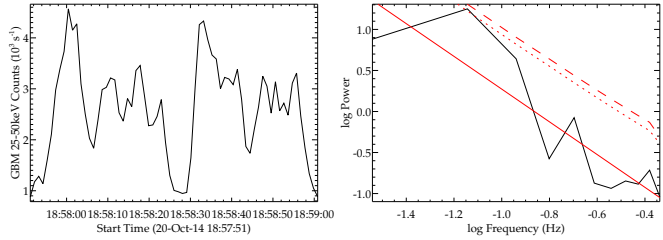}
	\caption{Similar to Fig. \ref{152goes}, with \emph{Fermi}/GBM 25--50\,keV data for flare 056.}
	\label{056ferm}
\end{figure}

\begin{figure}
	\centering
	\includegraphics[width=\linewidth]{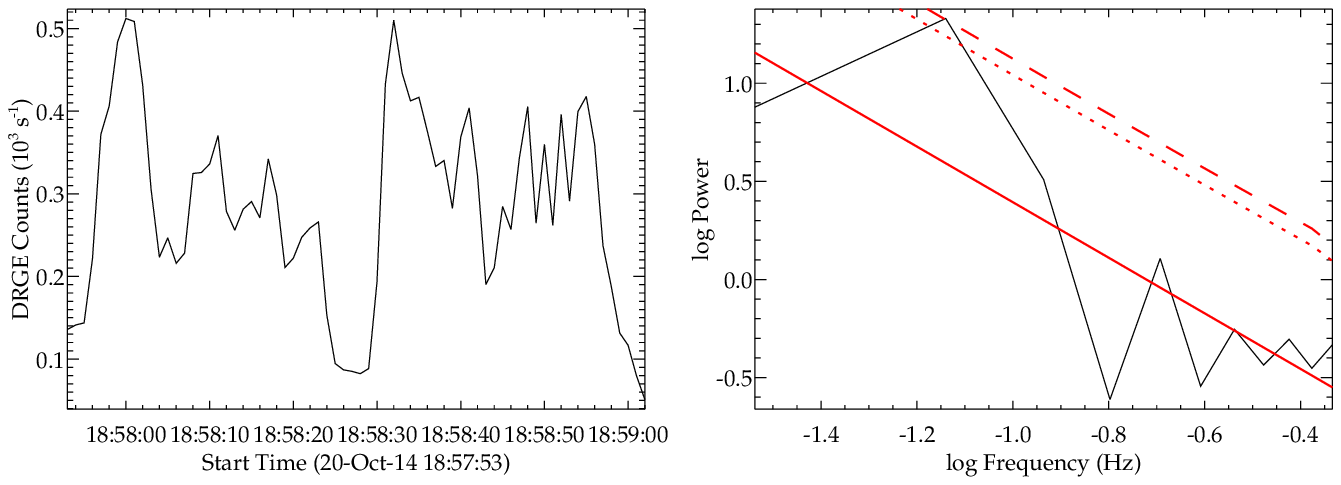}
	\caption{Similar to Fig. \ref{152goes}, with \emph{Vernov}/DRGE data for flare 056.}
	\label{056vern}
\end{figure}

\begin{figure}
	\centering
	\includegraphics[width=\linewidth]{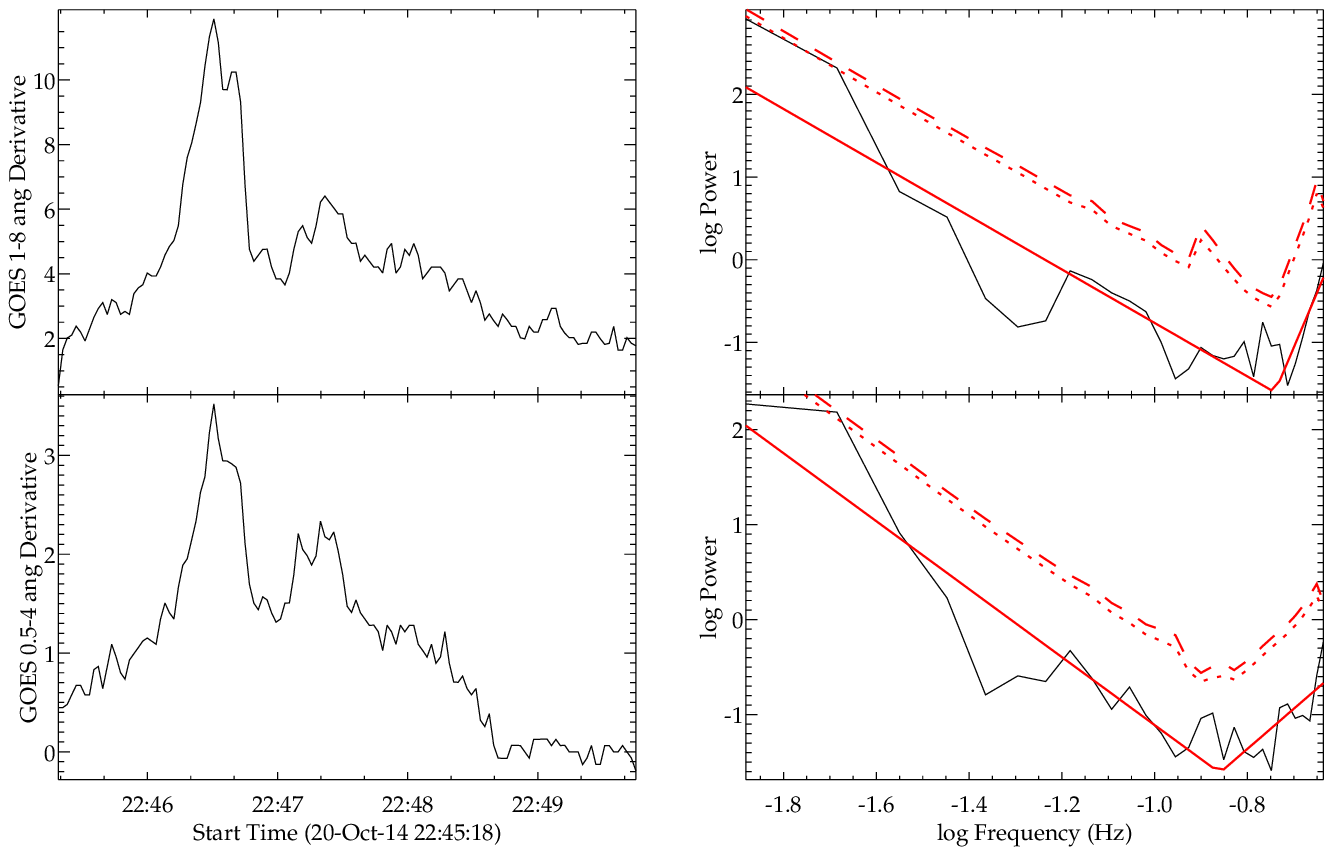}
	\caption{Similar to Fig. \ref{152goes}, with GOES/XRS data for flare 058.}
	\label{058goes}
\end{figure}

\begin{figure}
	\centering
	\includegraphics[width=\linewidth]{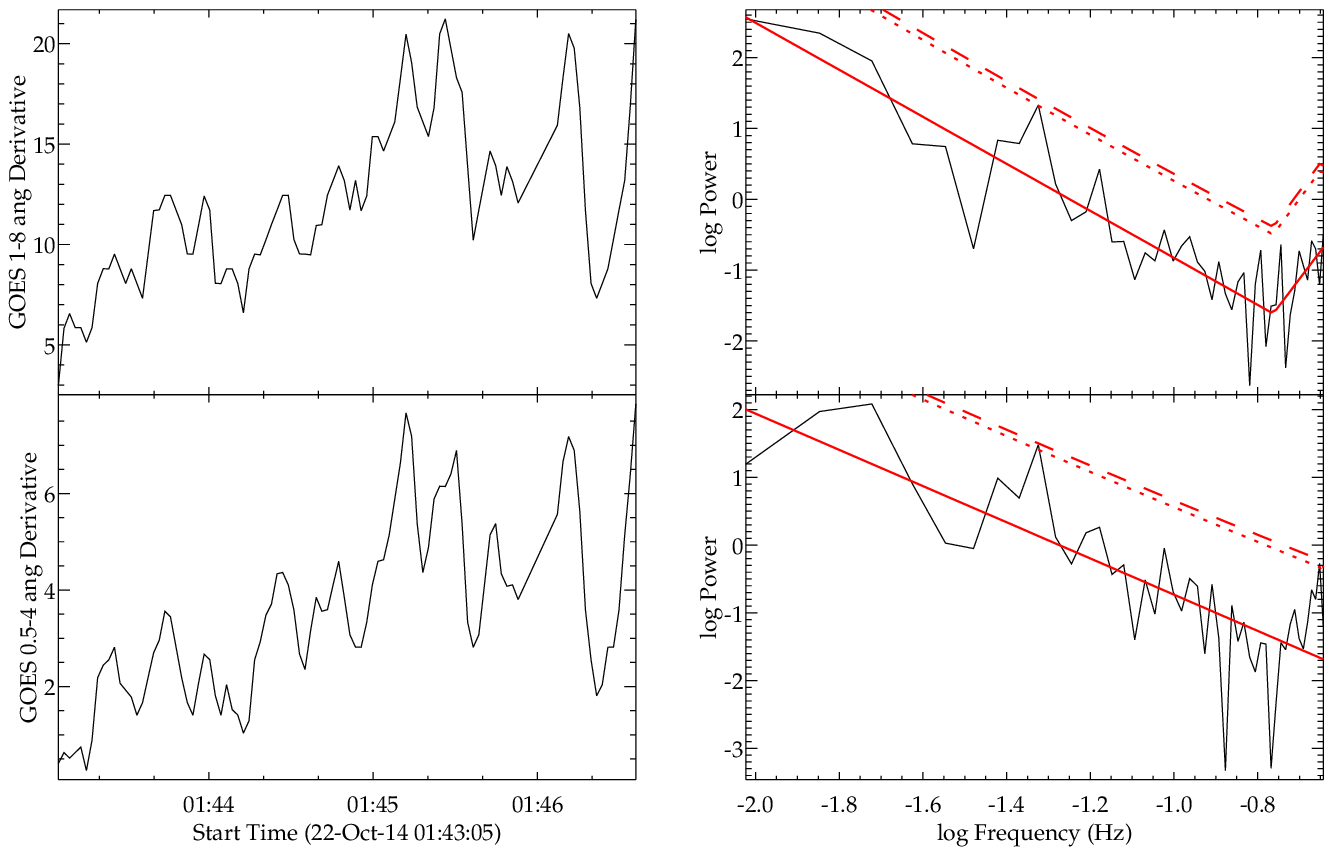}
	\caption{Similar to Fig. \ref{152goes}, with GOES/XRS data for flare 068.}
	\label{068goes}
\end{figure}

\begin{figure}
	\centering
	\includegraphics[width=\linewidth]{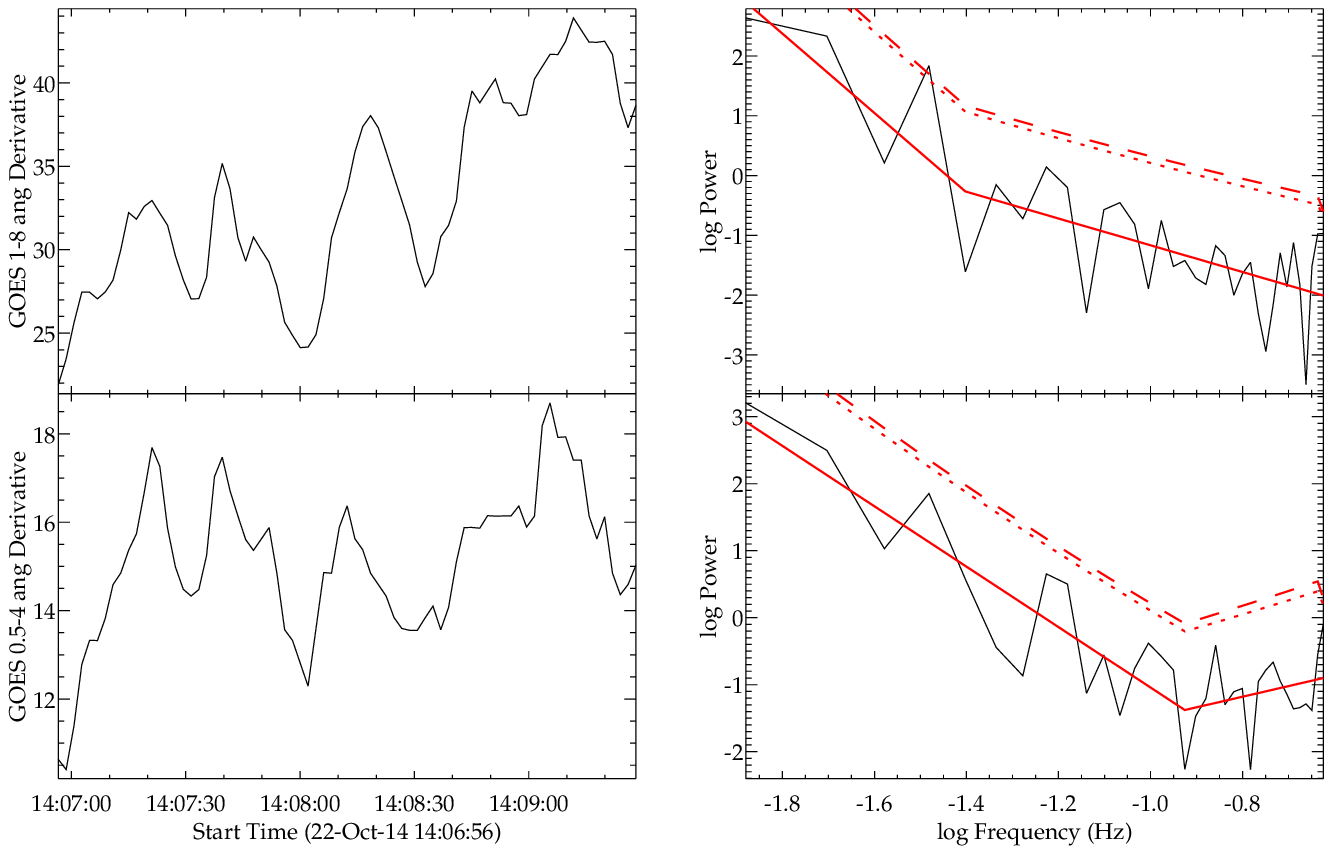}
	\caption{Similar to Fig. \ref{152goes}, with GOES/XRS data for flare 072.}
	\label{072goes0}
\end{figure}

\begin{figure}
	\centering
	\includegraphics[width=\linewidth]{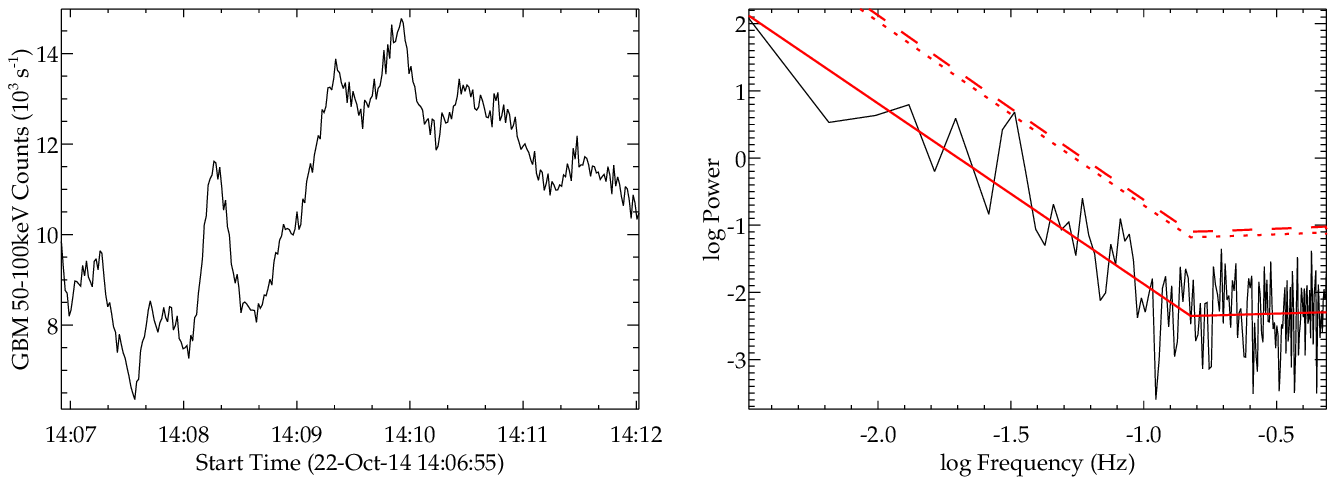}
	\caption{Similar to Fig. \ref{152goes}, with \emph{Fermi}/GBM data for flare 072.}
	\label{072ferm}
\end{figure}

\begin{figure}
	\centering
	\includegraphics[width=\linewidth]{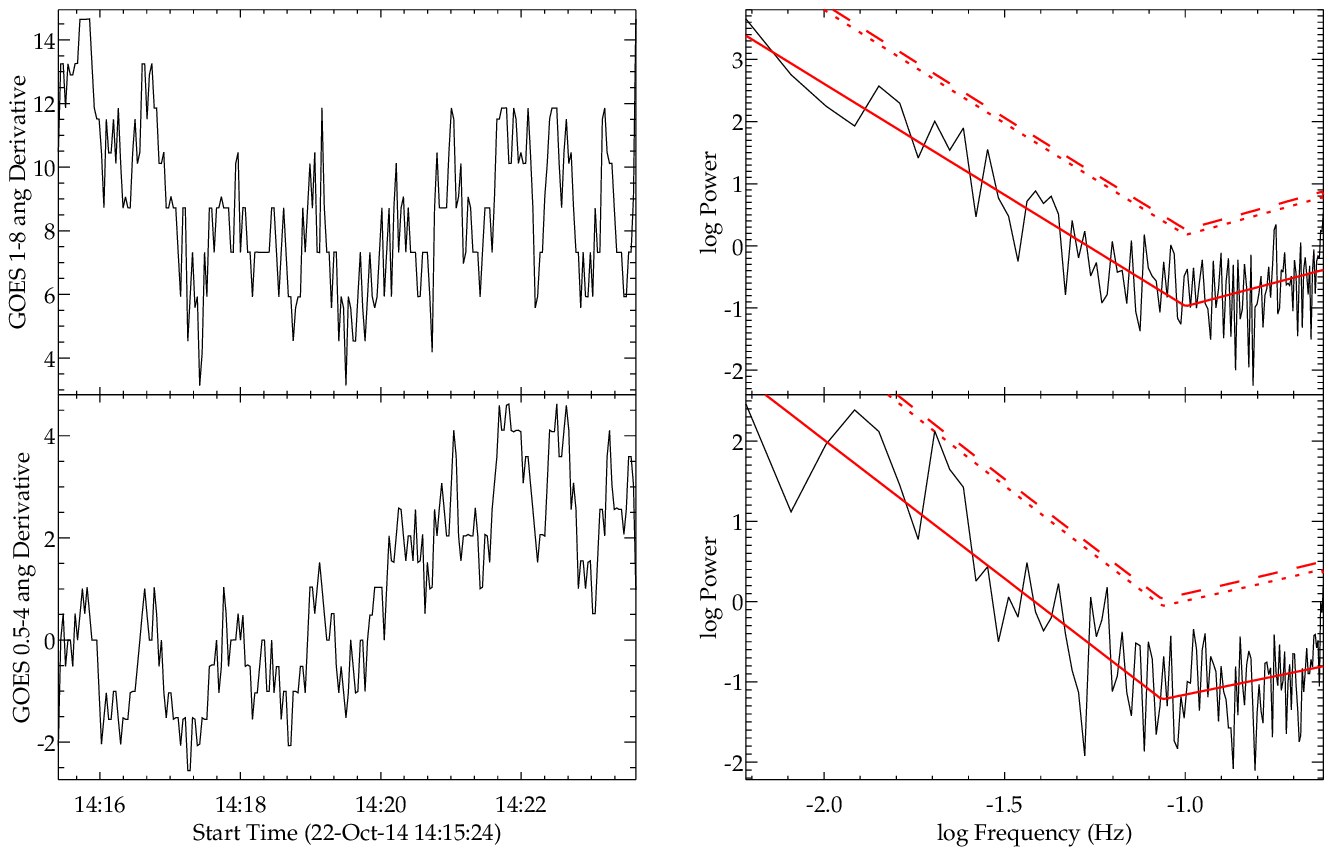}
	\caption{Similar to Fig. \ref{152goes}, with GOES/XRS data for flare 072.}
	\label{072goes1}
\end{figure}

\begin{figure}
	\centering
	\includegraphics[width=\linewidth]{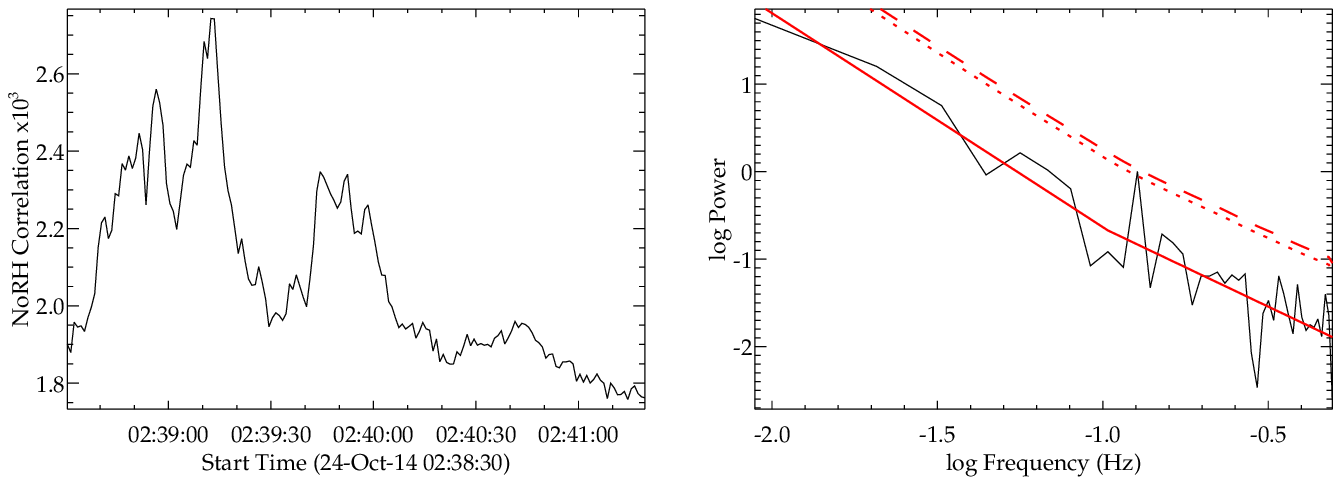}
	\caption{Similar to Fig. \ref{152goes}, with NoRH data for flare 079.}
	\label{079norh}
\end{figure}

\begin{figure}
	\centering
	\includegraphics[width=\linewidth]{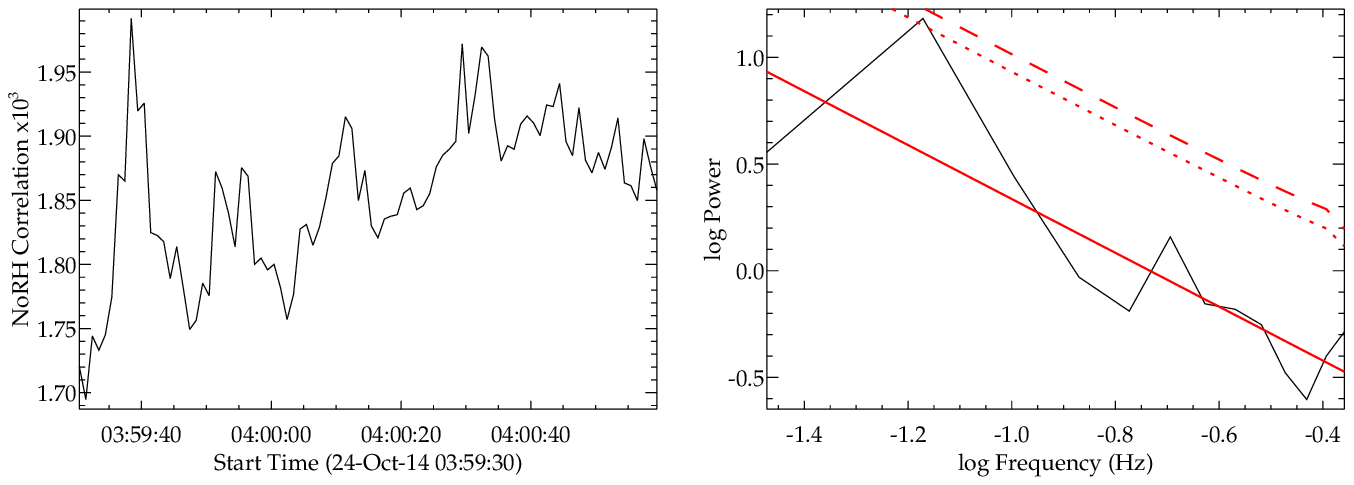}
	\caption{Similar to Fig. \ref{152goes}, with NoRH data for flare 081.}
	\label{081norh}
\end{figure}

\begin{figure}
	\centering
	\includegraphics[width=\linewidth]{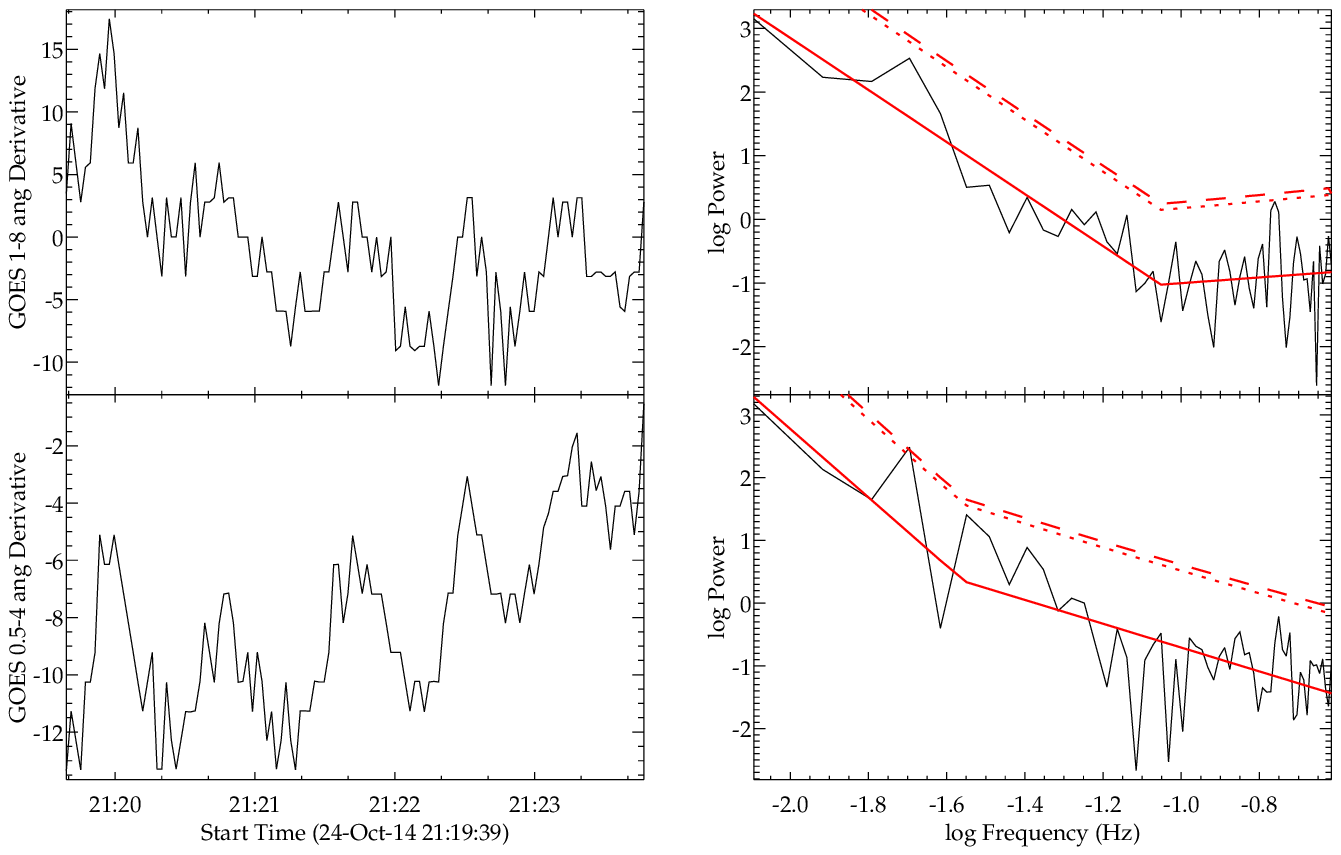}
	\caption{Similar to Fig. \ref{152goes}, with GOES/XRS data for flare 085.}
	\label{085goes0}
\end{figure}

\begin{figure}
	\centering
	\includegraphics[width=\linewidth]{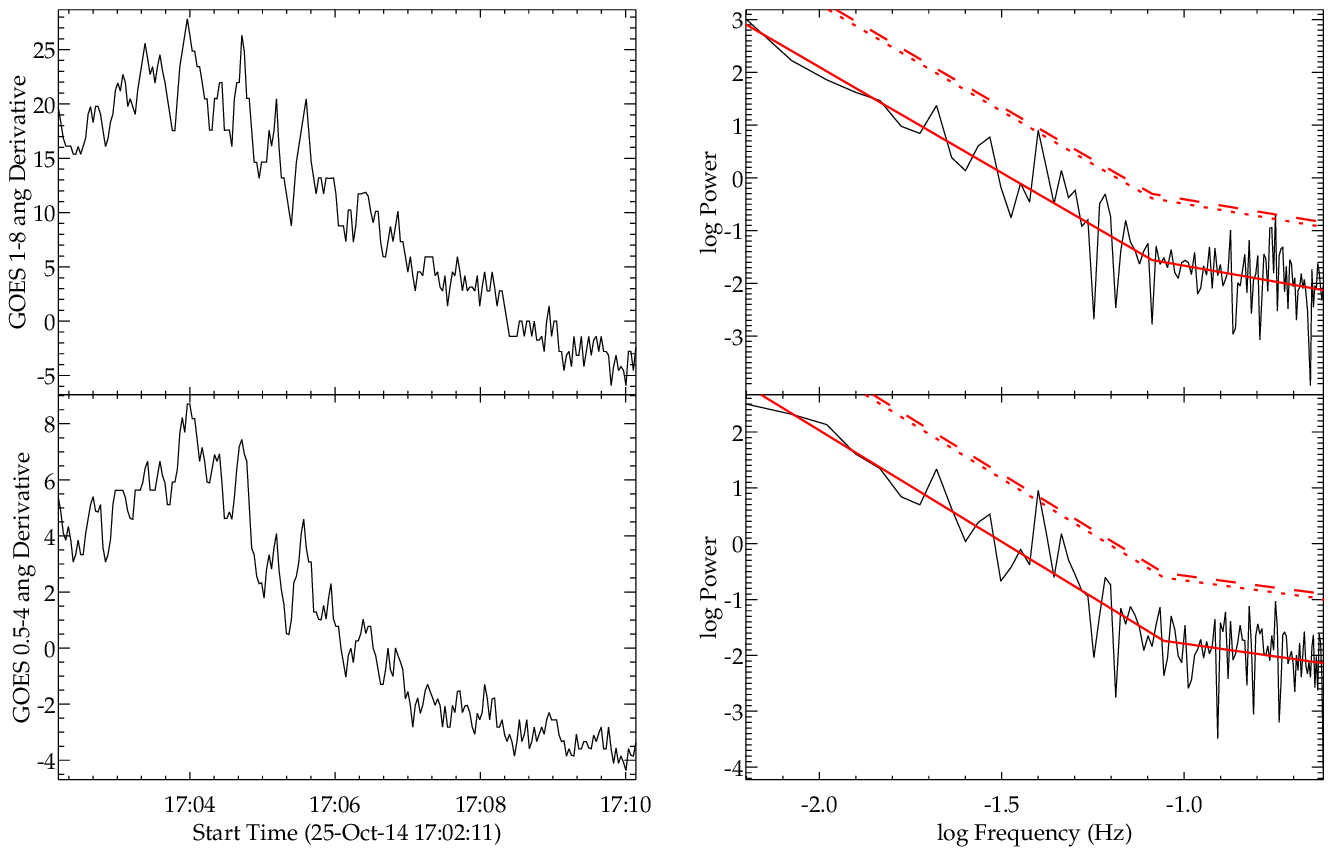}
	\caption{Similar to Fig. \ref{152goes}, with GOES/XRS data for flare 092.}
	\label{092goes}
\end{figure}

\begin{figure}
	\centering
	\includegraphics[width=\linewidth]{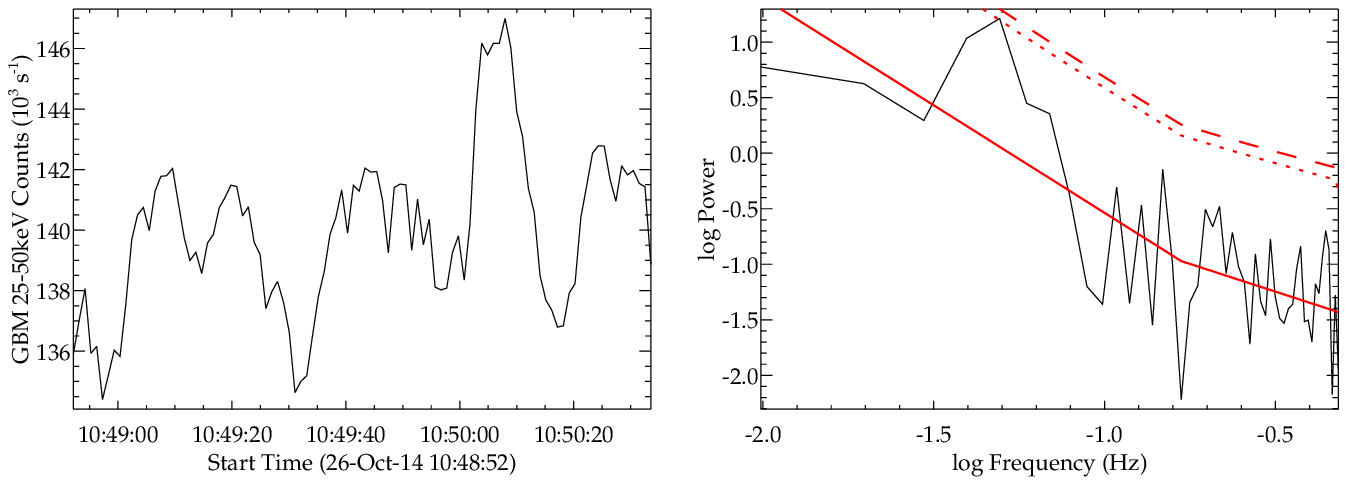}
	\caption{Similar to Fig. \ref{152goes}, with \emph{Fermi}/GBM data for flare 098.}
	\label{098ferm}
\end{figure}

\begin{figure}
	\centering
	\includegraphics[width=\linewidth]{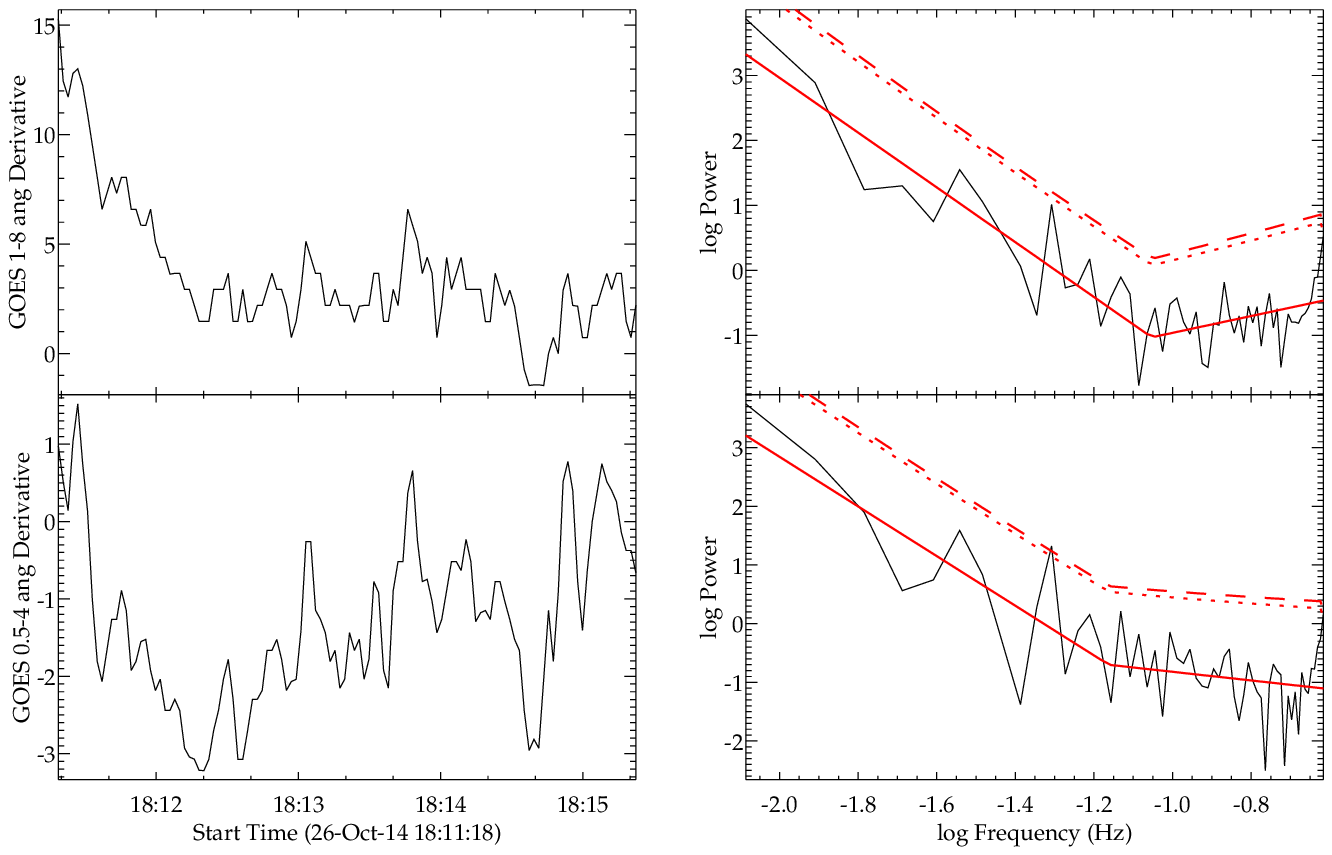}
	\caption{Similar to Fig. \ref{152goes}, with GOES/XRS data for flare 104.}
	\label{104goes}
\end{figure}

\begin{figure}
	\centering
	\includegraphics[width=\linewidth]{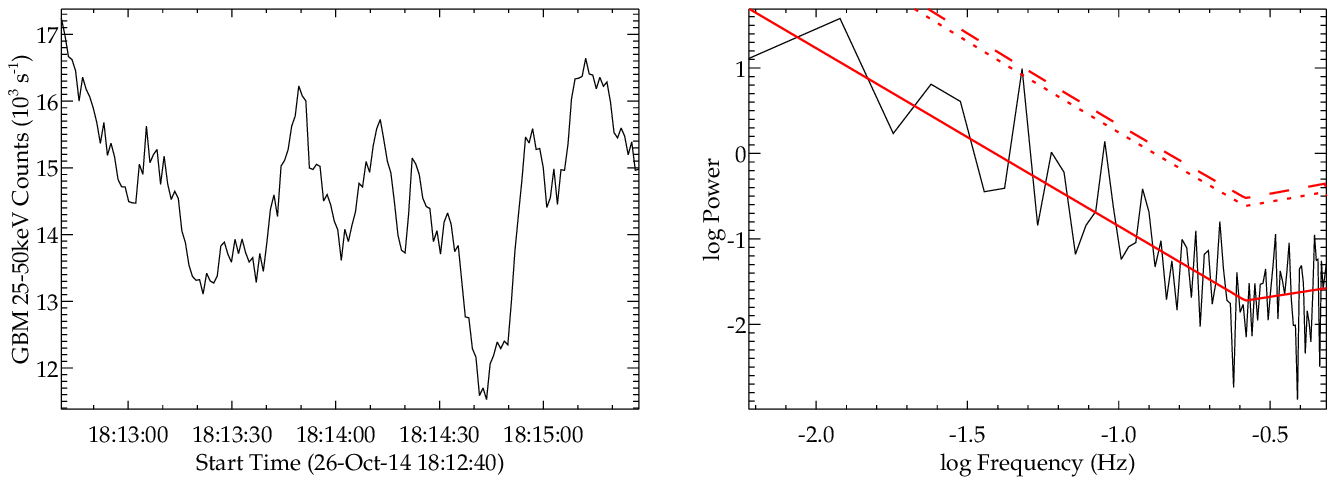}
	\caption{Similar to Fig. \ref{152goes}, with \emph{Fermi}/GBM data for flare 104.}
	\label{104ferm}
\end{figure}

\begin{figure}
	\centering
	\includegraphics[width=\linewidth]{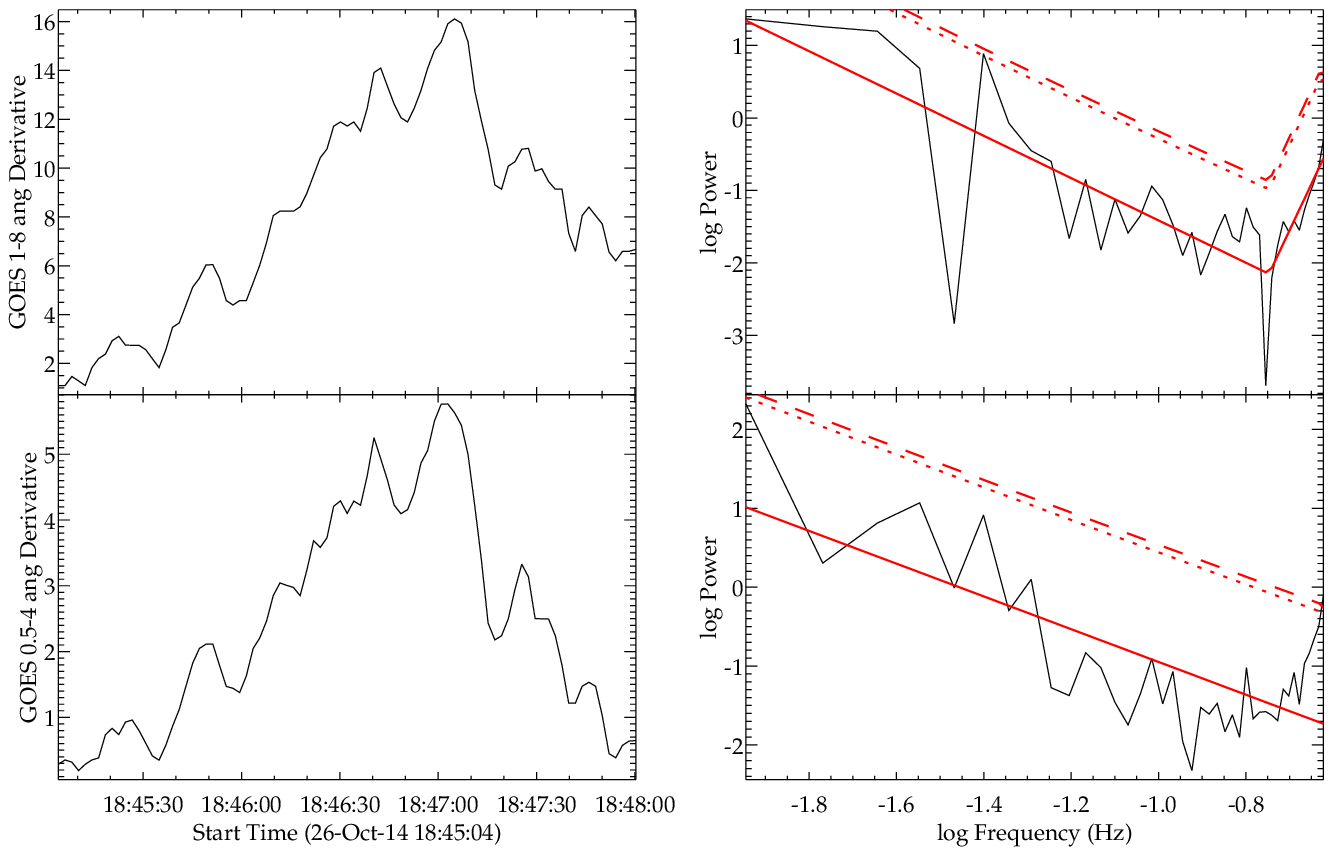}
	\caption{Similar to Fig. \ref{152goes}, with GOES/XRS data for flare 105.}
	\label{105goes}
\end{figure}

\clearpage

\begin{figure}
	\centering
	\includegraphics[width=\linewidth]{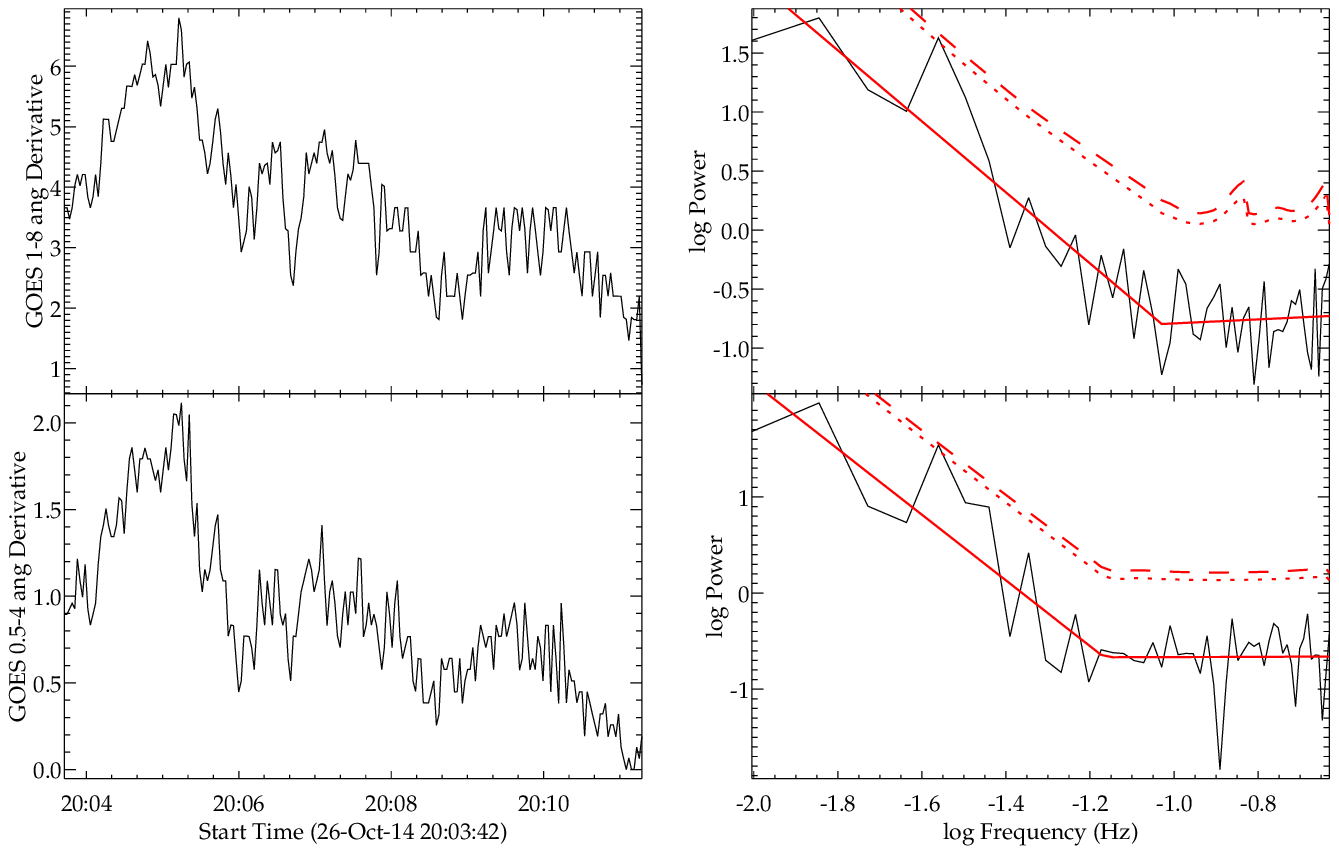}
	\caption{Similar to Fig. \ref{152goes}, with GOES/XRS data for flare 106.}
	\label{106goes}
\end{figure}

\begin{figure}
	\centering
	\includegraphics[width=\linewidth]{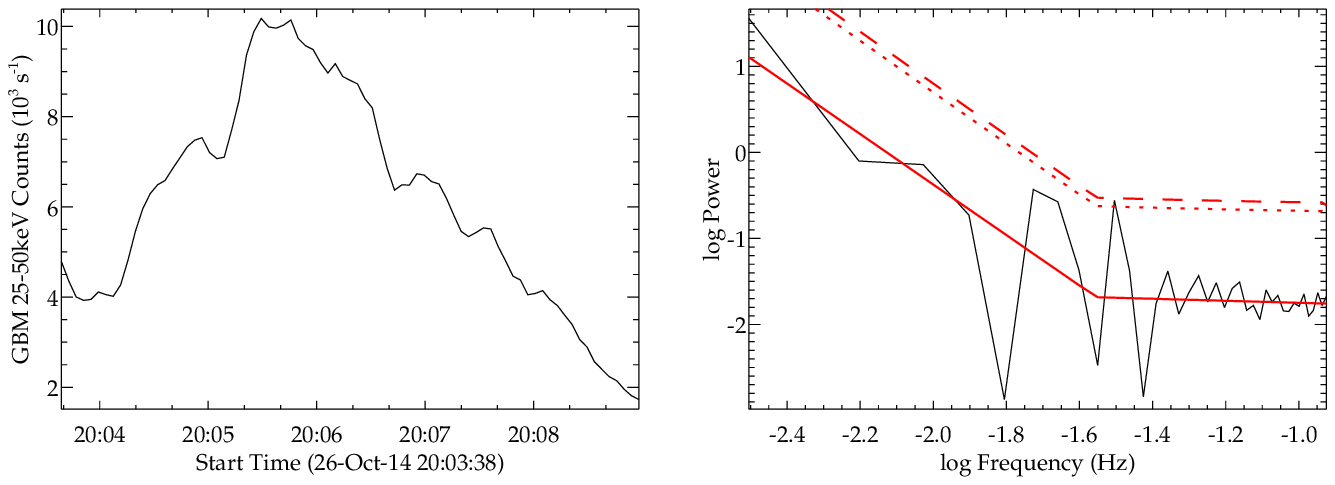}
	\caption{Similar to Fig. \ref{152goes}, with \emph{Fermi}/GBM data for flare 106.}
	\label{106ferm}
\end{figure}

\begin{figure}
	\centering
	\includegraphics[width=\linewidth]{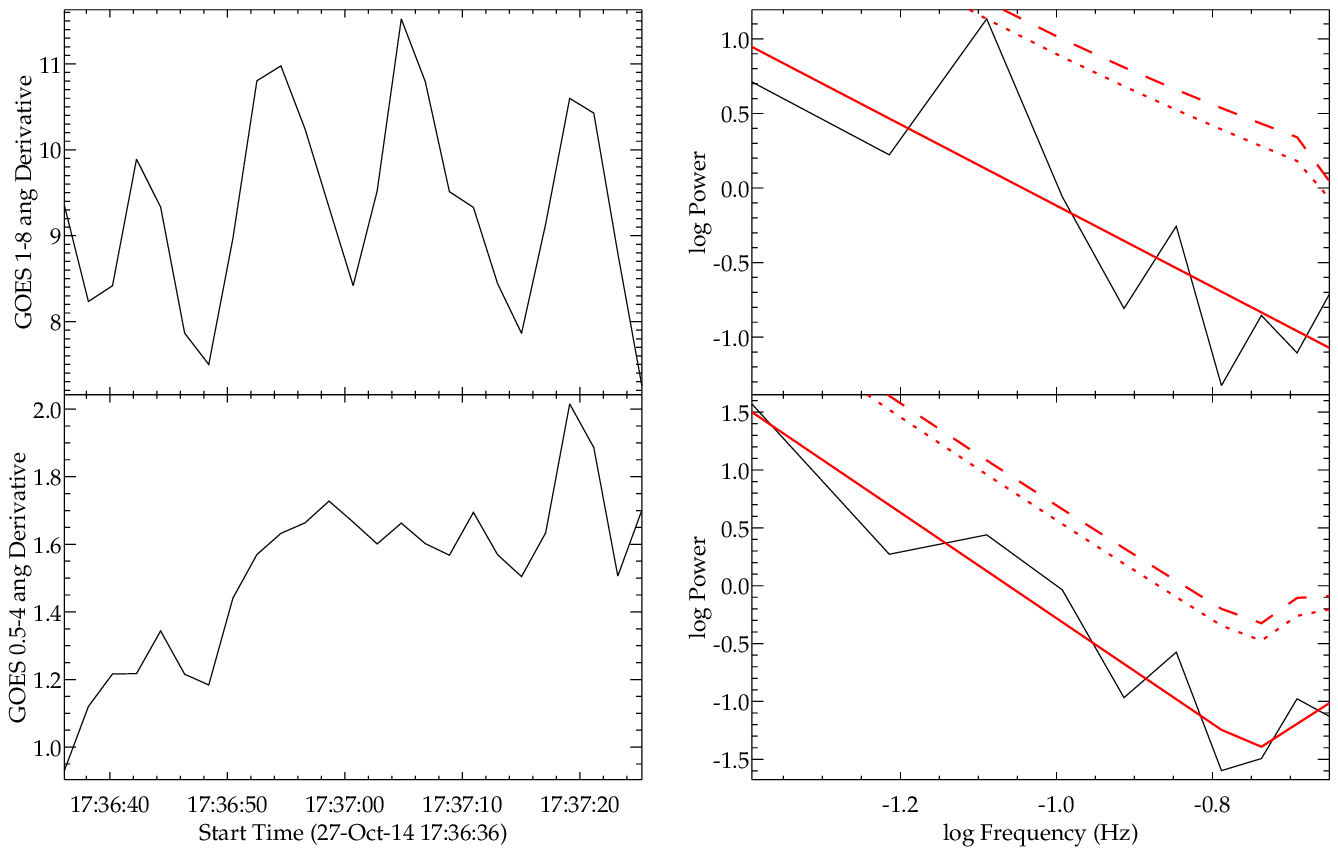}
	\caption{Similar to Fig. \ref{152goes}, with GOES/XRS data for flare 117.}
	\label{117goes}
\end{figure}

\begin{figure}
	\centering
	\includegraphics[width=\linewidth]{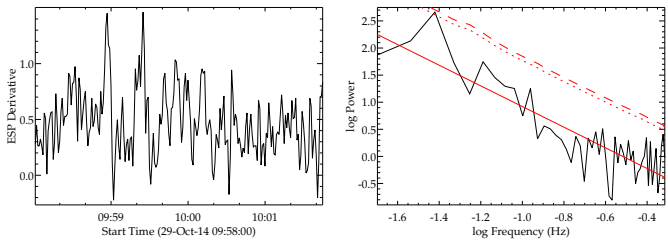}
	\caption{Similar to Fig. \ref{152goes}, with EVE/ESP data for flare 129.}
	\label{129eve}
\end{figure}

\begin{figure}
	\centering
	\includegraphics[width=\linewidth]{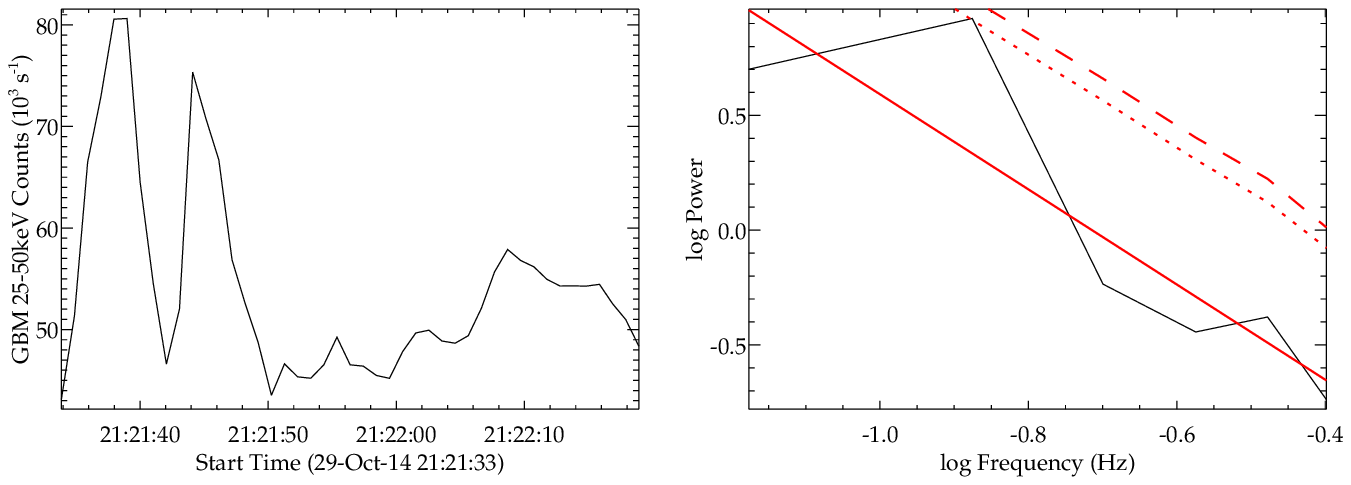}
	\caption{Similar to Fig. \ref{152goes}, with \emph{Fermi}/GBM data for flare 135.}
	\label{135ferm}
\end{figure}

\begin{figure}
	\centering
	\includegraphics[width=\linewidth]{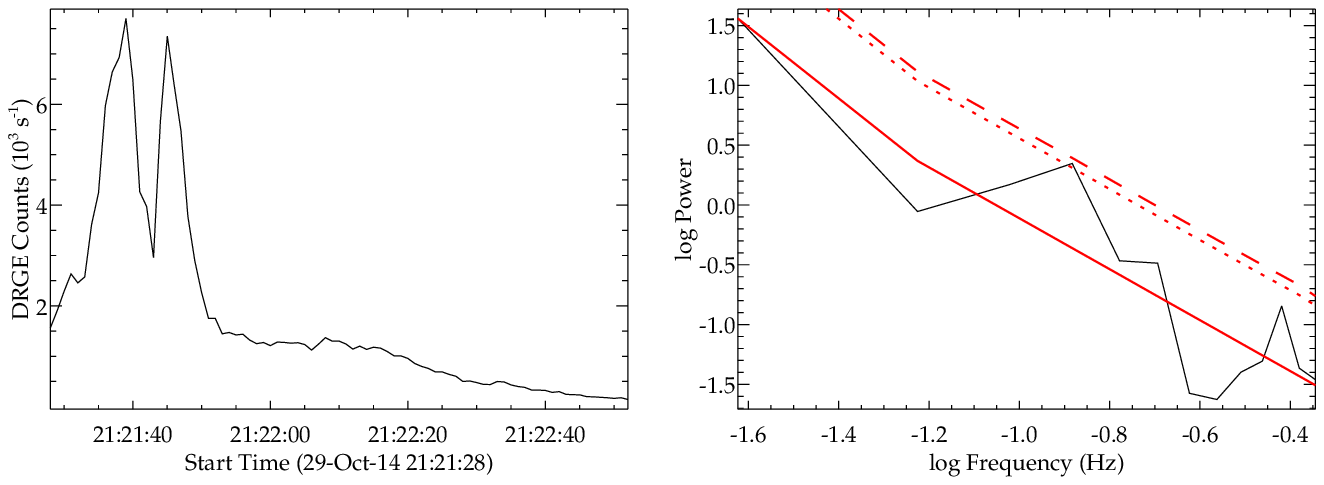}
	\caption{Similar to Fig. \ref{152goes}, with \emph{Vernov}/DRGE data for flare 135.}
	\label{135vern}
\end{figure}

\begin{figure}
	\centering
	\includegraphics[width=\linewidth]{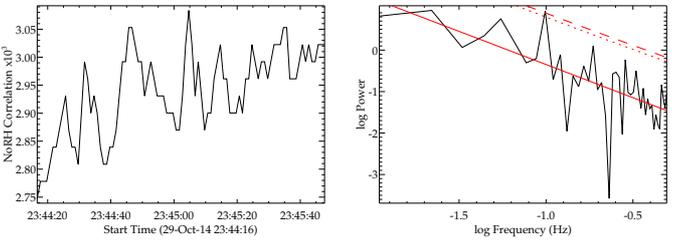}
	\caption{Similar to Fig. \ref{152goes}, with NoRH data for flare 138.}
	\label{138norh}
\end{figure}

\begin{figure}
	\centering
	\includegraphics[width=\linewidth]{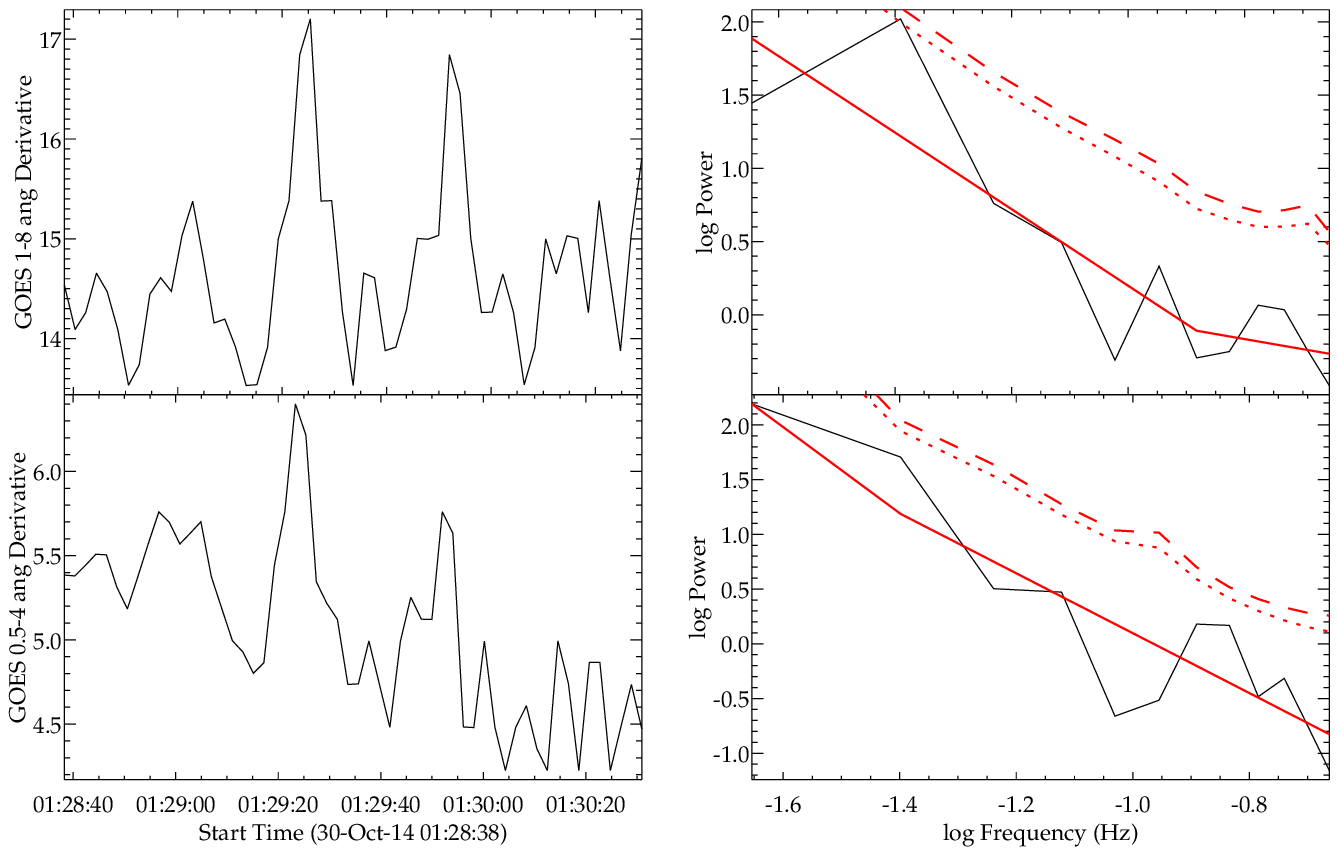}
	\caption{Similar to Fig. \ref{152goes}, with GOES/XRS data for flare 140.}
	\label{140goes}
\end{figure}

\clearpage

\begin{figure}
	\centering
	\includegraphics[width=\linewidth]{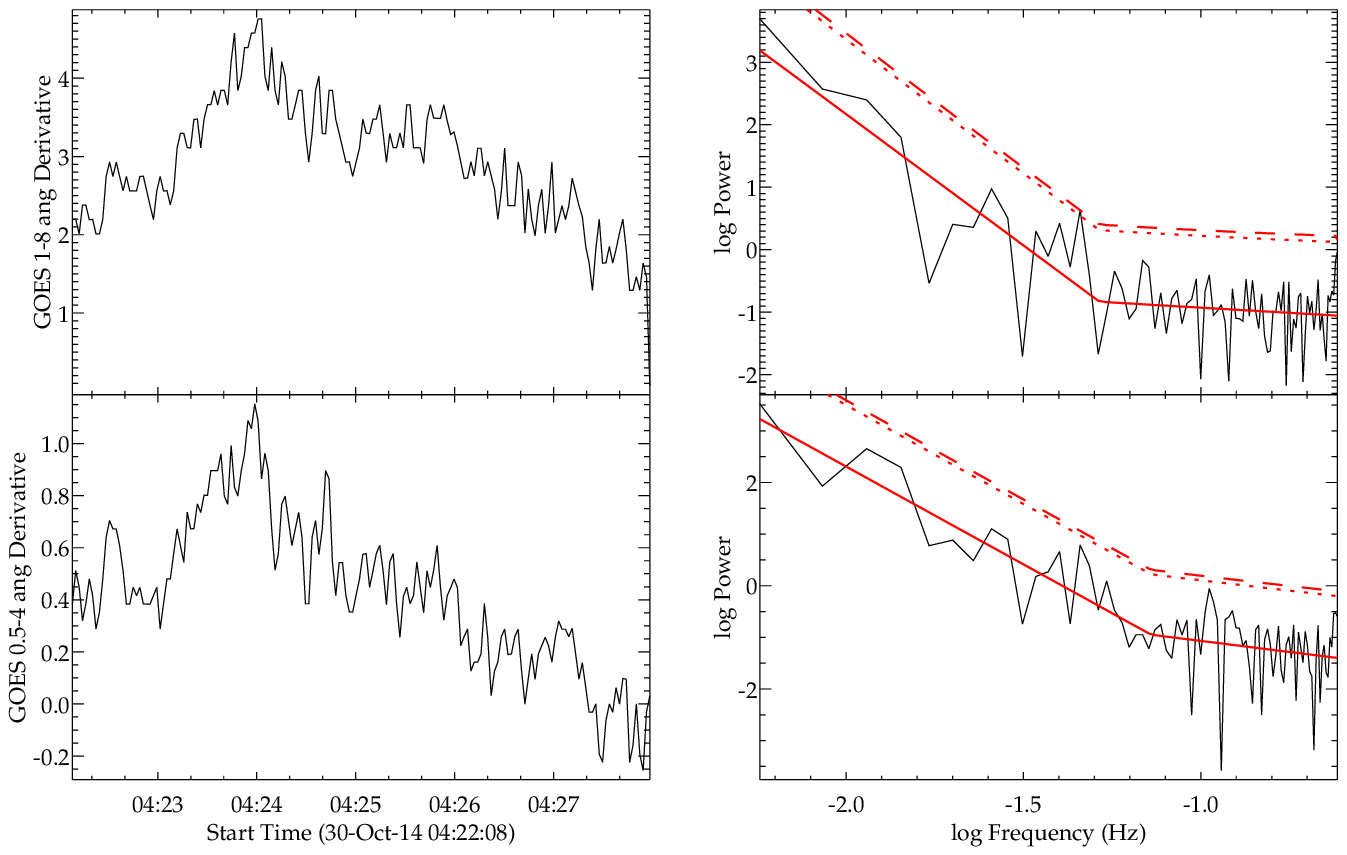}
	\caption{Similar to Fig. \ref{152goes}, with GOES/XRS data for flare 141.}
	\label{141goes}
\end{figure}

\begin{figure}
	\centering
	\includegraphics[width=\linewidth]{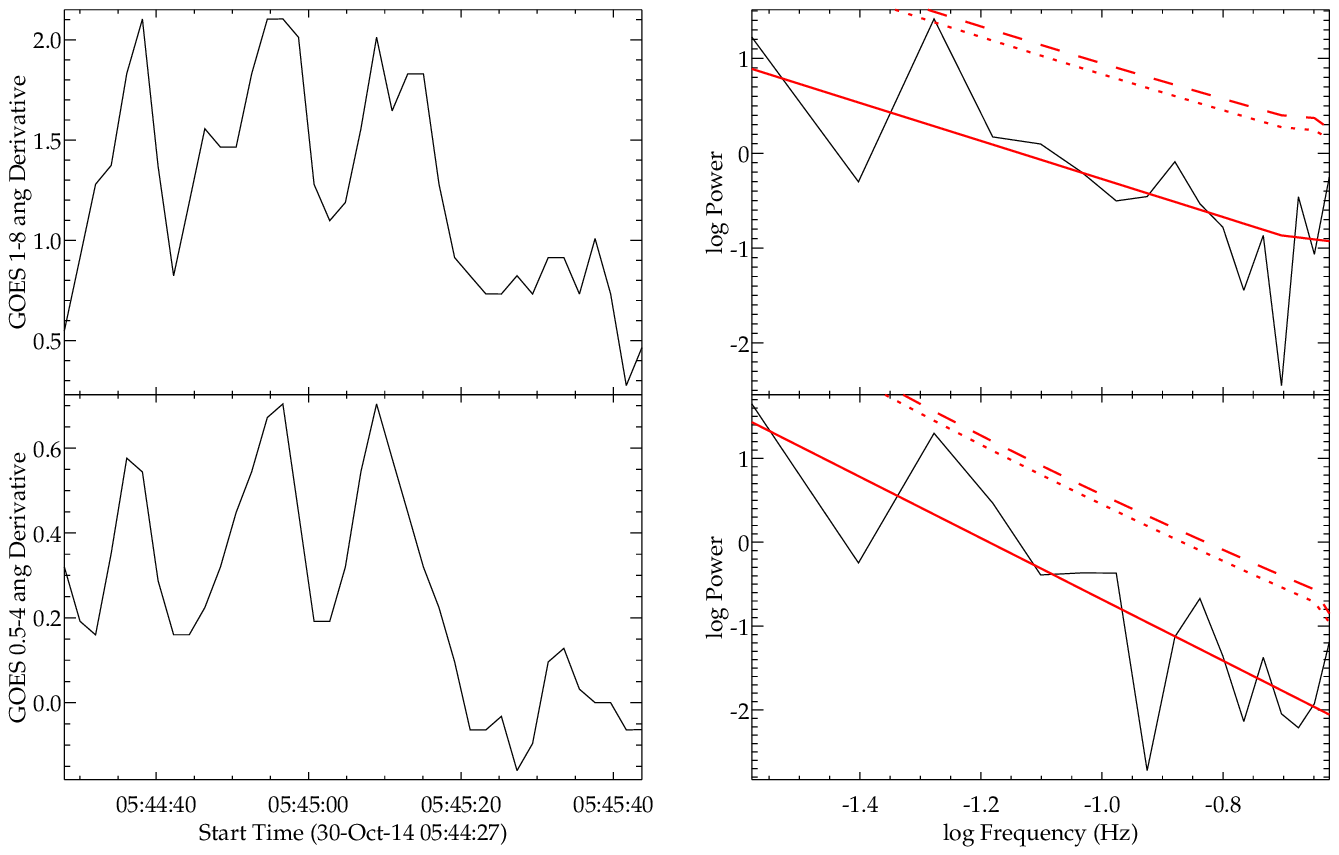}
	\caption{Similar to Fig. \ref{152goes}, with GOES/XRS data for flare 142.}
	\label{142goes}
\end{figure}

\begin{figure}
	\centering
	\includegraphics[width=\linewidth]{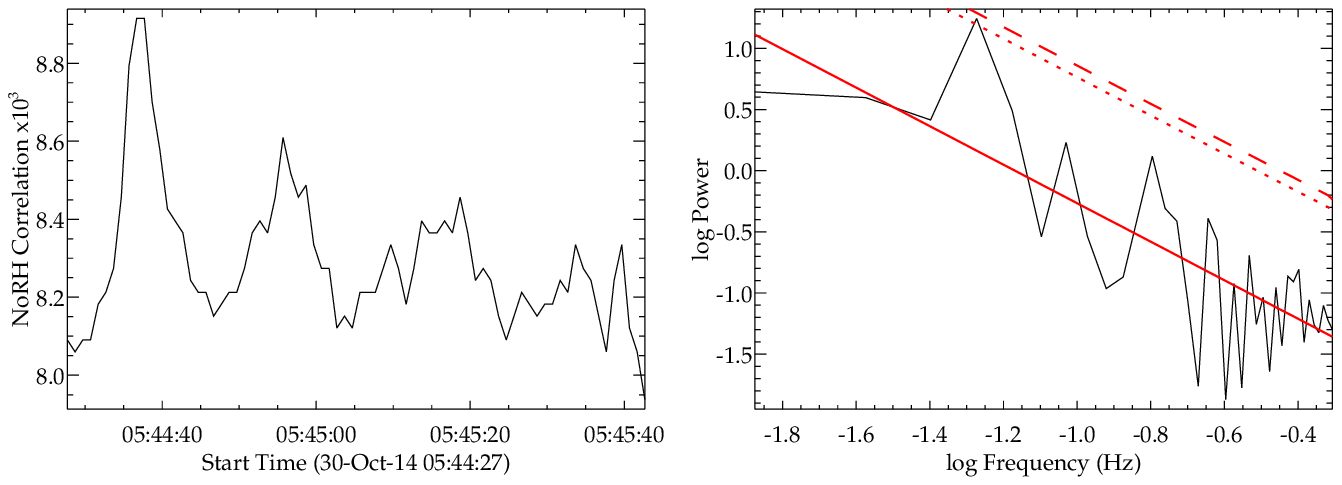}
	\caption{Similar to Fig. \ref{152goes}, with NoRH data for flare 142.}
	\label{142norh}
\end{figure}

\begin{figure}
	\centering
	\includegraphics[width=\linewidth]{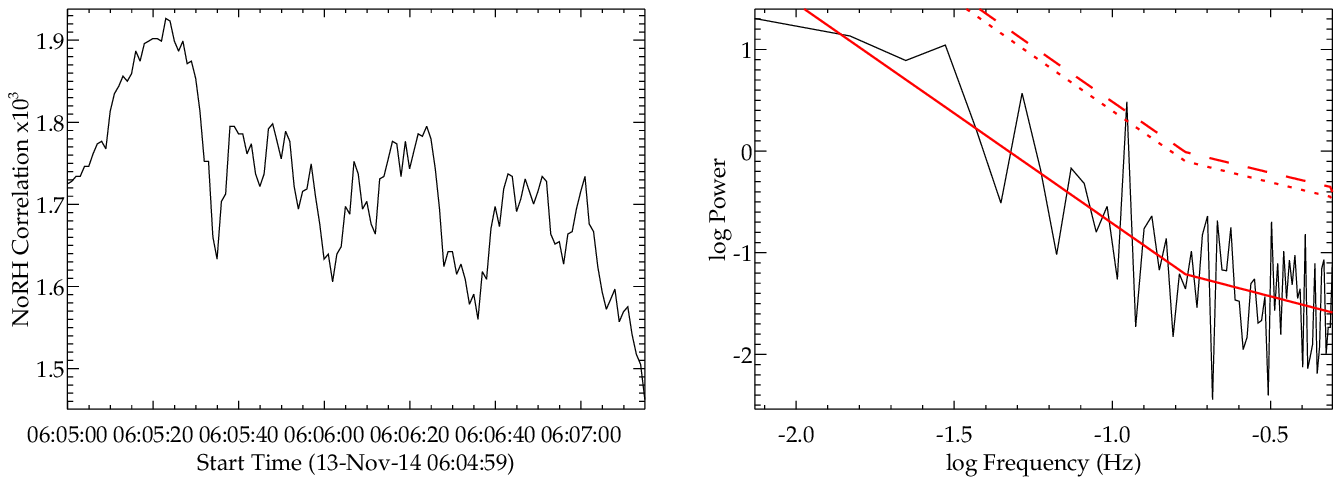}
	\caption{Similar to Fig. \ref{152goes}, with NoRH data for flare 147.}
	\label{147norh}
\end{figure}

\begin{figure}
	\centering
	\includegraphics[width=\linewidth]{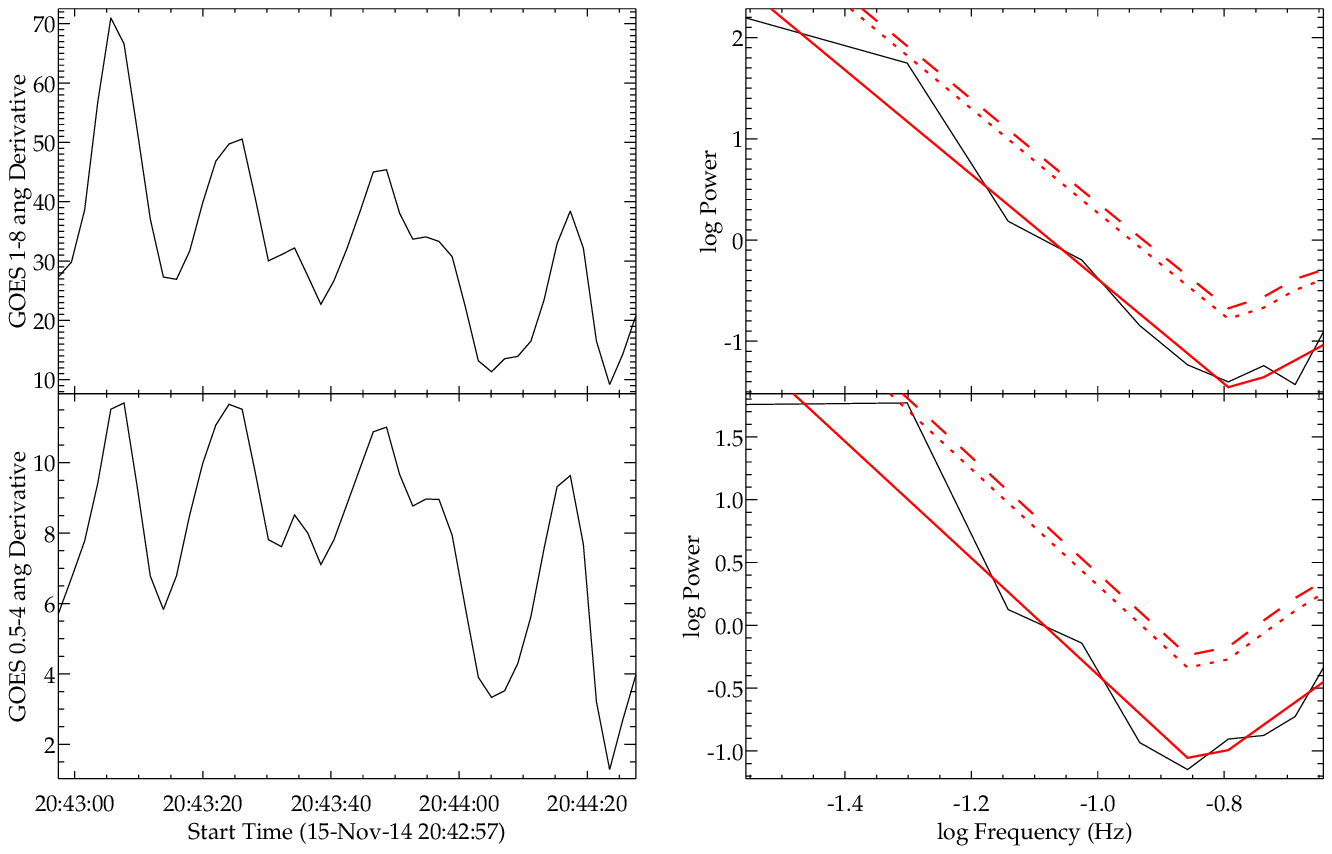}
	\caption{Similar to Fig. \ref{152goes}, with GOES/XRS data for flare 153.}
	\label{153goes}
\end{figure}

\begin{figure}
	\centering
	\includegraphics[width=\linewidth]{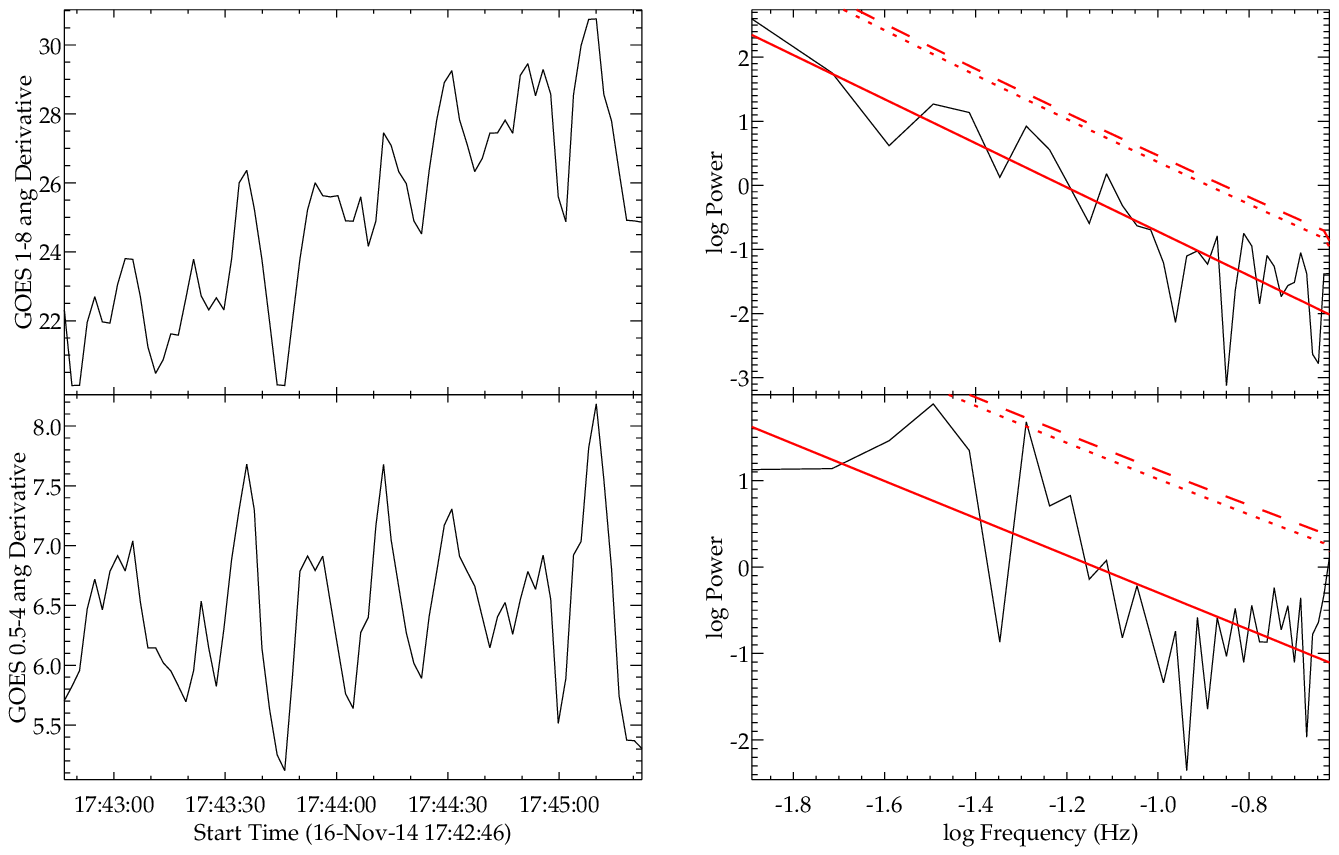}
	\caption{Similar to Fig. \ref{152goes}, with GOES/XRS data for flare 161.}
	\label{161goes}
\end{figure}

\begin{figure}
	\centering
	\includegraphics[width=\linewidth]{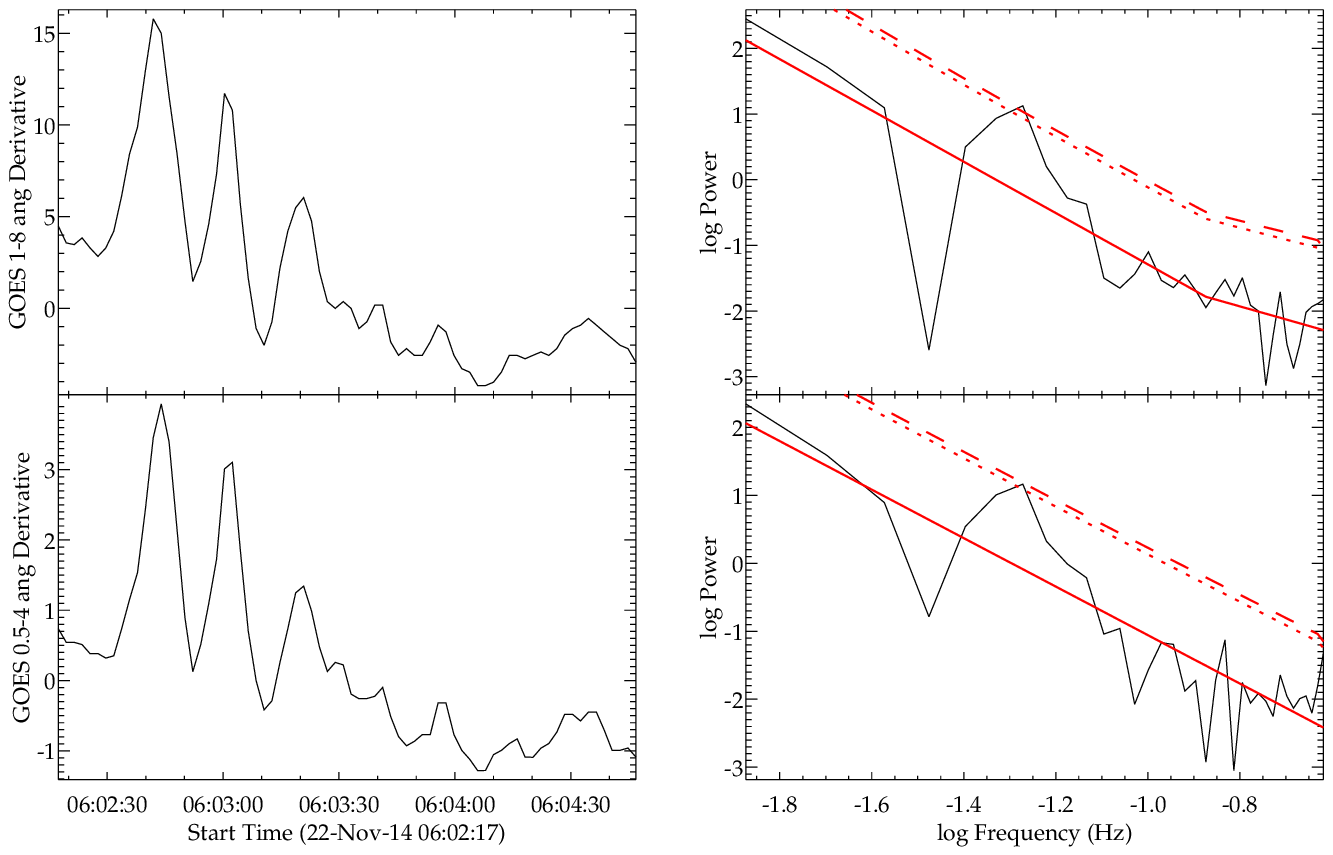}
	\caption{Similar to Fig. \ref{152goes}, with GOES/XRS data for flare 177.}
	\label{177goes}
\end{figure}

\end{appendix}

\end{document}